\documentclass[preprintnumbers,amsmath,amssymb,twocolumn,prb]{revtex4}

\usepackage{graphicx}
\usepackage{times}
\usepackage{amsmath}
\usepackage{amssymb}
\usepackage{bm}         
\usepackage{color}
\usepackage[utf8]{inputenc}
\usepackage[T1]{fontenc}

\begin{document}

\title{Flux-tunable supercurrent in full-shell nanowire Josephson junctions}

\author{G. Giavaras}\email{g.giavaras@gmail.com}
\author{R. Aguado}\email{ramon.aguado@csic.es}
\affiliation{Instituto de Ciencia de Materiales de Madrid, Consejo
Superior de Investigaciones Cient\'{i}ficas (ICMM-CSIC), Madrid,
Spain}

\begin{abstract}
Full-shell nanowires (a semiconducting core fully wrapped by an
epitaxial superconducting shell) have recently been introduced as
promising hybrid quantum devices. Despite this, however, their
properties when forming a Josephson junction (JJ) have not been
elucidated yet. We here fill this void by theoretically studying
the physics of JJs based on full-shell nanowires. In the
hollow-core limit, where the thickness of the semiconducting layer
can be ignored, we demonstrate that the critical supercurrent
$I^{\text{c}}$ can be tuned by an external magnetic flux $\Phi$.
Specifically, $I^{\text{c}}(\Phi)$ \emph{does not follow} the
Little-Parks modulation of the superconducting pairing
$\Delta(\Phi)$ and exhibits steps for realistic values of nanowire
radii. The position of the steps can be understood from the
underlying symmetries of the orbital transverse channels which
contribute to the supercurrent for a given chemical potential.
\end{abstract}

\maketitle

\emph{Introduction.--} The experimental demonstration of hybrid
semiconductor-superconductor Josephson junctions (JJs)
\cite{Doh272,Deng:12,Gharavi:14,Zuo:17,PhysRevB.100.155431,TiiraNC:17,
PhysRevB.100.064523,Carrad:20,Khan:20} has spurred a great deal of
research uncovering new physics of Andreev bound states (ABSs)
\cite{Sauls:PTRSA18}, including their spin splitting and
spin-orbit (SO) effects
\cite{PhysRevX.9.011010,Cayao:PRB15,Bargerbos:22,PhysRevLett.128.197702}
as well as their microwave response
\cite{PRXQuantum.3.030311,PhysRevLett.128.197702,PhysRevLett.129.227701,
PhysRevResearch.4.023170}. Moreover hybrid
semiconductor-superconductor JJs are being explored for novel
superconducting qubit applications \cite{Aguado:APL20} such as
gate-tunable transmon qubits
\cite{PhysRevLett.115.127001,PhysRevLett.115.127002,
PhysRevLett.116.150505,
PhysRevLett.126.047701,PhysRevLett.125.156804}, Andreev qubits
\cite{Hays:21,Pita-Vidal:22} and parity-protected qubits
\cite{PhysRevLett.125.056801,PRXQuantum.3.030303}. From a somewhat
different perspective, compatibility with high magnetic fields and
gate tunability hold promise for demonstrating topological quantum
computing based on Majorana zero modes
\cite{PhysRevX.6.031016,PhysRevB.95.235305,doi:10.1063/PT.3.4499}.

\begin{figure}
\includegraphics[width=8.0cm, angle=0]{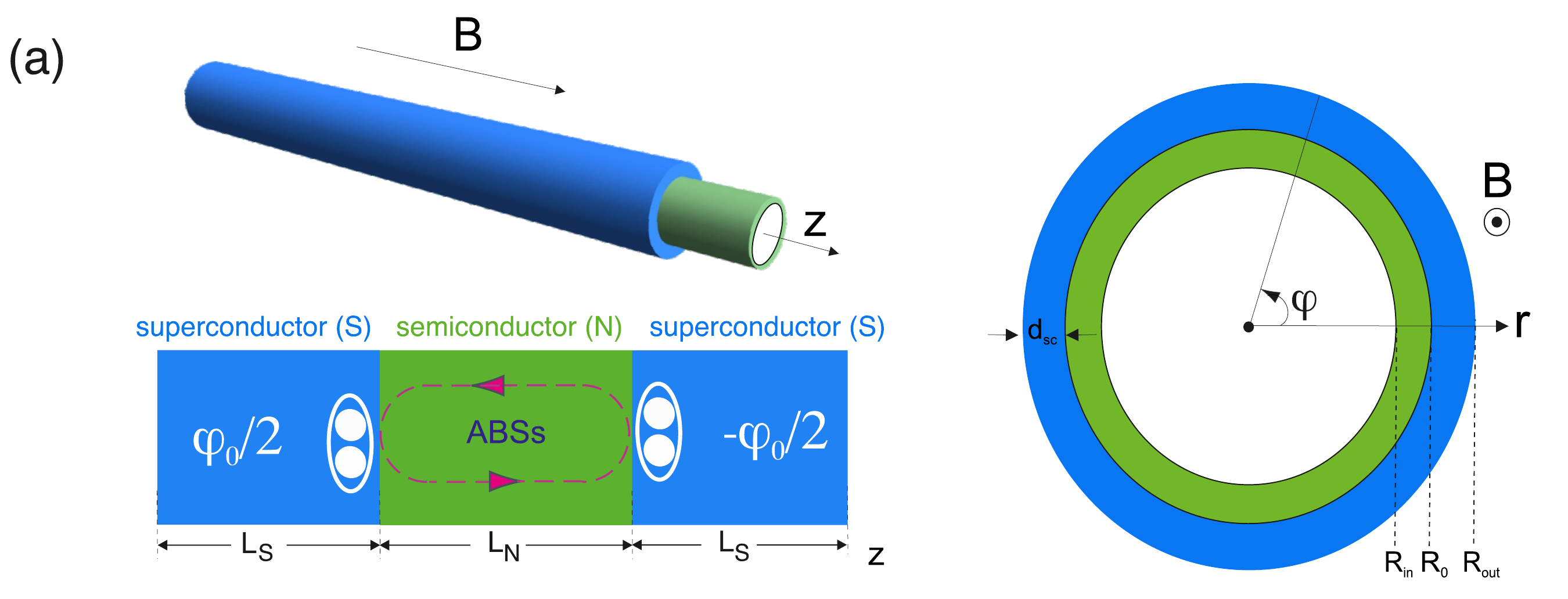}\\
\includegraphics[width=3.2cm, angle=270]{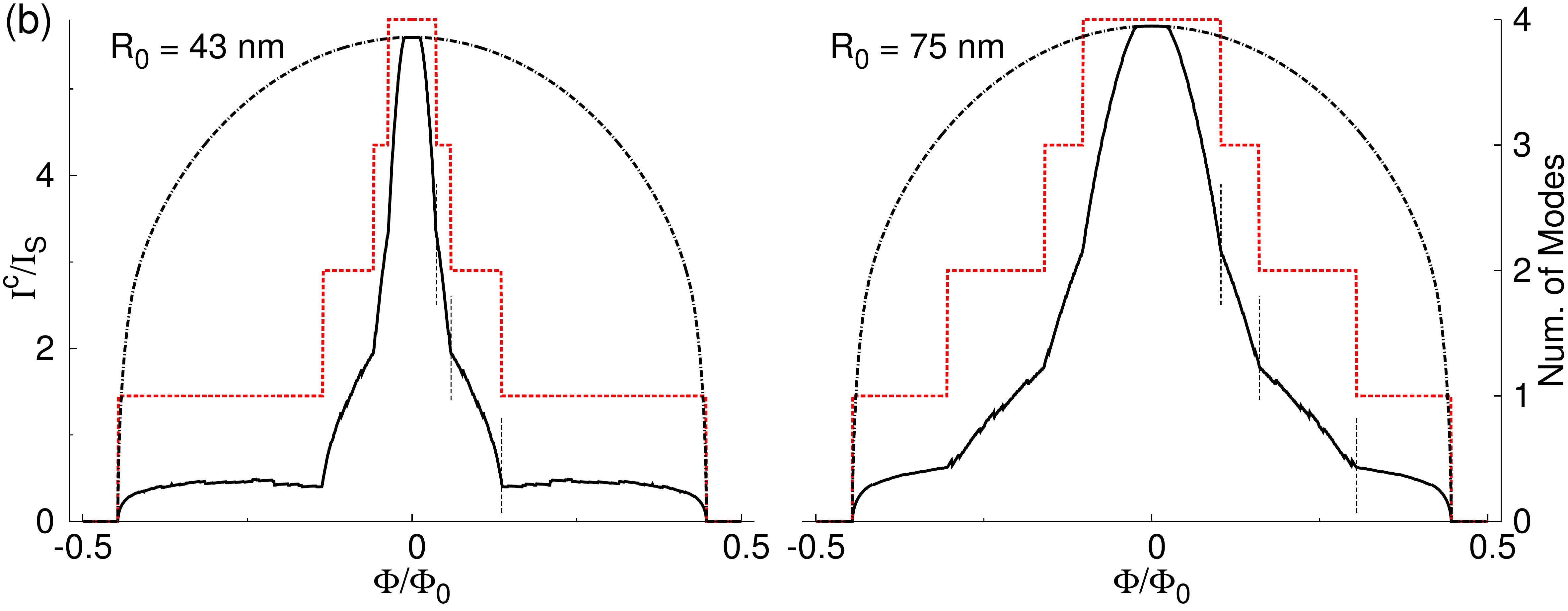}
\caption{(a) Left: Schematics of a semiconducting NW (green) fully
coated by a superconducting shell (blue) and threaded by a
longitudinal magnetic field $B$. Right: The cross section defines
an effectively insulating region (white) of radius $R_\text{in}$,
a semiconducting layer of thickness
$d_\text{semi}=R_0-R_\text{in}$ and a superconducting shell of
thickness $d_\text{sc}=R_\text{out}-R_0$. Within the hollow-core
approximation~\cite{Vaitiekenaseaav3392} the thickness of the
semiconducting layer is assumed to be $d_\text{semi}\approx 0$.
This fixes the radial coordinate $r=R_0$ and the external magnetic
flux $\Phi=\pi R_0^2B$. Bottom: Schematics of an SNS junction
where a normal (N) region of length $L_N$ is encapsulated between
two superconducting (S) regions of length $L_S$ and phase
difference $\varphi_0$. (b) Flux-tunable critical current,
$I^{\text{c}}$, (in units of $I_S = e \Delta_0/\hbar$) and number
of nondegenerate subgap modes (dotted lines, right axis) within
the zeroth lobe, $n=0$. By increasing the magnetic flux, $\Phi$,
the number of subgap modes contributing to $I^{\text{c}}$
decreases at each vertical line, `kink' point, giving rise to a
stepwise current profile. Dashdotted curves show
$I^{\text{c}}(\Phi)=I^{\text{c}}(0)\Delta(\Phi)/\Delta_0$. Similar
characteristics can be seen for $n\ne0$ lobes.}\label{fig1}
\end{figure}

Full-shell nanowires (NWs), where a semiconducting core fully
coated by an epitaxial superconductor \cite{Krogstrup:NM15} is
threaded by an external magnetic flux $\Phi$, have recently been
explored as a promising novel platform to generate Majorana zero
modes
\cite{Lutchyn:A18,Vaitiekenaseaav3392,PhysRevResearch.2.023171,Valentini:21,Valentini:22}.
Their interest, however, goes beyond Majorana physics since the
full-shell geometry gives rise to a great deal of new physics,
including nontrivial $\Phi$-dependent superconductivity
\cite{PhysRevB.101.060507} owing to the Little-Parks (LP) effect
\cite{PhysRevLett.9.9,PhysRev.133.A97}, as well as analogs of
subgap states in vortices \cite{PhysRevB.101.054515,San-Jose:22}.

While JJs based on full-shell NWs start to attract experimental
attention \cite{PhysRevLett.125.156804, PhysRevLett.125.116803,
PhysRevLett.126.047701, Ibabe:22}, a theoretical understanding is
still lacking. The purpose of this Letter is to fill this void by
presenting calculations of superconductor-normal-superconductor
(SNS) junctions based on hollow-core full-shell NWs [Figs.~1(a)].
Our main result is the demonstration of critical supercurrent,
$I^{\text{c}}$, tunability as a function of $\Phi$ [Fig.~1(b)].
Specifically, we find a stepwise decrease with $\Phi$-dependent
features which can be analytically understood in terms of the
underlying orbital degeneracies and symmetries of the ABS
spectrum. The $\Phi$-dependence reported here is \emph{completely
unrelated} to the LP modulation of the superconducting pairing
gap, and can be observed even when there is no LP modulation. In
stark contrast to previously reported flux-induced supercurrents
which are LP-dominated~\cite{PhysRevLett.125.156804, vekris21}.
Our findings have important implications in recently proposed
transmon qubit designs~\cite{PhysRevLett.125.156804}, where flux
tunability could offer further functionalities.

\emph{Nanowire model.}-- We first consider a cylindrical
semiconducting NW, unit vectors $({\hat e_r},{\hat
e_\varphi},{\hat e_z})$, in the presence of an axial magnetic
field ${\vec B}=B{\hat e_z}$ and with finite SO coupling. Assuming
that the electrons are strongly confined near the surface of the
NW (hollow-core approximation~\cite{Vaitiekenaseaav3392}), we fix
the radial cylindrical coordinate $r=R_0$ [Fig.~1(a)], thus the
flux that threads the NW cross-section is $\Phi=\pi B R_0^2$ and
the vector potential is $\vec{A}=A_{\varphi}{\hat
e_\varphi}=\frac{\Phi}{2\pi R_0} {\hat e_\varphi}$. The
Hamiltonian then reads
\begin{equation}\label{NW}
H_0(\vec{A})=\frac{(\vec{p}+e A_{\varphi}{\hat e_\varphi}
)^2}{2m^*}-\mu+H_{\text{SO}},
\end{equation}
with
$\vec{p}=(p_\varphi,p_z)=(-\frac{i\hbar}{R_0}\partial_\varphi,-i\hbar\partial_z)$
being the momentum operator, $m^{*}$ the effective mass and $\mu$
the chemical potential. Assuming radial inversion symmetry
breaking, namely $\vec{\alpha}=\alpha{\hat e_r}$, the Rashba SO
Hamiltonian is \cite{PhysRevB.83.115305} $H_{\text{SO}}=
H^z_{\text{SO}}+H^\perp_{\text{SO}}=\frac{\alpha}{\hbar}
\left[p_z\sigma_\varphi-(p_\varphi+eA_{\varphi})\sigma_z\right]$,
with the spin-1/2 Pauli matrices $\sigma_\varphi=\sigma_y
\cos(\varphi)-\sigma_x \sin(\varphi)$, $\sigma_z$ and the SO
coupling $\alpha$.

Owing to the proximity effect, the semiconducting core acquires
superconducting pairing terms. Importantly, they are modulated by
$\Phi$ through the LP effect~\footnote{Here, we implicitly assume
that the coupling between the semiconductor and the superconductor
is strong, such that the proximity-induced pairing terms in the
semiconductor inherit the LP effect of the superconducting
shell.}, which induces a winding of the superconducting phase in
the shell around the nanowire axis $\bm\Delta=\Delta e^{i
n\varphi}$. Both the amplitude $\Delta$ and the winding (fluxoid)
number $n$ depend implicitly on $\Phi$ (Appendix A). 
Defining the normalized flux $n_\Phi=\Phi/\Phi_0$, with
$\Phi_0=h/2e$, the winding number reads $n = \left
\lfloor{n_\Phi}\right \rceil$. Thus, we measure deviations from
integer fluxes through the variable $\phi = n -n_\Phi$, with
$\phi=0$ corresponding to the middle of a so-called LP
lobe~\cite{PhysRevB.101.060507}. In what follows
$\Delta(\Phi=0)\equiv \Delta_0$.

In the Nambu basis $\Psi=(\psi_{\uparrow},
\psi_{\downarrow},\psi^\dag_{\downarrow}, -\psi^\dag_{\uparrow})$,
the Bogoliubov-de-Gennes (BdG) Hamiltonian $H_{\rm BdG}$ can be
decomposed as a set of decoupled one-dimensional
models~\cite{PhysRevB.100.155431,
PhysRevB.74.245327,Richter:NL08,PhysRevB.83.115305,PhysRevB.91.045422}
labelled in terms of the eigenvalues $m_j$ of a generalized
angular momentum operator~\cite{Lutchyn:A18} $\hat
J_z(n)=-i\partial_\varphi+\frac{1}{2}\sigma_z+\frac{1}{2}n
\tilde{\tau}_z$, with $\tilde{\tau}$ acting in Nambu space.
Physically acceptable wavefunctions require
that~\cite{Vaitiekenaseaav3392} $m_j=\pm 1/2$, $\pm 3/2$, $\ldots$
for even $n$ and $m_j=0$, $\pm1$, $\ldots$ for odd $n$. The
resulting BdG Hamiltonians~\footnote{Without SO coupling and LP
effect, Eq.~(\ref{Hmatrix}) reduces to the model used in
Ref.~\onlinecite{PhysRevB.100.155431} to study supercurrent
interference in JJs based on cylindrical semiconducting NWs.
Without pairing Eq.~(\ref{Hmatrix}) is the BdG (electron-hole
redundant) analog of the model used to study conductance
oscillations in semiconducting core-shell NWs
\cite{PhysRevB.74.245327,Richter:NL08,PhysRevB.83.115305,PhysRevB.91.045422}.}
can be conveniently written as:
\begin{equation}\label{Hmatrix}
H_{\rm BdG} = \left(\begin{array}{cc}
   H_A  & H^z_{\text{SO}} \\
  -H^z_{\text{SO}} & H_B  \\
\end{array}\right),
\end{equation}
with $H^z_{\text{SO}}=-\alpha\partial_z\tilde{\tau}_z$, and
\begin{equation}\label{HA}
H_A = \left(\begin{array}{cc}
  \frac{p_z^2}{2m^*} + V_1 & \Delta\\
\Delta& -\frac{p_z^2}{2m^*} + V_2    \\
\end{array}\right).
\end{equation}
The Hamiltonian $H_B$ is obtained by substituting $V_1\rightarrow
V_3$, $V_2\rightarrow V_4$, and the potential terms are
($\hbar=1$)
\begin{eqnarray}
V_{1(2)}(\phi) &=& V^0_{1(2)}+ \delta^+_{m_j}(\phi) \pm \frac{1}{8m^{*}R^{2}_{0}} \phi^2  ,\nonumber\\
V_{3(4)}(\phi) &=& V^0_{3(4)}+ \delta^-_{m_j}(\phi) \pm
\frac{1}{8m^{*}R^{2}_{0}} \phi^2.
\end{eqnarray}
At $\phi=0$, the effective potentials are $V^0_{1(3)} = - \mu +
\frac{(1\mp 2m_j)^2}{8m^{*}R_0^{2}} + \frac{(1\mp
2m_j)\alpha}{2R_0}$, with $V^0_2 = -V^0_1$ and $V^0_4 = -V^0_3$.
Small deviations from integer fluxes ($\phi=0$) are captured by
the linear terms
\begin{equation}\label{competition}
\delta^\pm_{m_j}(\phi)  = \frac{-2m_j \pm 1}{4m^{*}R^{2}_0}\phi
\pm \frac{\alpha}{2 R_0}\phi.
\end{equation}

At $\phi=0$, the terms $V^0_{1}(m_j)$ and $V^0_{3}(m_j)$ govern
the chemical potential at which the levels belonging to $H_A$ and
$H_B$ respectively cross zero energy. The important quantity is
the chemical potential difference which depends on the angular
motion and is equal to $V^0_{3}(m_j)-V^0_{1}(m_j) = m_j/m^{*}R_0^2
+ 2m_j \alpha/R_0$. Interestingly, when $\alpha=0$ the energy
levels of $H_{A}(m_j+1)$ and $H_{B}(m_j)$ are degenerate because
$V^0_{1}(m_j+1)=V^0_{3}(m_j)$, therefore, levels belonging to
different $m_j$ modes cross zero energy simultaneously.

\begin{figure}
\includegraphics[width=3.0cm, angle=270]{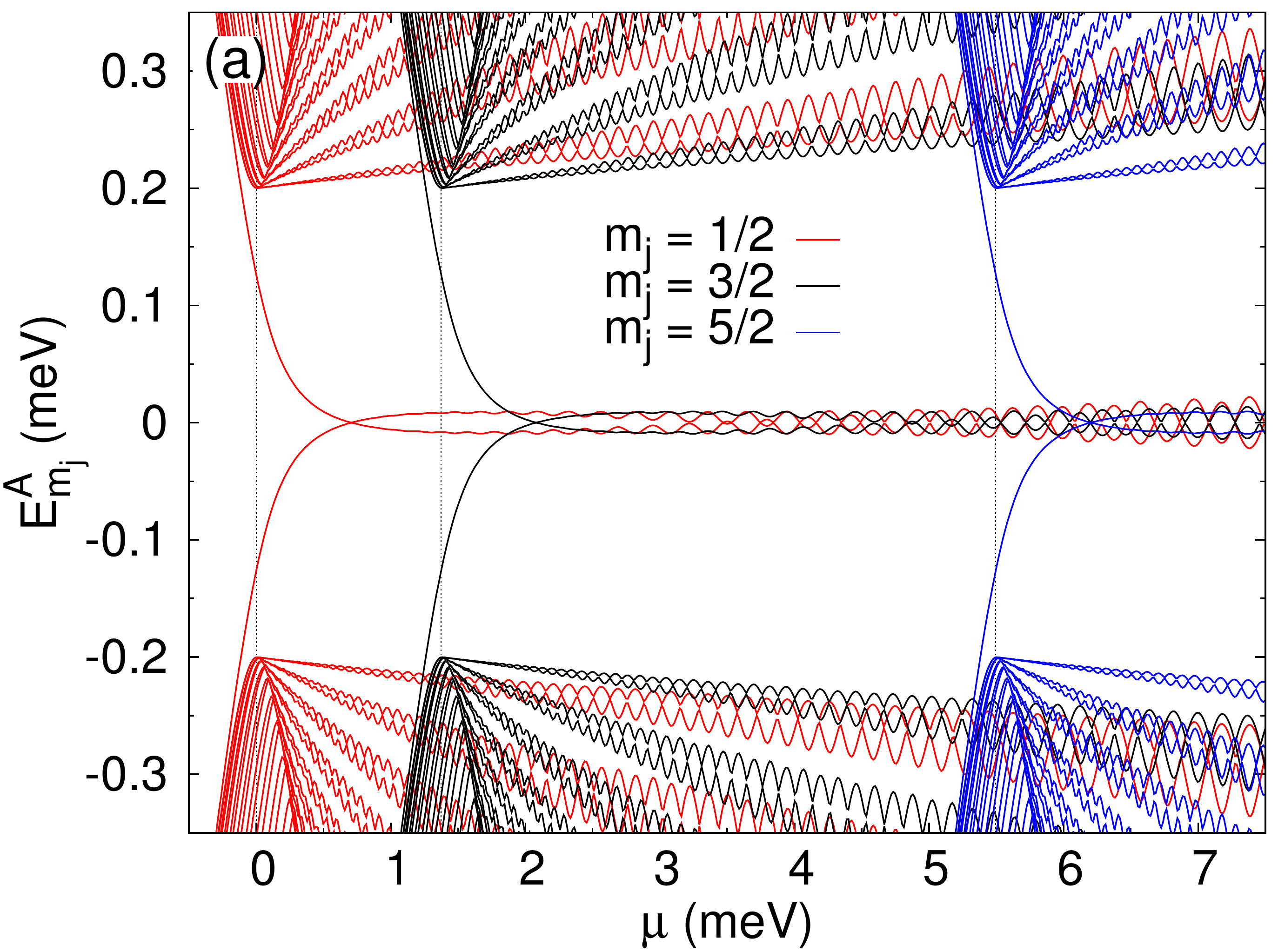}
\includegraphics[width=3.0cm, angle=270]{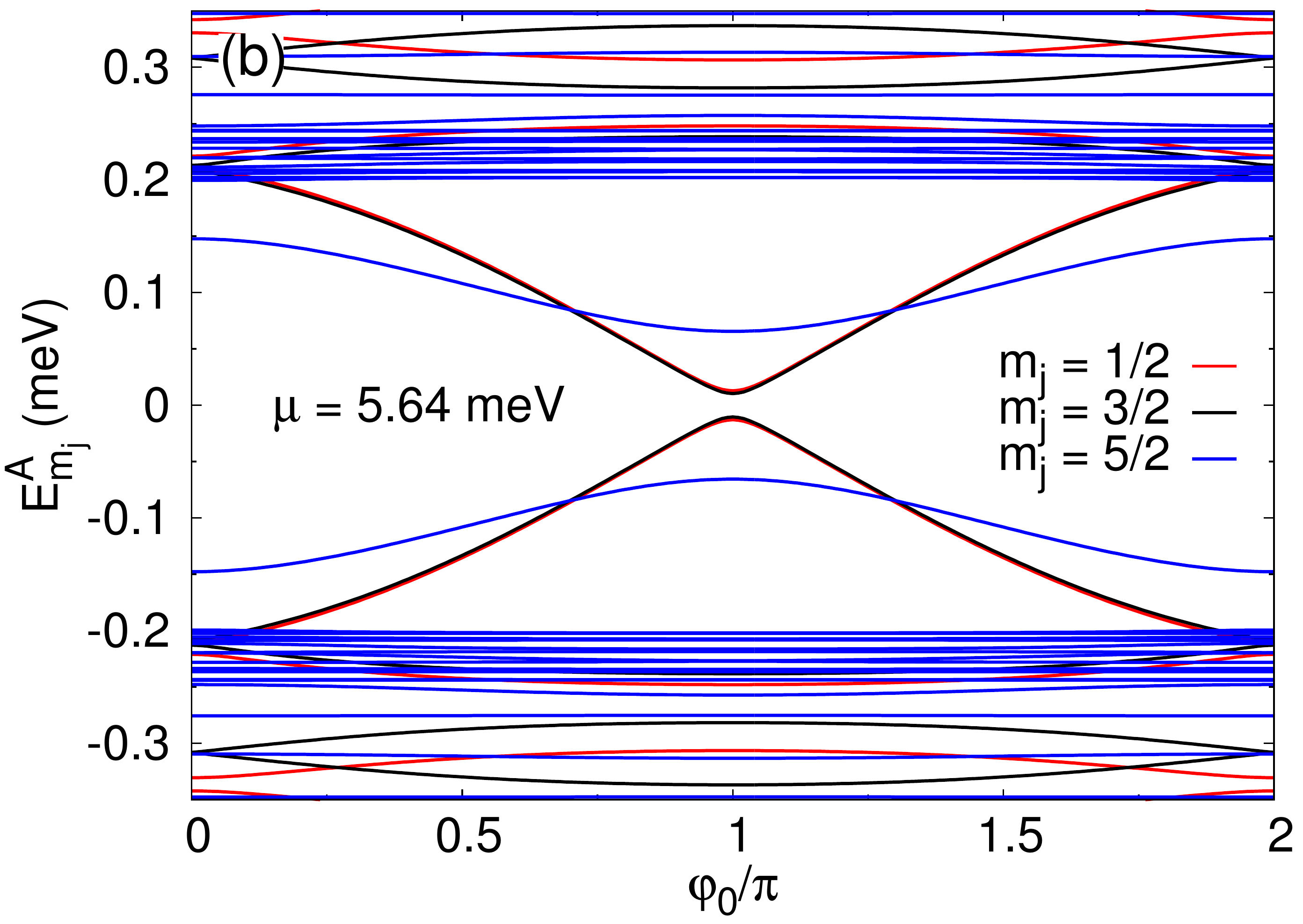}\\
\includegraphics[width=4.0cm, angle=270]{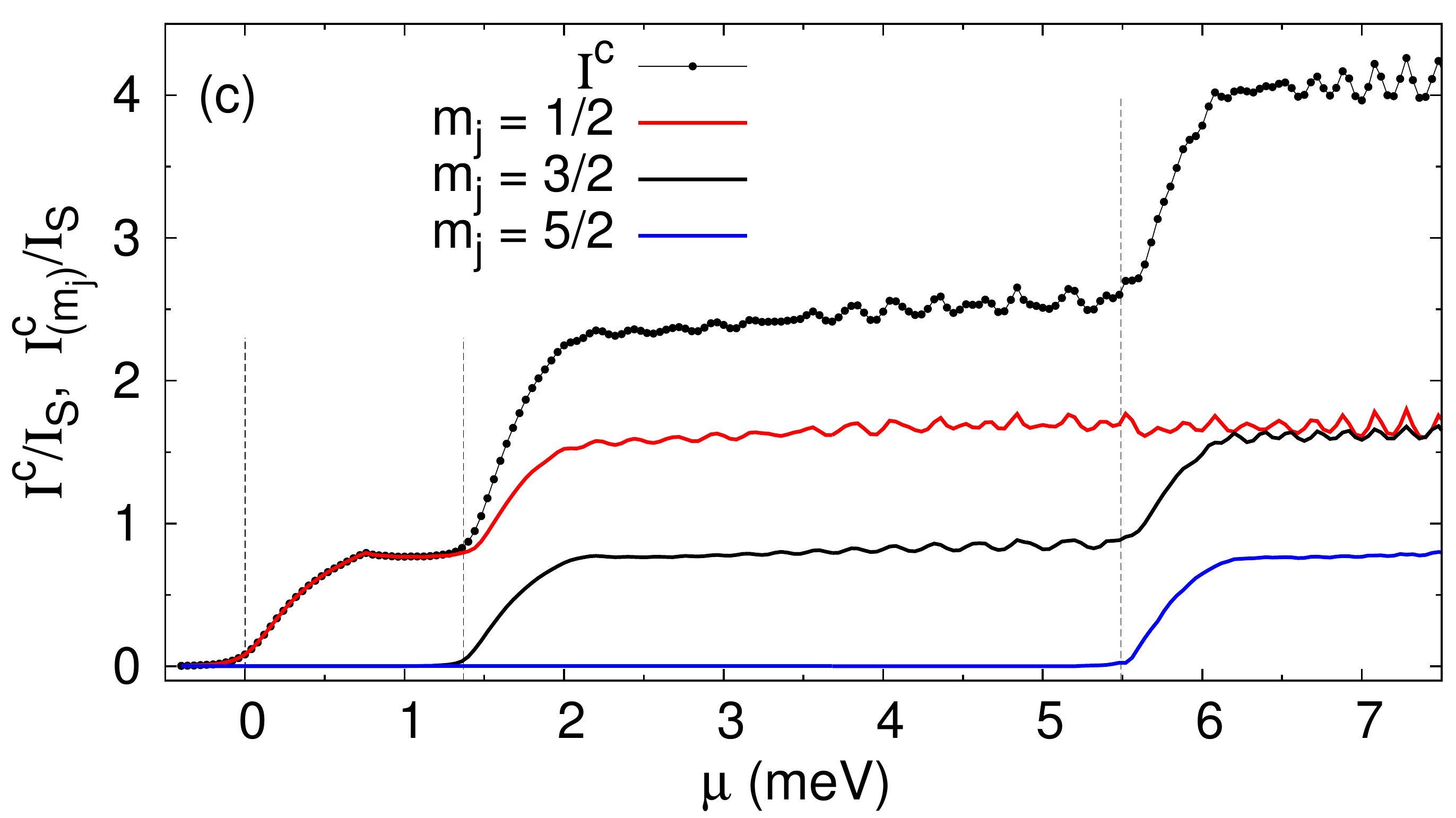}
\caption{(a) Zero-flux energies of $H_A(m_j)$ as a function of
chemical potential at $\varphi_0=\pi$ with $E^{A}_{-m_j} = -
E^{B}_{m_j}$ and $E^{B}_{m_j} = E^{A}_{m_j+1}$. As $\mu$ increases
the vertical lines define the effective potential
$V^{0}_1(m_j)\approx 0$ for $m_j=1/2$, 3/2, 5/2 respectively. The
small oscillatory dependence around $E^{A}_{m_j}=0$ is generic in
SNS junctions governed by the BdG equation. (b) Energies of
$H_A(m_j)$ as a function of $\varphi_0$. There are in total 5
positive subgap levels dispersing with $\varphi_0$: $E^{A}_{1/2}$,
$E^{B}_{1/2}=E^{A}_{3/2}$ and $E^{B}_{3/2}=E^{A}_{5/2}$. Any other
pair of levels has a non-zero spacing but this can be too small to
resolve it. The degeneracies are lifted for $\alpha\neq 0$
(Appendix E). (c) Critical current, $I^{\text{c}}$, and $m_j$
contributions $I^{\text{c}}_{(m_j)} = I^{\text{c}}_{-m_j} +
I^{\text{c}}_{m_j}$. Vertical lines are the same as in (a).
Parameters: $L_S=2000$ nm, $L_N=100$ nm, $R_0=43$ nm, $\alpha=0$,
$\Delta=\Delta_0=0.2$ meV and $I_S = e
\Delta_0/\hbar$.}\label{fig2}
\end{figure}

\begin{figure}
\includegraphics[width=3.0cm, angle=270]{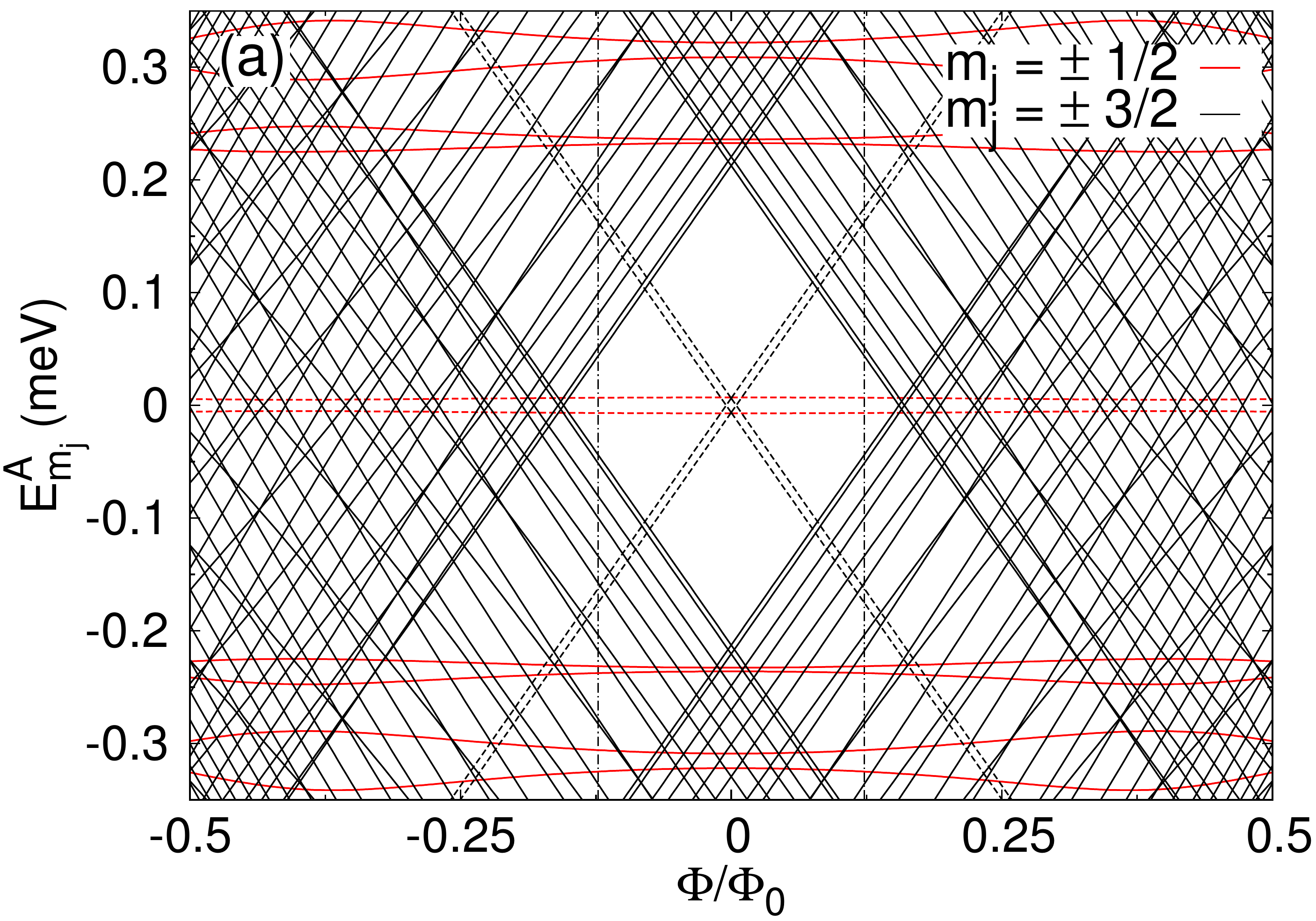}
\includegraphics[width=3.0cm, angle=270]{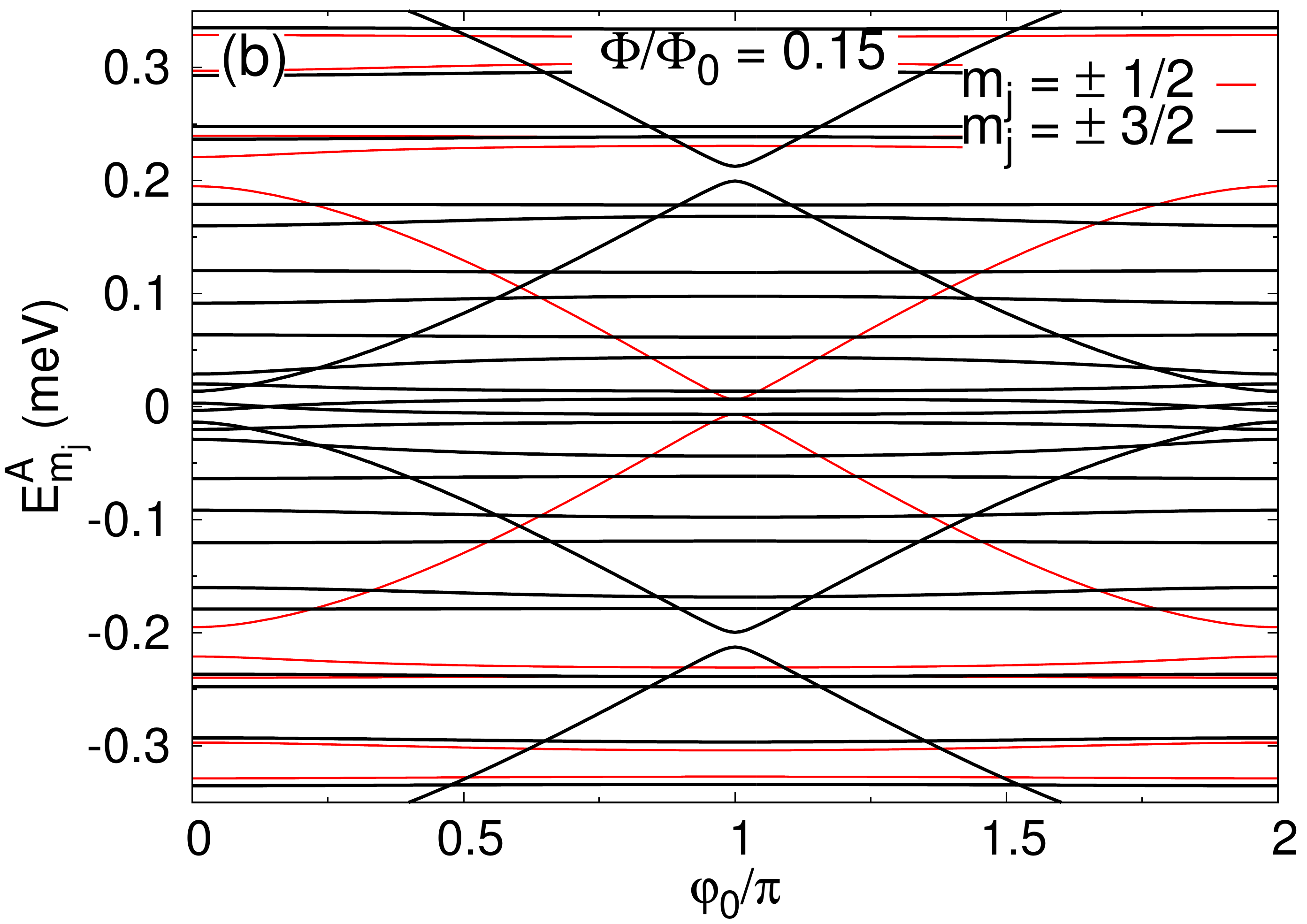}\\
\includegraphics[width=4.0cm, angle=270]{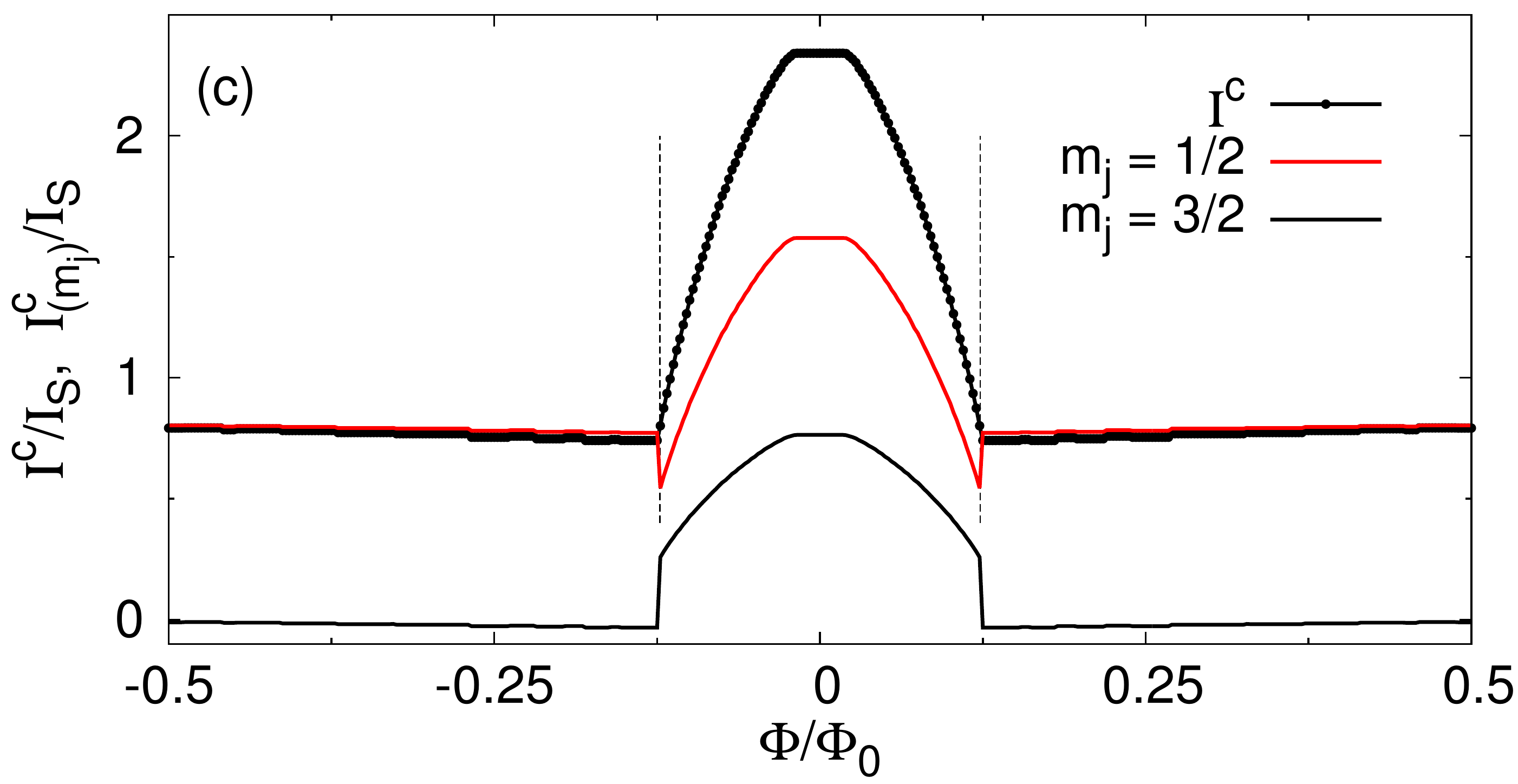}
\caption{(a) Energies of $H_A(m_j)$ as a function of magnetic flux
at $\varphi_0=\pi$ with $E^{A}_{-m_j} = - E^{B}_{m_j}$ and
$E^{B}_{m_j} = E^{A}_{m_j+1}$. Dotted lines show how the
anticrossing points at $E^{A}_{m_j}\approx 0$, $\Phi=0$ evolve
with flux. Vertical lines define the flux, $\Phi/\Phi_0 = \pm
0.125$, where the contribution of a subgap mode of $H_B(m_j)$ or
$H_A(m_j+1)$ is suppressed. (b) Energies of $H_A(m_j)$ as a
function of $\varphi_0$ at $\Phi/\Phi_0=0.15$. (c) Critical
current, $I^{\text{c}}$, and $m_j$ contributions
$I^{\text{c}}_{(m_j)} = I^{\text{c}}_{-m_j} + I^{\text{c}}_{m_j}$.
Vertical lines are the same as in (a). Parameters: $L_S=2000$ nm,
$L_N=100$ nm, $R_0=43$ nm, $\alpha=0$, $\Delta=\Delta_0=0.2$ meV,
$\mu=2.5$ meV and $I_S = e \Delta_0/\hbar$.}\label{fig3}
\end{figure}

\emph{SNS junction.--} The SNS junction is defined by including a
spatial dependence of the pairing potential of the form
$\Delta_{R/L}=\Delta e^{i (n\varphi\pm\varphi_0/2)}$, where
$\varphi_0$ is the superconducting phase difference and R/L
denotes two right/left superconducting (S) regions of length
$L_S$. The normal (N) region is defined as $\Delta_{N}=0$ within a
length $L_N$ \cite{Cayao2018}. For simplicity, we assume in the
main text that $\mu$ is position-independent (uniform) along the
$z$ direction. In a realistic experimental implementation,
however, the superconducting full shell is expected to screen any
external electric field making gating only effective in the N
region. This configuration can be modelled as a smooth spatial
variation of the chemical potential in the N region (Appendix B).
A uniform chemical potential results in the
maximum critical current, whereas the current is reduced as the
potential offset between the N and S regions increases.

The supercurrent-phase relationship $I(\varphi_0)$ can be written
in terms of independent contributions for each angular number
$m_j$. Assuming zero temperature, it reads
\begin{equation}\label{formula}
I(\varphi_0) = \sum_{m_j}
I_{m_j}(\varphi_0)=-\frac{e}{\hbar}\sum_{m_j}\sum_{k>0}\frac{dE_{k,m_j}}{d\varphi_0},
\end{equation}
where $E_{k,m_j}$ are the positive BdG eigenvalues which are
computed numerically by discretizing the SNS junction
\footnote{Discretizing on a lattice with uniform spacing, the
continuum BdG eigenvalue problem is transformed to a matrix
eigenvalue problem which is solved by standard numerical
routines.}. The critical current is
$I^{\text{c}}=\text{max}[I(\varphi_0)]$ and can be decomposed into
different orbital components $I^{\text{c}}_{m_j}$.

\emph{Zero-flux SNS junction.--} We start the analysis of the SNS
junction for $\Phi=0$, and plot the energies of $H_A(m_j)$ in
Fig.~\ref{fig2}(a). BdG levels are induced in a systematic way in
the superconducting gap by increasing the chemical potential
$\mu$, hence tuning the number of active modes in the junction.
The vertical lines in Fig.~\ref{fig2}(a) correspond to the
effective potential $V^{0}_1(m_j)\approx 0$ specifying the
required $\mu$ that shifts an extra mode into the gap. Because of
the relatively small normal region considered here, $L_N=100$ nm,
each of $H_A(m_j)$ and $H_B(m_j)$ can contribute a single subgap
mode. The degree of $\varphi_0$-dispersion depends on the exact
value of $\mu$ and some subgap levels can be quasi-degenerate
[Fig.~\ref{fig2}(b)].

Figure~\ref{fig2}(c) illustrates the critical current together
with the $m_j$ contributions. The vertical lines have the same
meaning as in Fig.~\ref{fig2}(a), so when an extra energy level
shifts into the gap the critical current exhibits a noticeable
increase. Small fluctuations of the current, which are more
pronounced at larger $\mu$ values, are due to the small variations
of the subgap levels at $E^{A}_{m_j}\approx 0$ as shown in
Fig.~\ref{fig2}(a). The details of the current profile depend on
the characteristic length scales ($R_0$, $L_S$, $L_N$) of the SNS
junction. A smooth barrier-like local potential $\mu(z)$ on top of
the global $\mu$ allows us to precisely control the number of
active modes by depleting the N region (Appendix B).

\emph{Flux tunable critical current.--} We proceed to study finite
flux effects for an SNS junction governed by Eq.~(\ref{Hmatrix}).
At small fluxes the linear terms $\delta^{\pm}_{m_{j}}(\phi)$
[Eq.~(\ref{competition})] are the dominant ones, and produce a
shift of the corresponding zero-flux energies. The main physics is
illustrated in Fig.~\ref{fig3}(a), where we plot the energies of
$H_A(m_j)$ as a function of $\Phi$ when only $m_j=1/2$, $3/2$ are
relevant. The key feature here is that by increasing $\Phi$ the
$\Phi=0$ subgap levels which anticross ($\varphi_0=\pi$,
$E^{A}_{m_j}\approx0$) shift in the quasi-continuum. Although,
these levels still anticross when $\Phi\ne 0$, the anticrossing
point gradually shifts outside the gap. This is demonstrated
clearer in Fig.~\ref{fig3}(b) where the energies of $H_A(m_j)$ are
plotted versus $\varphi_0$. The anticrossing lying outside the gap
is due to $E^{B}_{1/2}$, $E^{A}_{3/2}$,
[$\delta^{-}_{1/2}(\phi)=\delta^{+}_{3/2}(\phi)\ne0$] whereas the
anticrossing lying in the gap is due to $E^{A}_{1/2}$ for which
$\delta^{+}_{1/2}(\phi)=0$. The required magnetic flux to suppress
the contribution of a subgap mode of $H_B(m_j)$ or $H_A(m_j+1)$ is
of the order of $(2m_j + 1)|\phi| \approx 4 \Delta m^{*}R^{2}_0$,
therefore, larger $m_j$ subgap modes are suppressed at smaller
flux values. This simplified approach assumes that $\mu$ is large
enough so that the corresponding anticrossing point lies at zero
energy. In Fig.~\ref{fig3}(c), $\Phi/\Phi_0 \approx \pm 0.125$
defines a crossover, kink point, which is formed so long as a
subgap mode is suppressed, and then the current versus flux drops
at a smaller overall rate. For the particular example in
Fig.~\ref{fig3}(c) $\delta^{+}_{1/2}(\phi)=0$, so this rate is
zero but as shown below the physics is more interesting for larger
$m_j$ values.

\begin{figure}[t]
\includegraphics[width=10.5cm, angle=270]{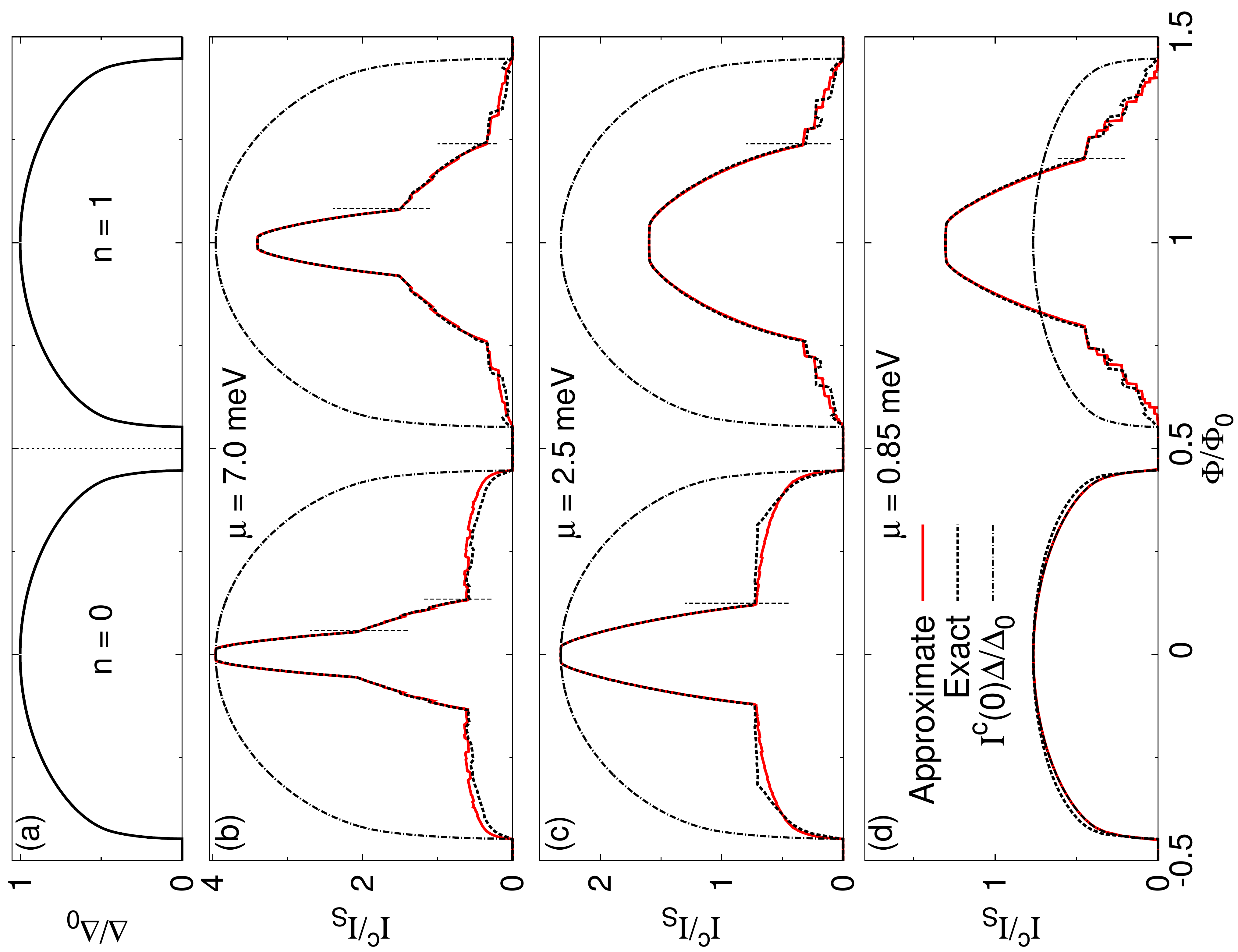}
\caption{(a) Pairing potential versus flux for lobes $n=0$ and
$n=1$. (b), (c), (d) Exact and approximate currents calculated
with $\Delta$ as in (a). Vertical lines indicate kink points where
the number of non-degenerate subgap modes decreases by one. For
$n=0$, $\mu=7$ meV and $\Phi=0$, there are 3 subgap modes coming
from $m_j=1/2$, $3/2$, $5/2$. For $\Phi\ne0$ two kink points are
formed when $E^{A}_{5/2}$ and $E^A_{3/2}$ shift successively
outside the gap. For $n=0$, $\mu=2.5$ meV and $\Phi=0$, there are
2 subgap modes coming from $m_j=1/2$, $3/2$, thus, for $\Phi\ne0$
one kink point is formed when $E^A_{3/2}$ shifts outside the gap.
For $n=0$, $\mu=0.85$ meV the single mode current ($m_j=1/2$) is
LP-dominated because $\delta^{+}_{1/2}=0$. For the same reason the
mode $E^{A}_{1/2}$ remains to a good approximation in the gap
independent of $\Phi$ for any $\mu$. Approximate results for
$n=0$, $\mu=15$ meV and $m_j=1/2$, $3/2$, $5/2$, $7/2$ are shown
in Fig.~1(b). Similar arguments are valid for $n=1$ but now the
current is never LP-dominated. At $\Phi=\Phi_0$ and $\mu=7$ meV
the angular numbers $m_j=0$, 2 are relevant while for $\mu=2.5$,
0.85 meV only $m_j=0$ is relevant. Because $\delta^{+}_{m_j}\ne0$
all modes shift outside the gap leading to a more `noisy' current
profile compared to $n=0$. A weak SO coupling, $\alpha\neq 0$,
does not alter the basic characteristics (Appendix C). Parameters:
$L_S=2000$ nm, $L_N=100$ nm, $R_0=43$ nm, $\alpha=0$,
$\Delta_0=0.2$ meV, $\xi= 80$ nm, $d_\text{sc}=0$ and $I_S = e
\Delta_0/\hbar$.}\label{fig4}
\end{figure}

We now include the LP modulation of the pairing
amplitude (Appendix A)
$\Delta = \Delta(n,\Phi, R_0,
d_\text{sc}, \xi)$, where $\xi$ is the coherence length of the
superconducting shell. When the shell thickness
$d_\text{sc}\rightarrow 0$, then $\Delta$ depends on $\phi$
instead of $\Phi/\Phi_0$. We consider a destructive regime in
which $\Delta = 0$ near the boundaries of the lobes and
$\Delta=\Delta_0$ at the center of the lobes [Fig.~\ref{fig4}(a)].

The flux tunable critical current is presented in Fig.~\ref{fig4}.
The current at $\phi=0$ (namely $n_\Phi=\Phi/\Phi_0=0$, 1) depends
on $\mu$ which specifies the number and position of subgap modes
in the superconducting gap. Once this number is fixed, by varying
$\Phi$ with respect to the centre of the lobe the current is
gradually reduced, and a kink point is formed at the flux where
the contribution of a subgap mode vanishes. The current for $n=1$
exhibits similar characteristics to that for $n=0$ but with
noticeable differences, e.g., the $\phi=0$ currents in the two
lobes are different even when $\Delta$ is the same. The reason is
the different potentials $V_{1(2)}$ involved in the BdG
Hamiltonian. According to Fig.~\ref{fig4}, the flux dependence of
$\Delta$ cannot be used to explain $I^{\text{c}}(\Phi)$. The
\textit{formula}~\cite{PhysRevLett.125.156804, vekris21}
$I^\text{c}(\Phi)\approx I^\text{c}(\Phi=0)\Delta(\Phi)/\Delta_0$
completely fails to capture the correct flux dependence of
$I^{\text{c}}$ in the multi-mode regime~\footnote{The agreement
for $n=0$, $\mu=0.85$ meV stems from $\delta^{+}_{1/2}=0$.}.
Instead, a good \textit{approximation} to the current is obtained
by assuming the approximate flux dependence of the BdG levels,
$E_{m_j}(\phi) \approx E_{m_j}(\phi=0)\Delta(\phi)/\Delta_0 +
\delta^{\pm}_{m_{j}}(\phi)$, with $E_{m_j}(\phi=0)$ being the
exact $\phi=0$ BdG levels, and using Eq.~(\ref{formula}) with all
positive levels included. When the terms
$\delta^{\pm}_{m_{j}}(\phi)$ become smaller, e.g., by increasing
$R_0$, the kink points shift at higher flux values [Fig.~1(b)].

\emph{Simplified SNS junction model.--} The approximate results
presented in Fig.~\ref{fig4} reveal the vital role of the linear
terms $\delta^{\pm}_{m_j}(\phi)$. To obtain further
\textit{qualitatively} insight we develop a simplified model where
$I(\varphi_0)$ is governed by the ABSs:
\begin{equation}\label{approx-model}
E_{\pm,k}(\varphi_0,\Phi) = \pm \Delta (\Phi) \sqrt{ 1 - \tau_k
\sin^2(\varphi_0/2) } + w_k \Phi/\Phi_0,
\end{equation}
$k=1$, 2, $\ldots$ $M$ is the number of ABSs and the parameters $0
< \tau_k \le 1$ model the transparency of the SNS junction. The
linear terms $w_k\Phi/\Phi_0$ with $w_k=(k-1)/2m^{*}R^{2}_0$ play
the same role as $\delta^{+}_{m_j}(\phi)$
[Eq.~(\ref{competition})] for $\alpha=0$ (Appendix C). 
The exact $\varphi_0$-dispersion is not important and we adopt
Eq.~(\ref{approx-model}) for simplicity and to illustrate the
crossover from the LP-regime to the stepwise regime.

For $\tau_k=1$ and when all $w_{k}$ are zero, the analytically
computed supercurrent [Eq.~(\ref{formula})] can be written as $ M
I_{Z}(\varphi_0)$, and for the critical current in a spinfull
junction we recover the standard result
\cite{PhysRevLett.66.3056,PhysRevB.45.10563} $I^{\text{c}}(\Phi) =
M e \Delta(\Phi)/\hbar = I^{\text{c}}(0) \Delta(\Phi)/\Delta_0$;
the flux dependence of $I^{\text{c}}$ is due solely to the LP
modulation of $\Delta$. A completely different situation occurs
when $w_{k}\ne0$. Now within the flux range $ 0 \le w_k
\Phi/\Phi_0 \le \Delta$ we define the corresponding flux dependent
phase $\theta_k=\theta_k(\Phi)$ satisfying $E_{-,k}(\theta_k)=0$
with $0\le \theta_k \le \pi$. As $\Phi$ increases and the levels
$E_{-,k}(\theta_k)$ shift gradually upwards a decrease in
$I^{\text{c}}(\Phi)$ is expected. A simple inspection shows that
$I^{\text{c}}(\Phi)$ is equal to the larger of $kI_{Z}(\theta_k)$
and $(k-1)I_{Z}(\theta_{k-1})$, with $k\rightarrow k-1$ as the
flux increases. For the flux
\begin{equation}
\frac{\Phi}{\Phi_0} = \Delta(\Phi) \sqrt{  \frac{k^2 - (k-1)^2 }{
k^2 w^{2}_{k} - (k-1)^2 w^{2}_{k-1} } },
\end{equation}
the currents satisfy
\begin{equation}\label{points}
kI_{Z}(\theta_k)=(k-1)I_{Z}(\theta_{k-1}),
\end{equation}
then a kink point is formed and the number of ABSs contributing to
the critical current decreases by one~\footnote{Because $w_1=0$,
as happens with $\delta^{+}_{1/2}$, the current $I_{Z}(\theta_1)$
should be replaced by $e\Delta(\Phi)/\hbar$. Details can be found
in Appendix C.}.
Because $\theta_{k-1}$ varies with flux slower than $\theta_{k}$,
Eq.~(\ref{points}) leads to a stepwise decrease of the current.
For very large $R_0$ ($>150$ nm), $w_k$ are vanishingly small and
the current steps/kink points are formed at flux values lying
(very) near the boundaries of the lobe; in this case
$I^{\text{c}}(\Phi)$ is trivially LP-dominated. In contrast,
experimentally
reported~\cite{Valentini:21,Valentini:22,vekris21,Ibabe:22} $R_0$
($\lesssim 100$ nm) guarantee a stepwise decrease.
In Appendix C 
we generalize the simplified model to $\tau_k<1$ and make a
connection with the exact BdG model in Appendix D.

\emph{Conclusion.--} The critical supercurrent in full-shell
nanowire Josephson junctions exhibits a stepwise dependence as a
function of an external magnetic flux. This flux dependence has
striking features, unrelated to the Little-Parks modulation of the
superconducting pairing. The position of the steps depends on the
underlying symmetries of the transverse channels contributing to
the supercurrent and it is thus gate-tunable. This prediction
should be robust for low-disordered samples and specially in the
low-flux range, $n=0$ with $\Phi\lesssim \Phi_0/2$, where
transverse channel interference due to $m_j$ mixing, not included
here, should be negligible. Such effects, similar to
Fraunhofer-like interference in diffusive many-channel planar
junctions~\cite{PhysRevLett.99.217002} and few-channel hybrid NW
junctions~\cite{Gharavi:14,Zuo:17,PhysRevB.100.155431}, could be
of relevance in the first LP lobe, $n=1$ at $\Phi\approx\Phi_0$,
and lead to further structures in $I^{\text{c}}$. Experiments able
to discriminate between the flux modulation of the gap and the
intrinsic subgap structure, for example, Joule heating
experiments~\cite{Ibabe:22}, could be an interesting platform to
explore the effects predicted here. Our findings could be also of
relevance for transmon qubits based on full shell
NWs~\cite{PhysRevLett.125.156804}, where flux tunability of the
Josephson coupling under an axial magnetic field (without
requiring split-junction geometries) could lead to novel
functionalities.

\acknowledgments This research was supported by Grants
PID2021-125343NB-I00 and TED2021-130292B-C43 funded by
MCIN/AEI/10.13039/501100011033, "ERDF A way of making Europe" and
European Union NextGenerationEU/PRTR. Support by the CSIC
Interdisciplinary Thematic Platform (PTI+) on Quantum Technologies
(PTI-QTEP+) is also acknowledged.

\appendix

\section{Superconducting pairing potential}

The superconducting pairing potential, $\Delta$, due to the
Little-Parks effect~\cite{PhysRevLett.9.9,PhysRev.133.A97} acquires a flux
dependence. If $\Delta_0$ is the value of $\Delta$ at $\Phi=0$
then according to Abrikosov-Gor'kov~\cite{abrikosov, PhysRev.136.A1500} a
pair-breaking term $\Lambda$ results in the following modulation
of $\Delta$:
\begin{widetext}
\begin{eqnarray}\label{LP}
\ln\frac{\Delta}{\Delta_0} &=&
-\frac{\pi}{4}\frac{\Lambda}{\Delta}, \quad \Lambda \le \Delta, \notag\\
\ln\frac{\Delta}{\Delta_0} &=& - \ln\left( \frac{\Lambda}{\Delta}
+ \sqrt{ (\Lambda/\Delta)^2 - 1}\right) +
\frac{\sqrt{(\Lambda/\Delta)^2-1}}{2(\Lambda/\Delta)}
-\frac{\Lambda}{2\Delta}\arctan\frac{1}{\sqrt{(\Lambda/\Delta)^2-1}},
\quad \Lambda \ge \Delta.
\end{eqnarray}
\end{widetext}
Within a Ginzburg-Landau theory~\cite{sternfeld, shah, dao,
schwiete} the magnetic flux dependence of $\Lambda$ can be
determined from the approximate expression
\begin{equation}
\Lambda(\Phi) \approx \frac{ \xi^2 k_{B} T_{c}}{\pi R^{2}_{0}}
\left[ 4\left( n - \frac{\Phi}{\Phi_0} \right)^{2} +
\frac{d^{2}_{\text{sc}}}{R^{2}_0} \left(
\frac{\Phi^2}{\Phi^{2}_{0}} + \frac{n^2}{3} \right) \right].
\end{equation}
The parameter $\xi$ denotes the coherence length of the
superconducting shell which has thickness $d_{\text{sc}}$, $n$ is
the lobe index, $T_{c}$ is the critical temperature at zero flux,
and $k_{B} T_{c} = \Delta_0/1.76$~\cite{PhysRev.136.A1500} where
$k_{B}$ is Boltzmann's constant. In our work, we focus on $R_0 \gg
d_{\text{sc}}$ and for simplicity we set $d_{\text{sc}}=0$; a
small non-zero $d_{\text{sc}}$ introduces only minor corrections
to the final magnetic flux dependence of $\Delta$. The numerical
solution to Eq.~(\ref{LP}) is well-known and can be found in the
literature, for example, in Ref.~\onlinecite{PhysRev.136.A1500}.
In the limit $\Lambda \rightarrow \Delta_0/2$ the pairing
potential nearly vanishes, $\Delta\rightarrow 0$, we then set
$\Delta=0$ for $\Lambda \ge \Delta_0/2$ to model a destructive
regime. One example of the pairing potential for $R=43$ nm,
$\xi=80$ nm, $d_{\text{sc}}=0$ is shown in Fig.~4(a) of the main
article.

\section{SNS junction with spatially dependent chemical potential}

In the main article the chemical potential, $\mu$, is taken to be
constant (spatially independent) along the SNS junction. This
configuration simplifies the theoretical analysis but it might be
difficult to realize experimentally. For this reason, we consider
one more configuration where the chemical potential is spatially
dependent, i.e., $\mu=\mu(z)$. In this context, we assume that an
electrostatic gate voltage tunes the chemical potential in the
normal (N) region with respect to the potential in the
superconducting (S) regions, thus, creating a potential offset
between the N and S regions. The spatial profile of the chemical
potential along the SNS junction is written as
\begin{equation}\label{mz}
\mu(z) = \mu_0 - \mu_{\text{pot}} f(z),
\end{equation}
where the function $f(z)$ is expected to depend on the exact
geometry of the junction, the microscopic details of the S-N
interfaces, and the charge distribution in the junction. We
consider a continuous variation across the S-N interfaces and
assume that
\begin{equation}\label{fz}
f(z) =  \frac{1}{ K\left(\frac{(z-z_0)^{2}}{2D^{2}} \right)},
\end{equation}
where for the function $K$ we take either $K=\exp$ or $K=\cosh$,
$z_0=0$ is the centre of the N region, and $\mu_{\text{pot}}$
($\ge0$) determines the potential offset. This offset is maximum
in the centre of the N region and the parameter $D$ controls the
length scale in which the chemical potential varies along the SNS
junction. As shown below, for a large $\mu_{\text{pot}}$ ($\approx
\mu_0$) the critical current vanishes whereas it is maximum when
$\mu_{\text{pot}}\approx 0$. The results in the main article are
for $\mu_{\text{pot}}=0$.

\begin{figure*}
\includegraphics[width=4.5cm, angle=270]{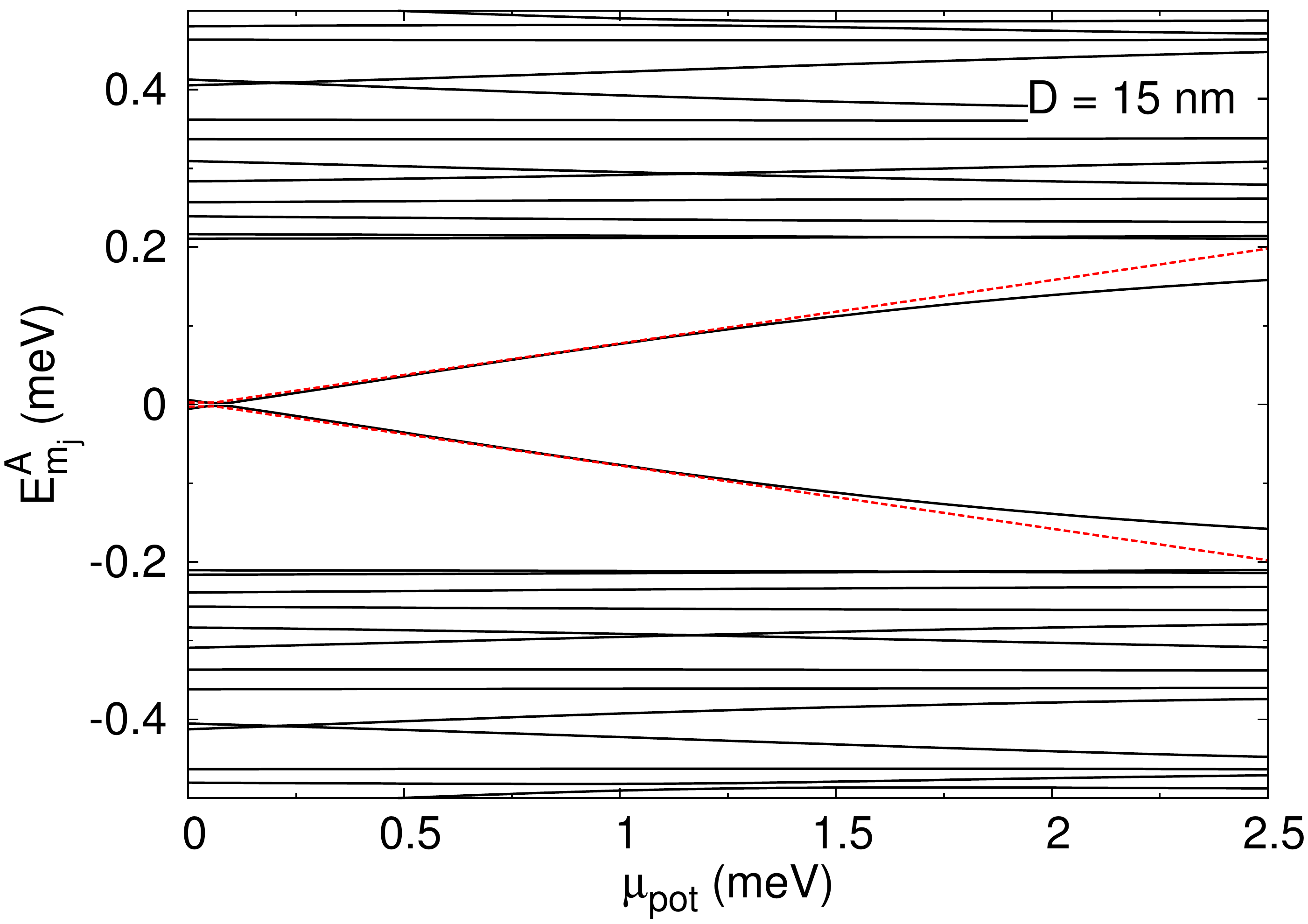}
\includegraphics[width=4.5cm, angle=270]{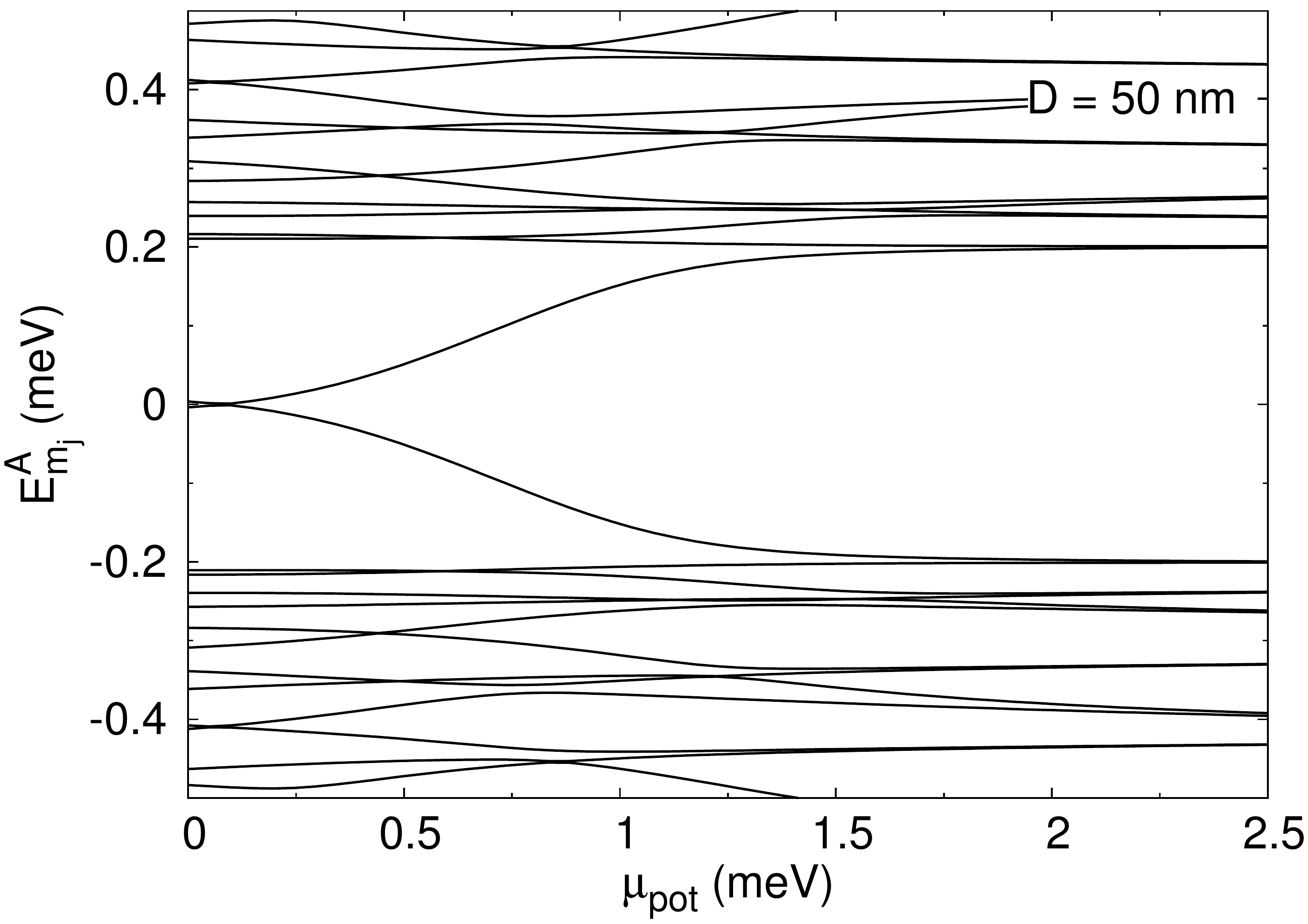}\\
\includegraphics[width=4.5cm, angle=270]{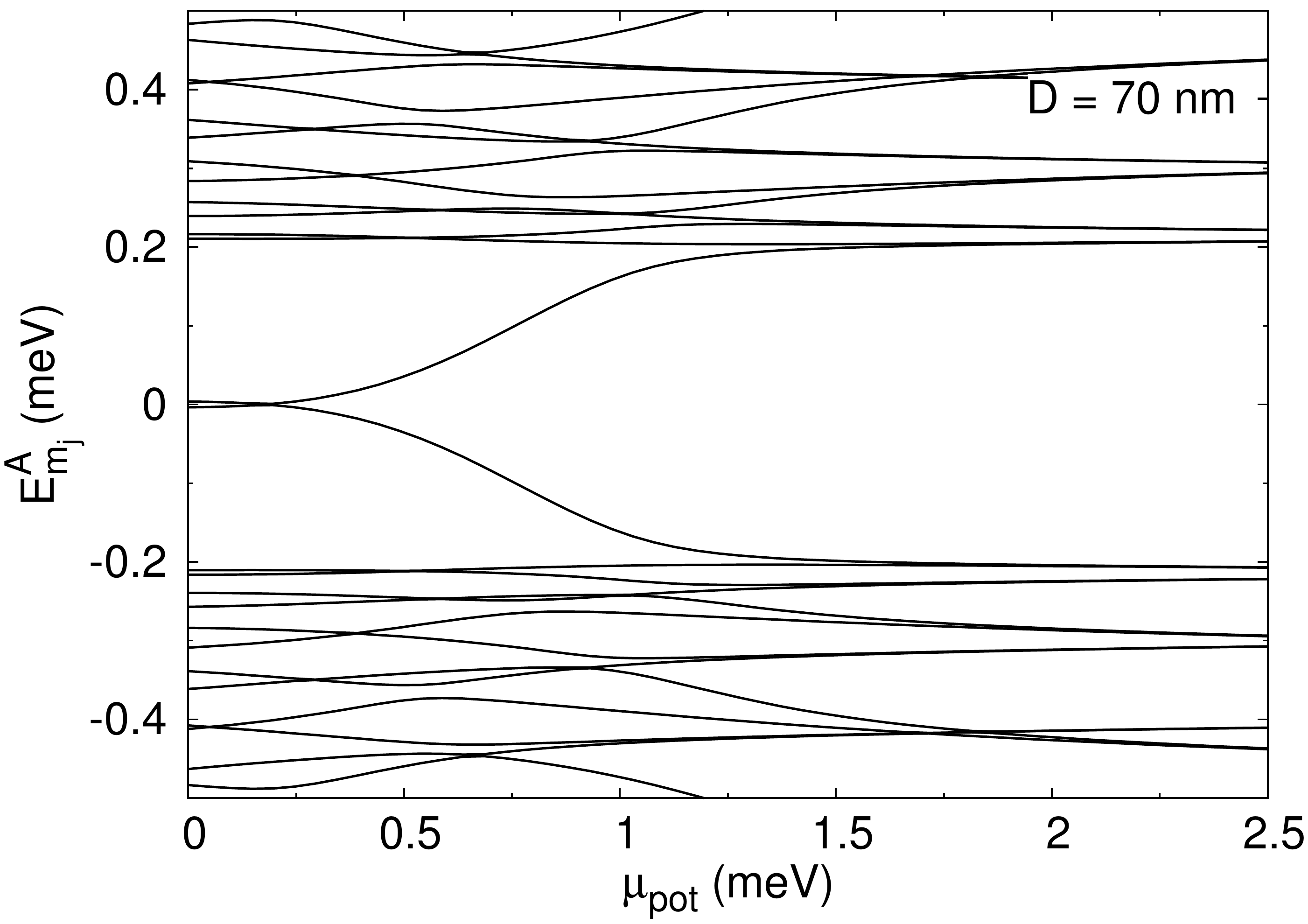}
\includegraphics[width=4.5cm, angle=270]{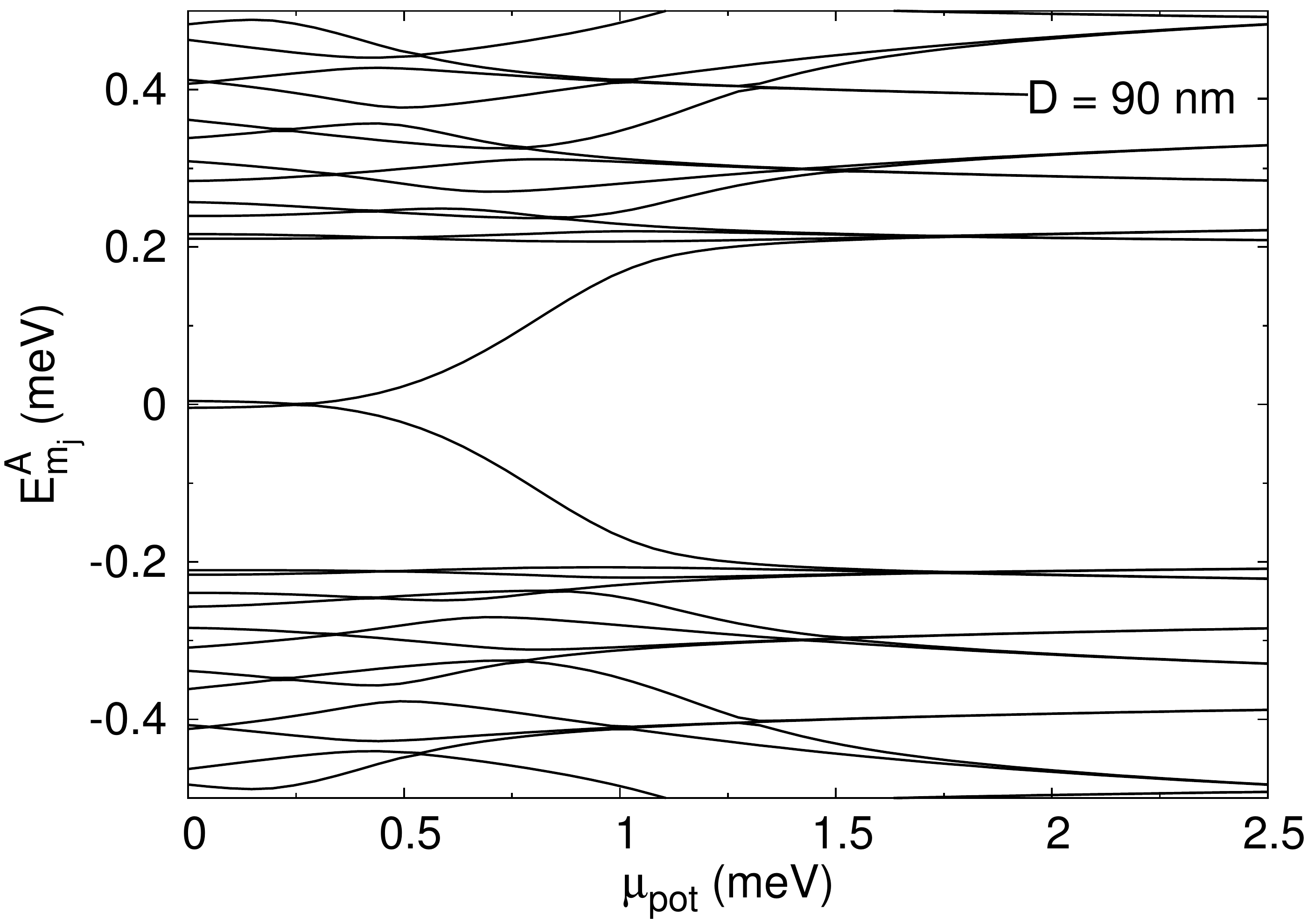}
\caption{Zero-flux energies of $H_{A}$ [Eq.~(3) main article] as a
function of $\mu_{\text{pot}}$ at phase difference $\varphi_0=\pi$
for $m_j=1/2$ and $K =\cosh$, with $E^{A}_{-m_j} = - E^{B}_{m_j}$
and $E^{B}_{m_j} = E^{A}_{m_j+1}$. Dashed curves for $D=15$ nm are
approximate energies derived from a two-level model described in
the text. Parameters: $\mu_0=0.9$ meV, $L_S=2000$ nm, $L_N=100$
nm, $R_0=43$ nm, $\alpha=0$, $\Delta=\Delta_0=0.2$
meV.}\label{enerD}
\end{figure*}

The effect of a nonzero $\mu_{\text{pot}}$ on the energy spectra
can be more easily understood in the regime of small $\mu_0$, so
that to a good approximation only the Hamiltonian $H_A$ [Eq.~(3)
main article] is relevant with $m_j=1/2$. One case illustrated in
Fig.~\ref{enerD} demonstrates that increasing $\mu_{\text{pot}}$
shifts the subgap levels near the edge of the superconducting gap.
This shift in turn reduces the overall degree of
$\varphi_0$-dispersion and consequently the critical current.
Provided $D$ and $\mu_{\text{pot}}$ are small the term
$-\mu_{\text{pot}} f(z)$ can be treated within a perturbative
two-level model, $H_A + \sigma_z \mu_{\text{pot}} f(z)$, using for
basis states the (two) subgap states at $\mu_{\text{pot}}=0$. Some
results of this approximate model are presented in
Fig.~\ref{enerD} for $D=15$ nm; the agreement with the exact
result is particularly good for small values of $\mu_{\text{pot}}$
and the correct linear behaviour is predicted. The two-level model
can also predict the correct $\varphi_0$-dispersion, however, by
increasing $D$ the model becomes quickly inaccurate. For example,
when $D=50$ nm and $\mu_{\text{pot}}\approx 1$ meV about 60 basis
states are needed to achieve the same convergence as for $D=15$
nm.

\begin{figure*}
\includegraphics[width=4.0cm, angle=270]{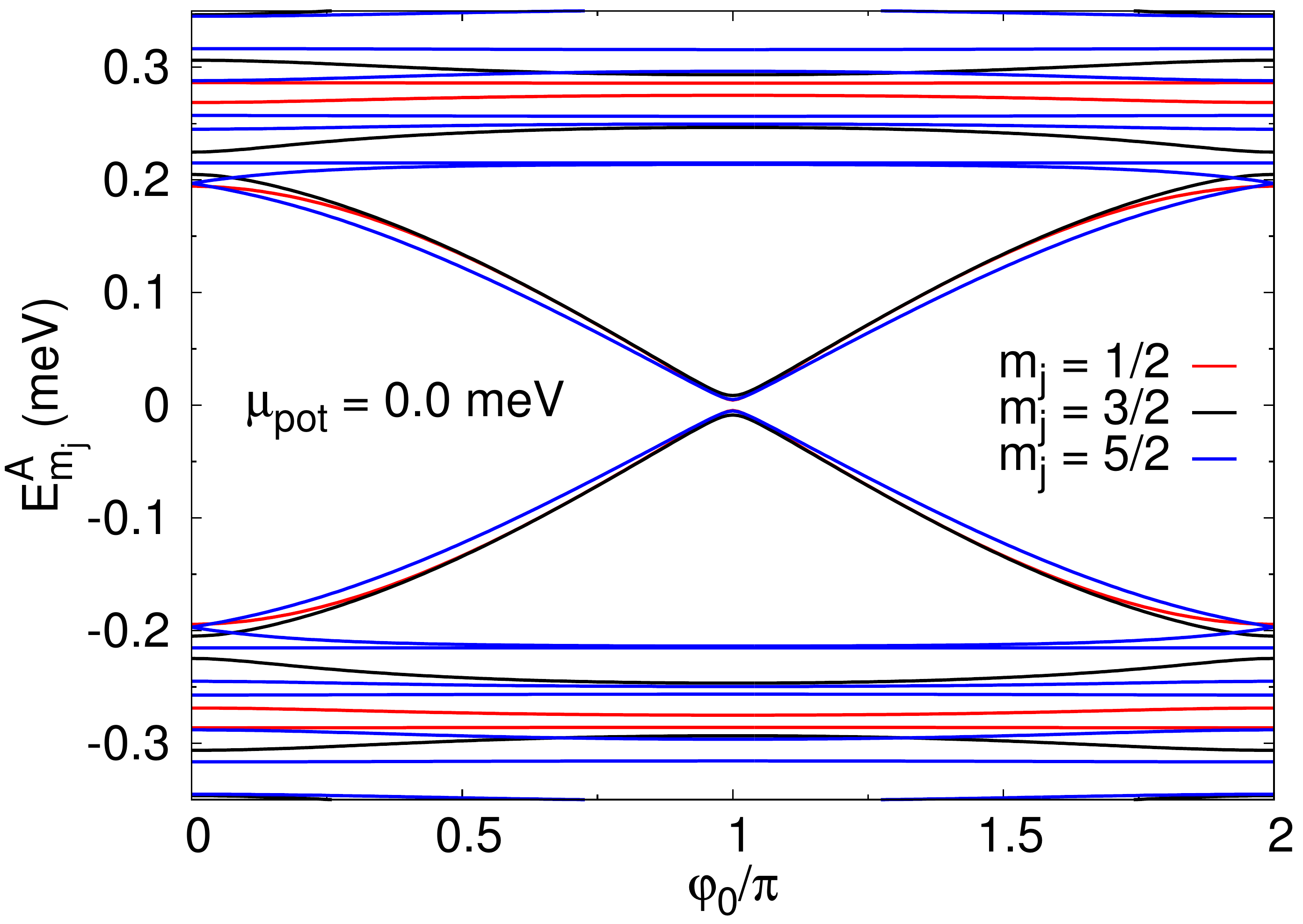}
\includegraphics[width=4.0cm, angle=270]{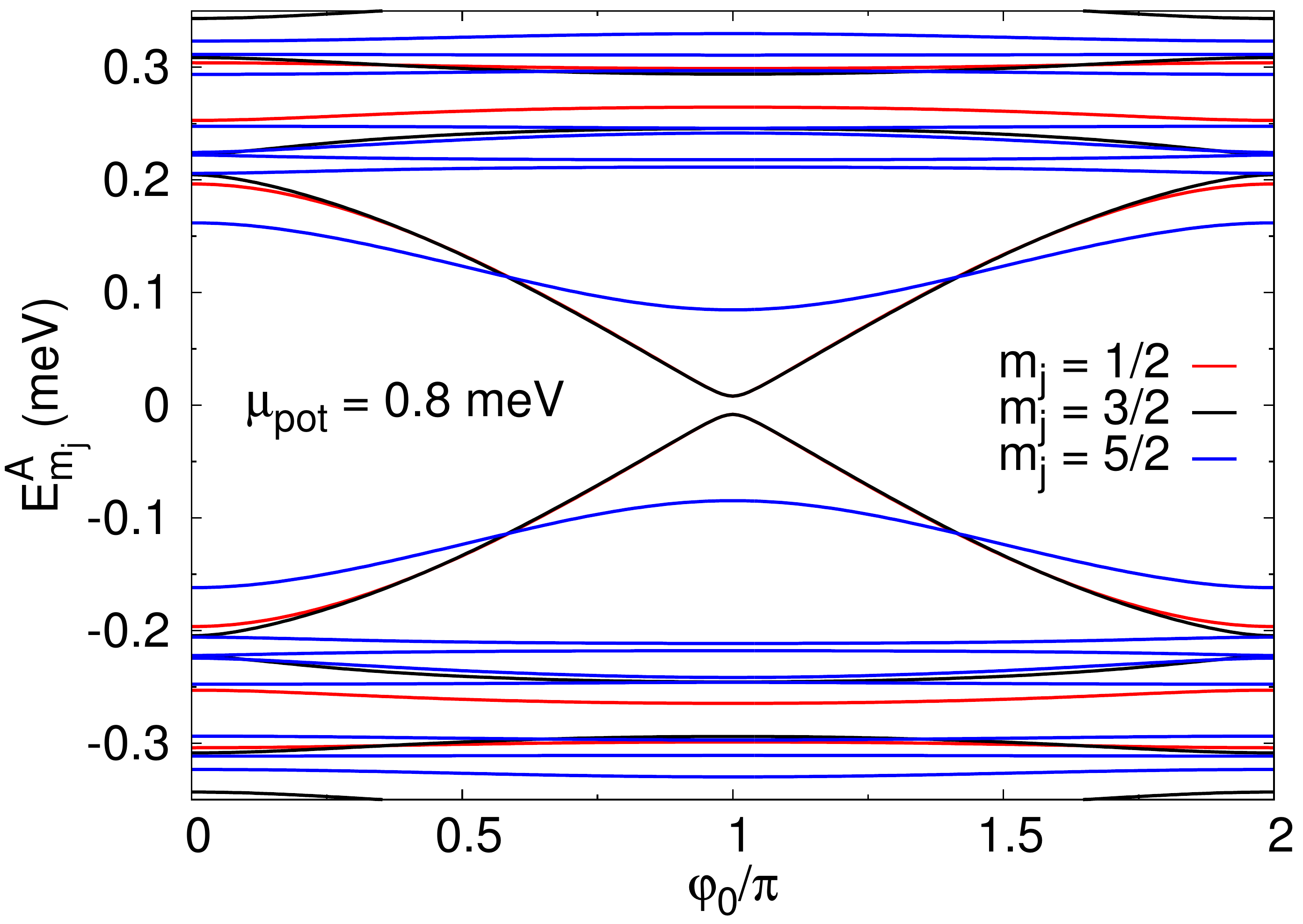}
\includegraphics[width=4.0cm, angle=270]{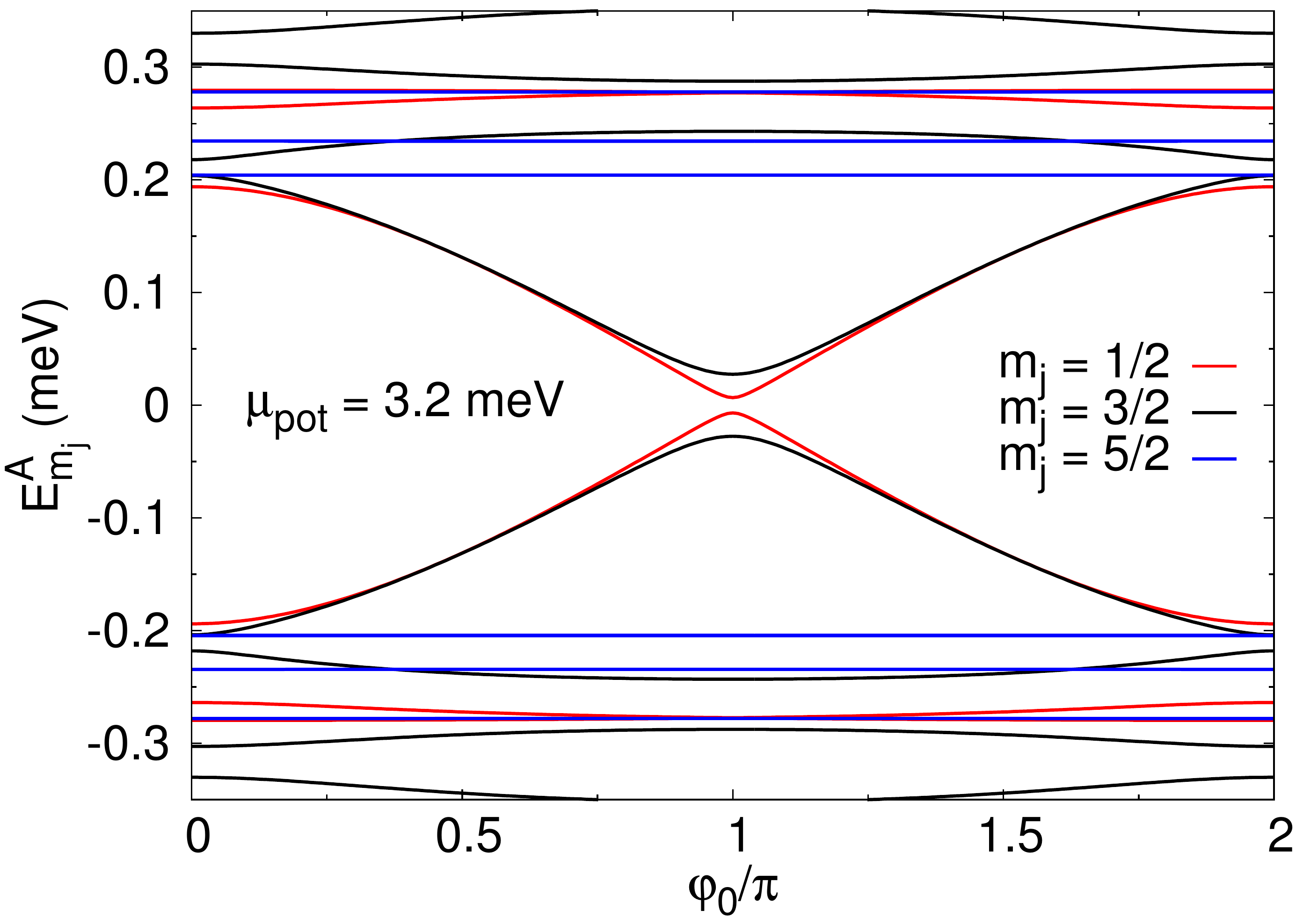}\\
\includegraphics[width=4.0cm, angle=270]{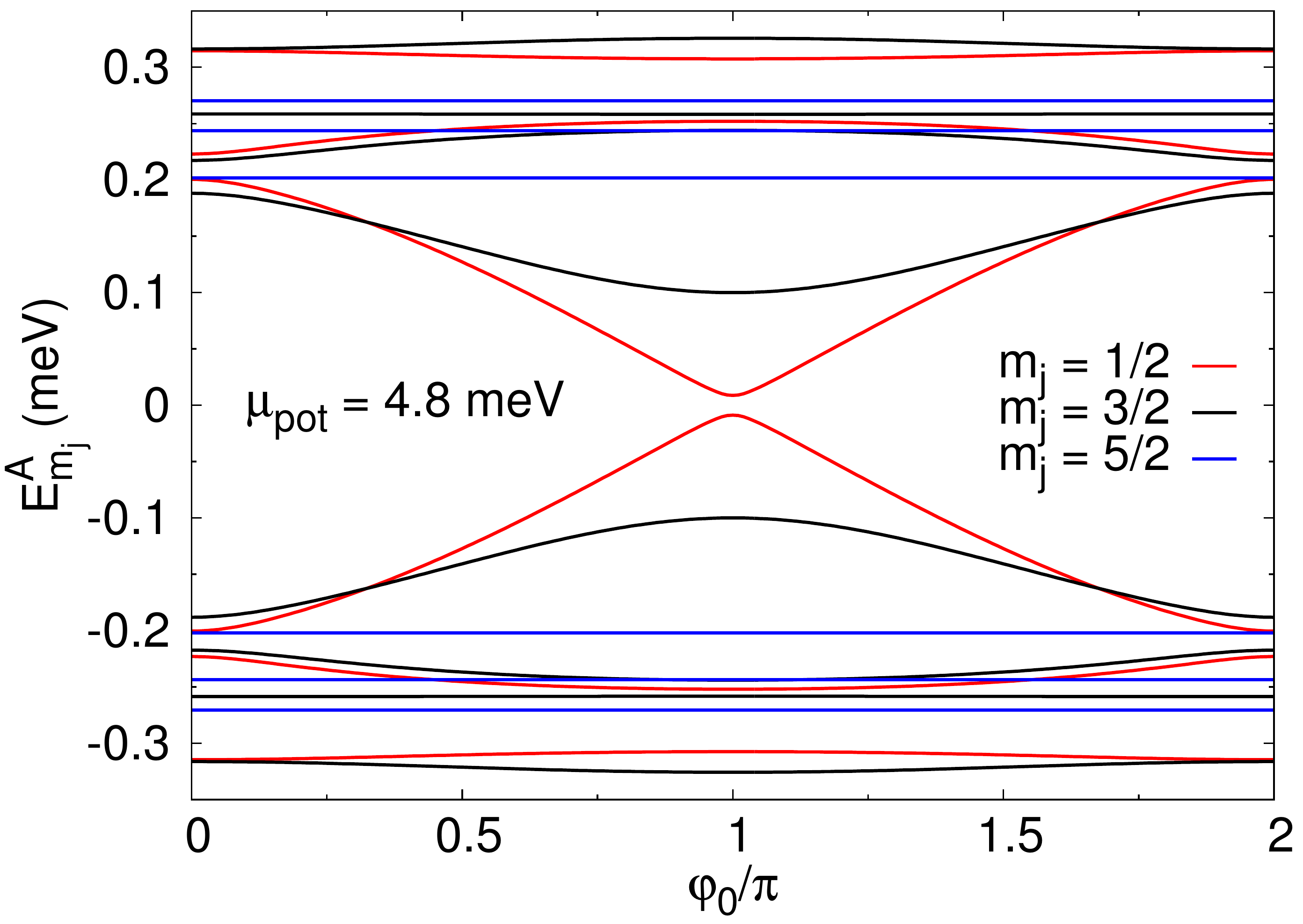}
\includegraphics[width=4.0cm, angle=270]{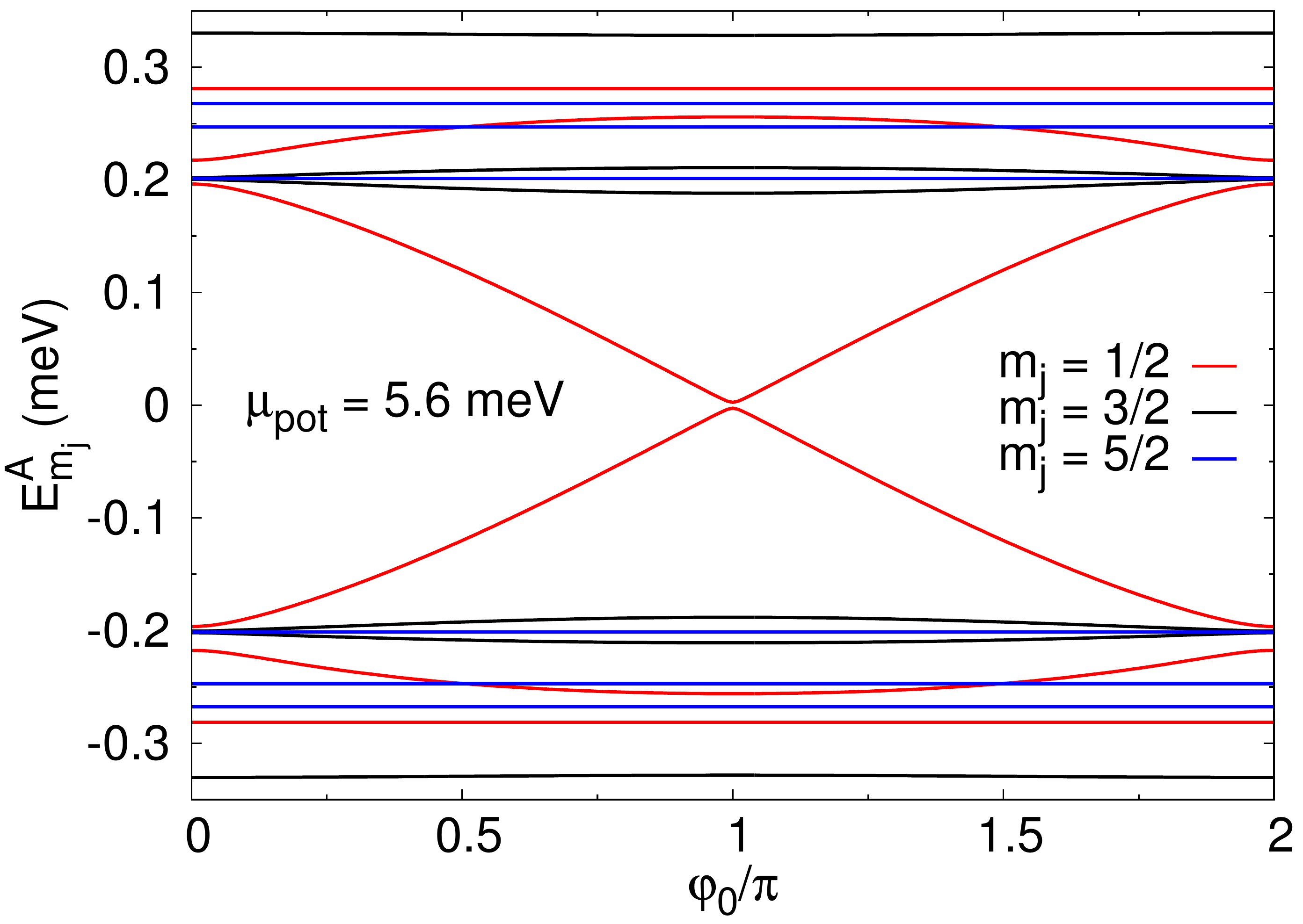}
\includegraphics[width=4.0cm, angle=270]{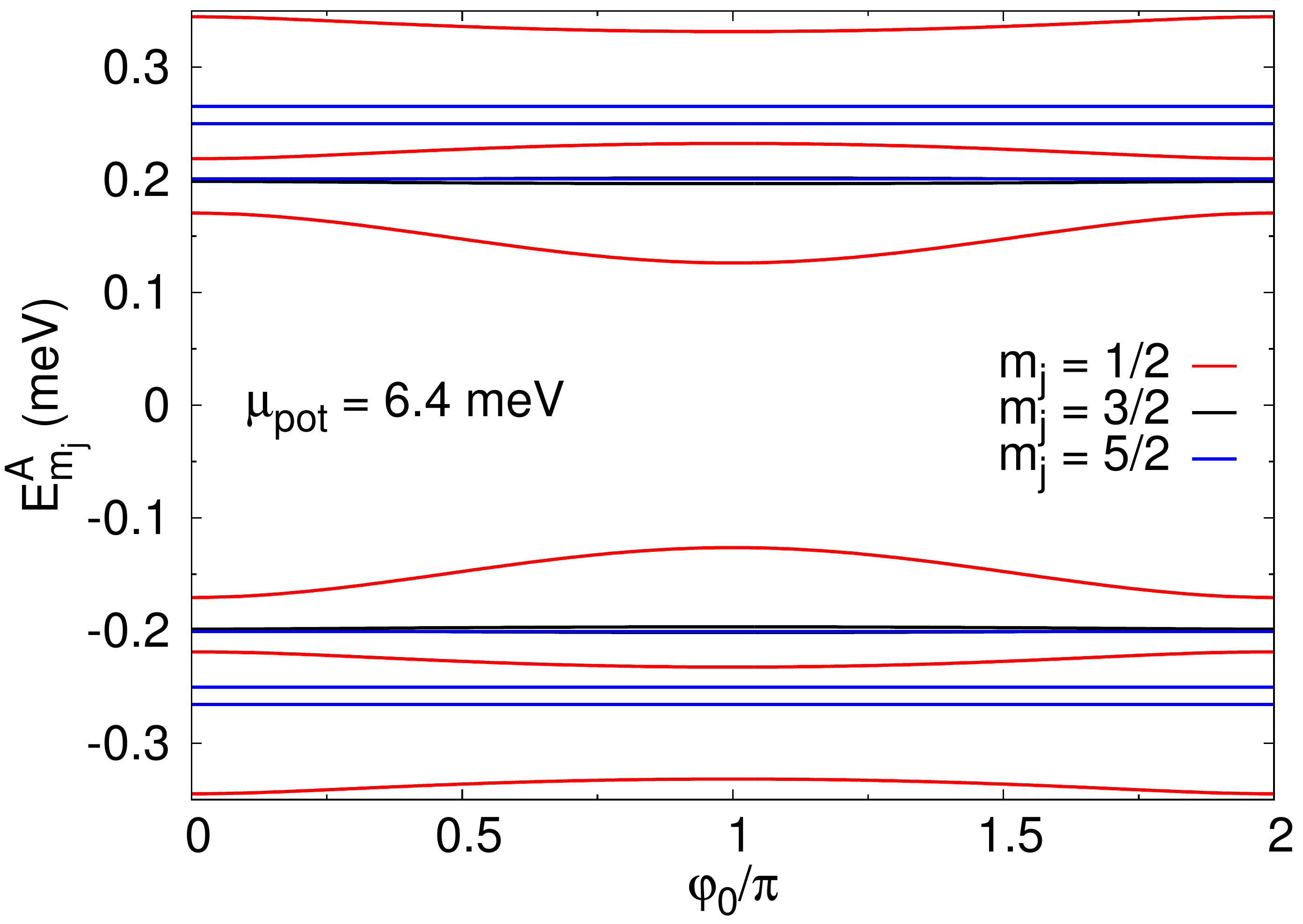}
\caption{Zero-flux energies of $H_A$  [Eq.~(3) main article] as a
function of phase difference $\varphi_0$ for $K=\cosh$, with
$E^{A}_{-m_j} = - E^{B}_{m_j}$ and $E^{B}_{m_j} = E^{A}_{m_j+1}$.
Parameters: $D=70$ nm, $\mu_0=6.5$ meV, $L_S=2000$ nm, $L_N=100$
nm, $R_0=43$ nm, $\alpha=0$, $\Delta=\Delta_0=0.2$
meV.}\label{EmuB}
\end{figure*}

A more general configuration is now examined when the energy
levels lying in the superconducting gap correspond to different
$m_j$ numbers. Some typical energy spectra of $H_{A}$ are plotted
in Fig.~\ref{EmuB}. At $\mu_{\text{pot}} = 0$ the subgap levels
correspond to $m_j=1/2$, $3/2$ and $5/2$, thus, there are 5
positive levels: $E^{A}_{1/2}$, $E^{B}_{1/2}$, $E^{A}_{3/2}$,
$E^{B}_{3/2}$ and $E^{A}_{5/2}$ with the degeneracies
$E^{B}_{1/2}=E^{A}_{3/2}$ as well as $E^{B}_{3/2}=E^{A}_{5/2}$. As
shown in Fig.~\ref{EmuB}, increasing $\mu_{\text{pot}}$ tends to
shift the subgap levels outside (near the edge of) the gap in a
systematic way, so levels which correspond to larger $m_j$ shift
outside the gap at smaller values of $\mu_{\text{pot}}$.
Consequently, the number of active subgap levels in the SNS
junction can be controlled at will.

\begin{figure*}
\includegraphics[width=5.cm, angle=270]{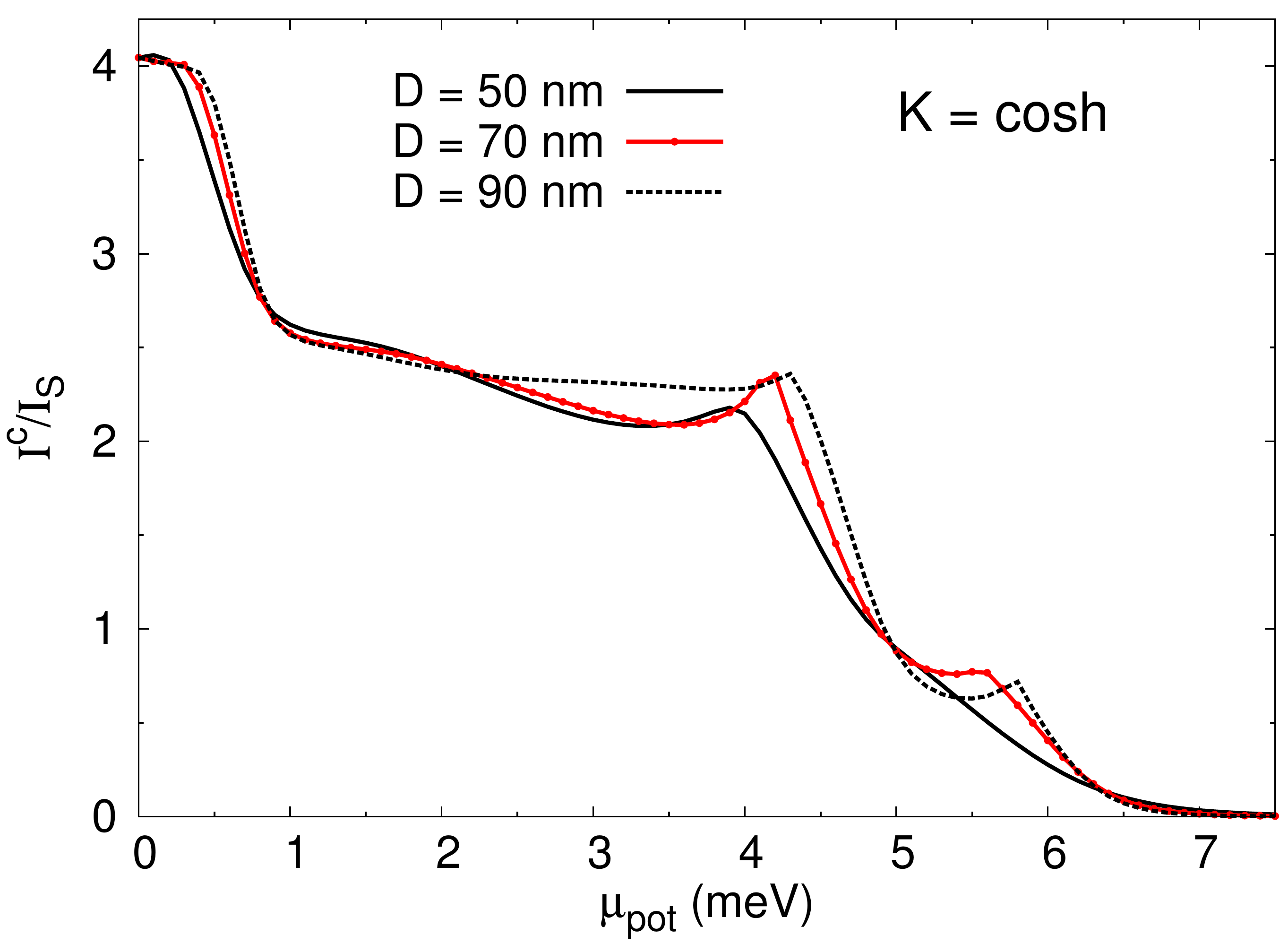}
\includegraphics[width=5.cm, angle=270]{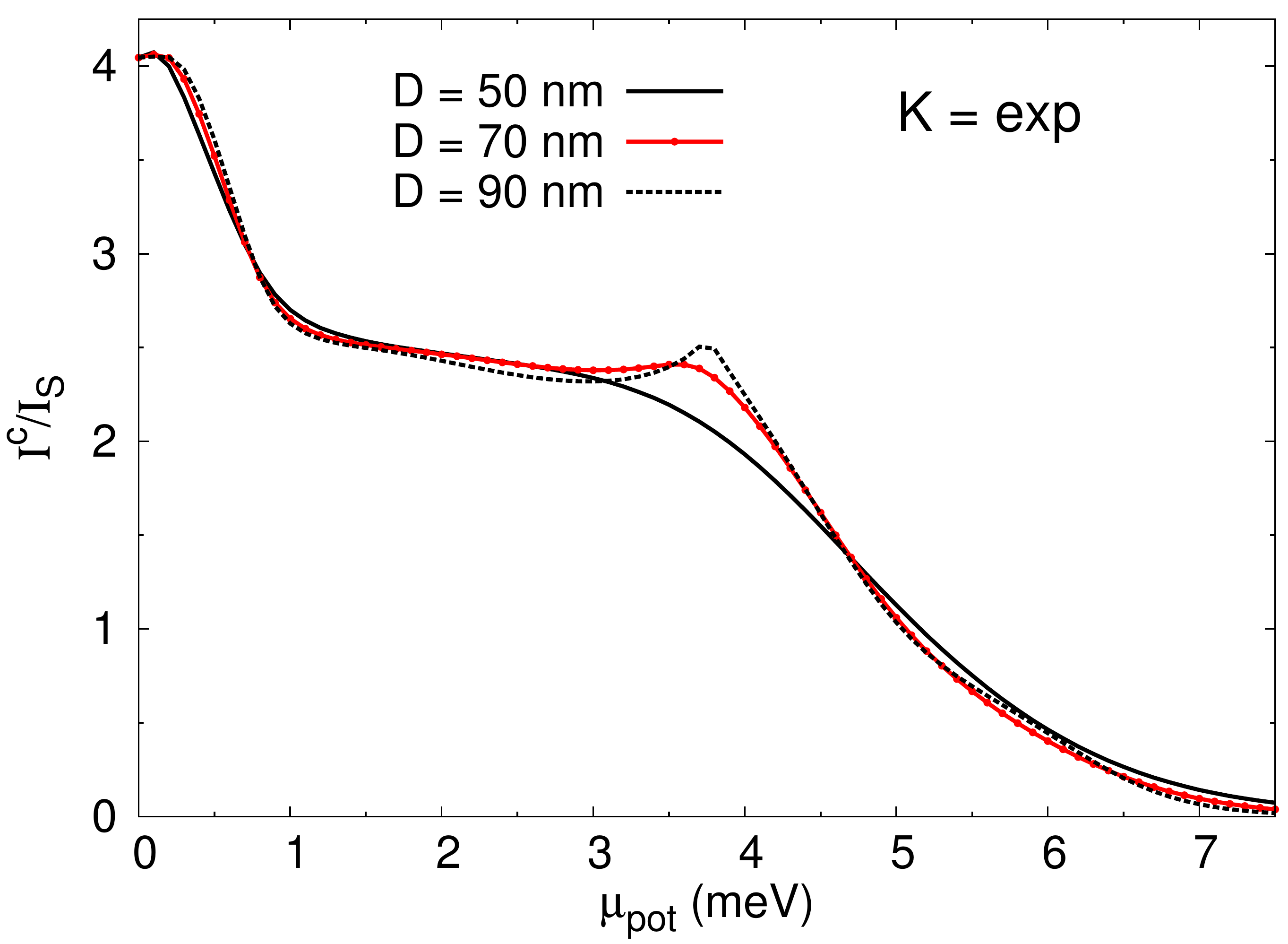}
\caption{Zero-flux critical current as a function of
$\mu_{\text{pot}}$. Parameters: $\mu_0=6.5$ meV, $L_S=2000$ nm,
$L_N=100$ nm, $R_0=43$ nm, $\alpha=0$, $\Delta=\Delta_0=0.2$ meV,
$I_S = e \Delta_0/\hbar$.}\label{ImuB}
\end{figure*}

When the levels shift outside the gap the critical current
decreases and eventually complete suppression occurs when
$\mu_{\text{pot}}\approx\mu_0$ (Fig.~\ref{ImuB}). The exact form
of $f(z)$ determines the details of the process. Specifically, the
decrease of the current is not necessarily monotonic and steps can
be formed because the different $m_j$ levels are not affected
equally by the potential term $\mu_{\text{pot}}$.

\section{Simplified SNS junction model}

In this section we present in some detail the simplified model
introduced in the main article. The subgap modes are written as
follows ($\hbar=1$)
\begin{equation}\label{ABS}
E_{\pm,k}(\varphi_0, \Phi) = \pm \Delta \sqrt{ 1 - \tau_k
\sin^2(\varphi_0/2) } + w_k \Phi/\Phi_0,
\end{equation}
with $k=1, 2 \ldots M$,
\begin{equation}
w_k = \frac{(k-1)}{2 m^{*} R^{2}_{0}},
\end{equation}
and $\Delta=\Delta(\Phi)$ everywhere. To avoid confusion we note a
few remarks. First, we focus on $w_k\ge0$ and multiply the
currents by 2 to account for $\pm m_j$. Second, the term
$w_k\Phi/\Phi_0$ results in the same flux dependence as
$\delta^{+}_{m_j}$ in the main article [Eq.~(5)] for the zeroth
lobe, $n=0$, and zero SO coupling, $\alpha=0$. The first lobe,
$n=1$, can be treated similarly. Finally, the degeneracy
$H_B(m_j)=H_A(m_j+1)$ is not considered in Eq.~(\ref{ABS}) since
this does not change qualitatively the final conclusions.

\begin{figure*}
\includegraphics[width=3.3cm, angle=270]{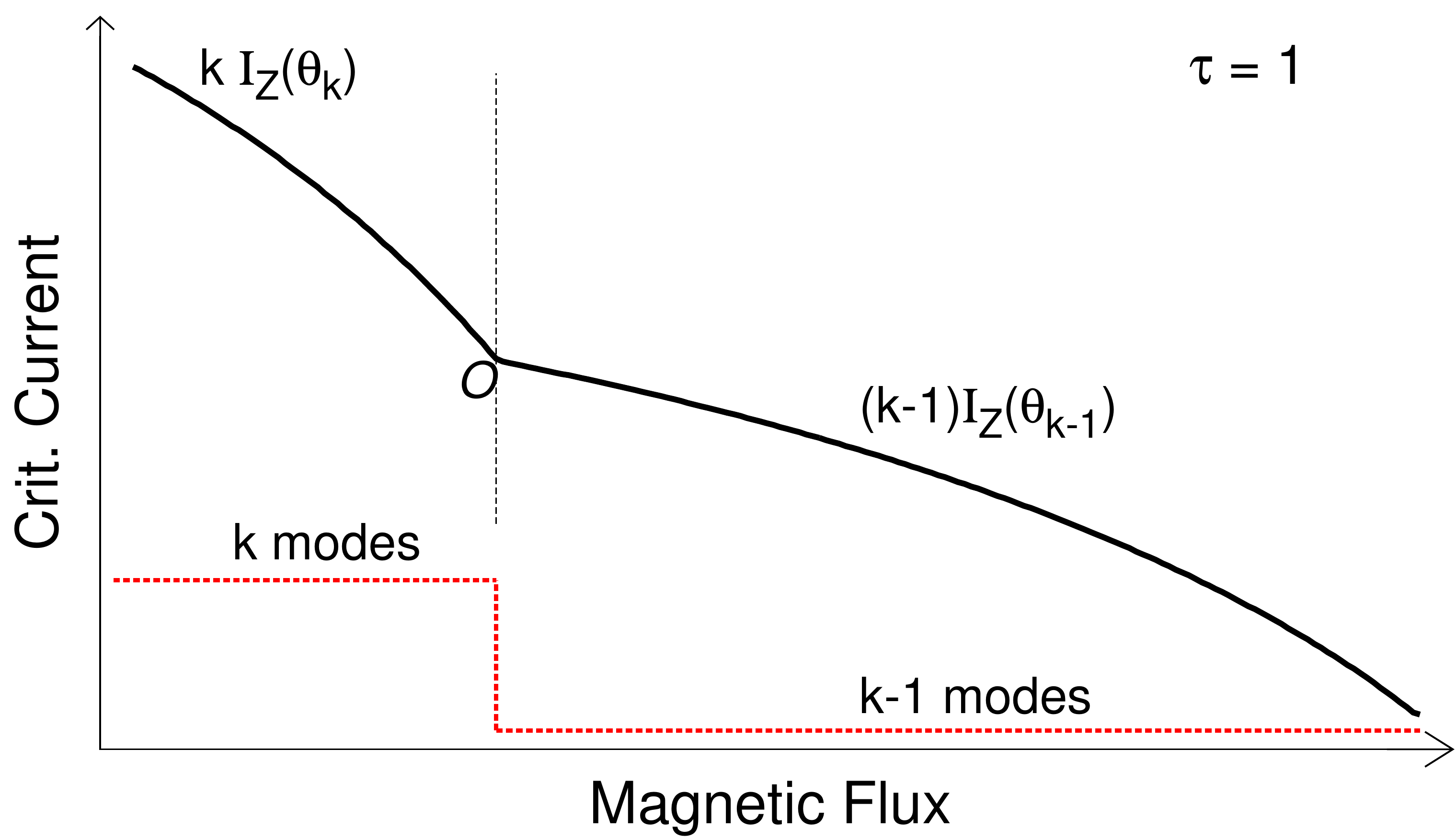}
\includegraphics[width=3.3cm, angle=270]{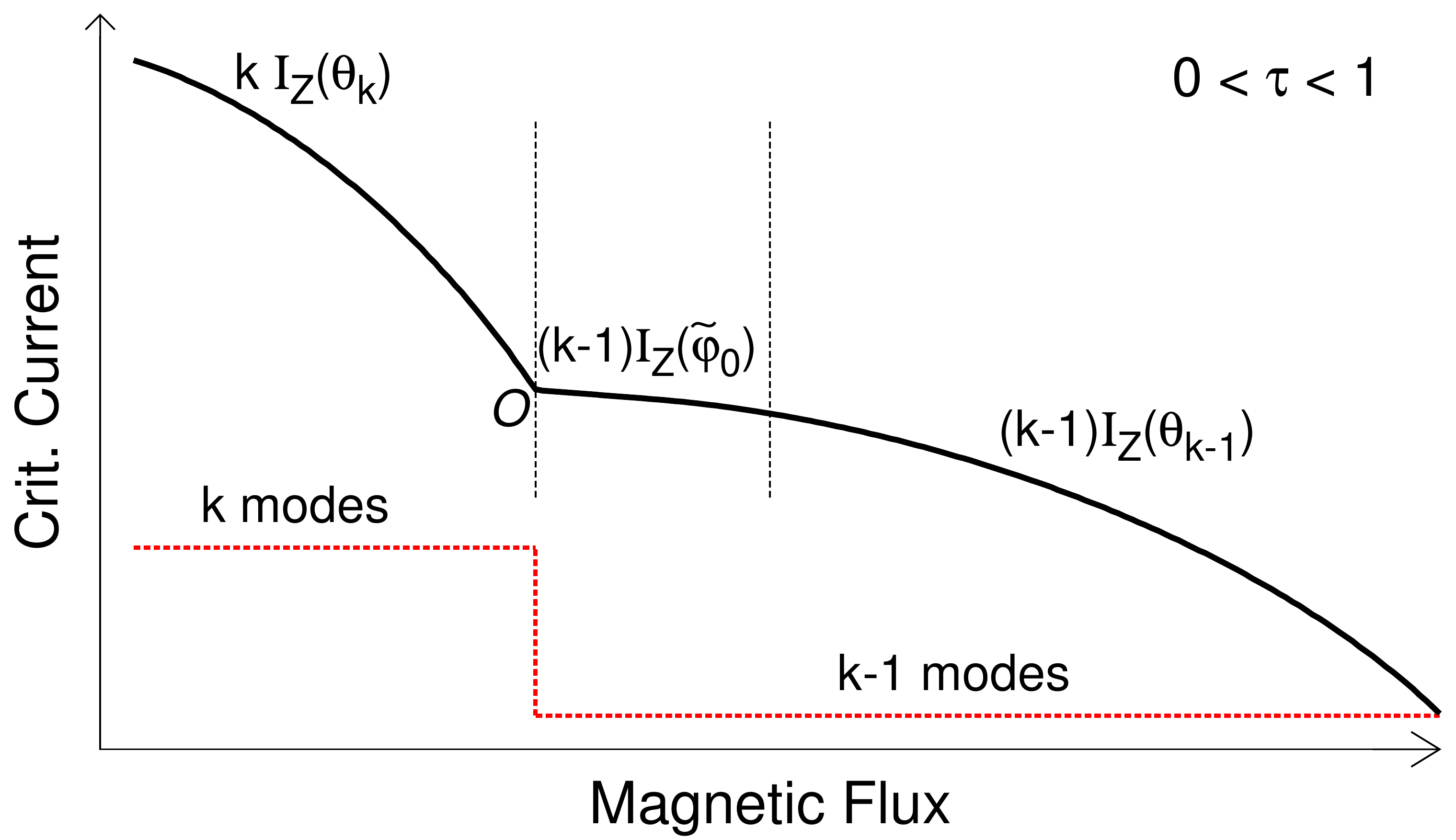}
\includegraphics[width=3.3cm, angle=270]{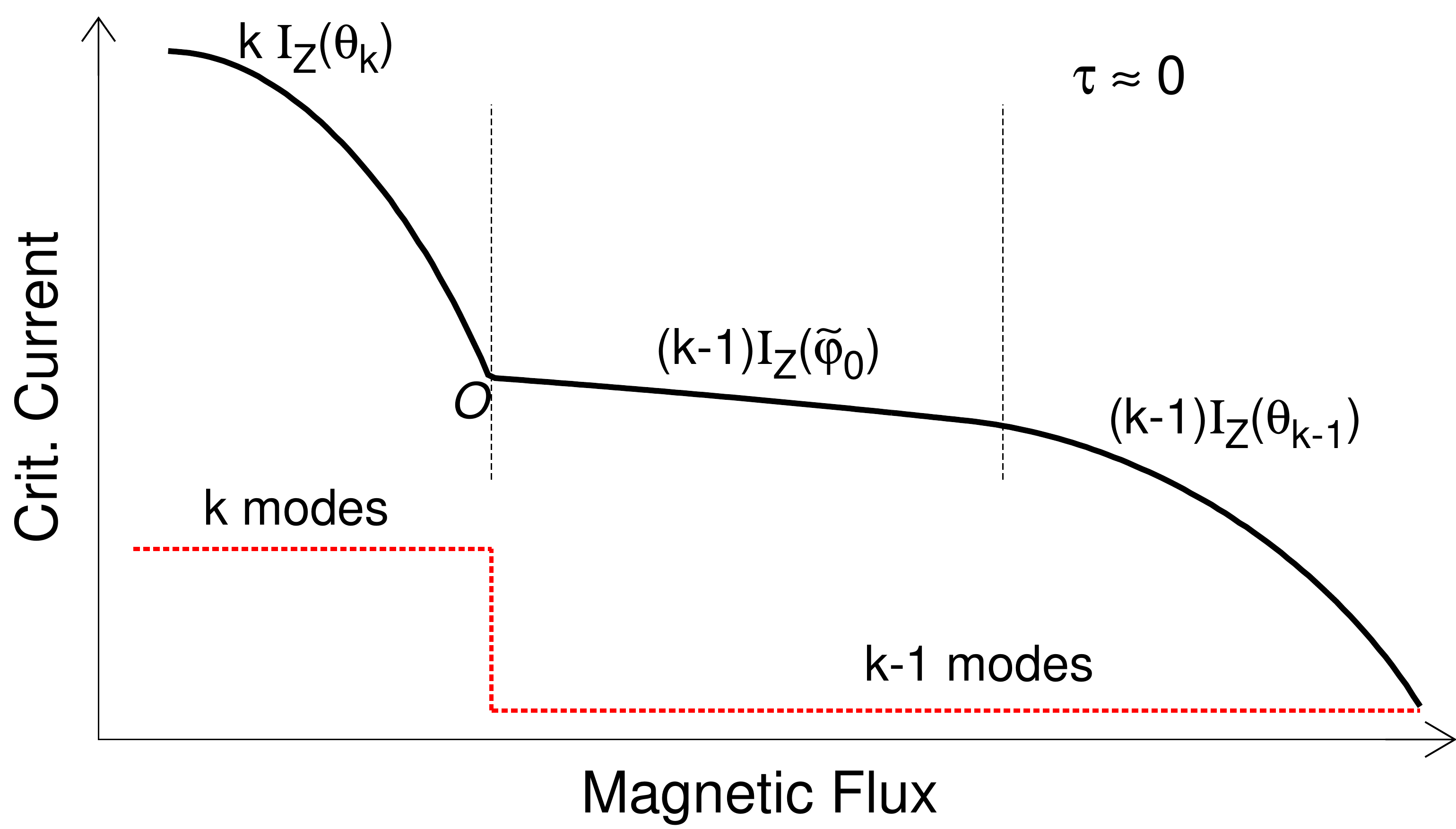}
\caption{Illustration of flux tunable critical current and number
of subgap modes predicted from simplified model, Eq.~(\ref{ABS}).
A single kink point ($\textit{O}$) is indicated in each frame. A
detailed explanation is given in the text.}\label{cartoon}
\end{figure*}

When all $w_k$ are zero and $\tau_k=\tau$ the supercurrent is
written as $M I_{Z}(\varphi_0)$ with
\begin{equation}
I_{Z}(\varphi_0) = - \frac{e}{\hbar} \frac{\Delta}{2} \frac{ \tau
\sin \varphi_0 }{ \sqrt{ 1 - \tau \sin^2(\varphi_0/2) } },
\end{equation}
while the phase $\tilde\varphi_0$ ($\le\pi$) giving the critical
current, $I^{\text{c}}$, can be readily extracted. An interesting
remark is that when $w_k\ne0$ the critical current can be
expressed with the help of the supercurrent $I_{Z}$. Specifically,
for $k\ne1$ we define within the flux range $\Delta \sqrt{1-\tau}
< w_k \Phi/\Phi_0 \le \Delta$ the corresponding flux-dependent
phase $\theta_k=\theta_k(\Phi)$ satisfying $E_{-,k}(\theta_k)=0$,
\begin{equation}\label{theta}
\theta_k = 2\sin^{-1} \sqrt{ \frac{1}{\tau} \left(1 - \frac{
w_{k}^2 \Phi^2}{\Delta^{2}\Phi_0^2} \right) },
\end{equation}
with $0\le \theta_k \le \pi$. As $\Phi$ increases, $I^{\text{c}}$
follows the $\Delta(\Phi)$ dependence until
$\theta_{M}=\tilde\varphi_0$, but when
$\theta_{M}<\tilde\varphi_0$ the decrease of $I^{\text{c}}$ due to
$w_{k}$ needs to be accounted for. Now $I^{\text{c}}$ is equal to
the largest of the three terms $kI_{Z}(\theta_k)$,
$(k-1)I_{Z}(\tilde\varphi_0)$ and $(k-1)I_{Z}(\theta_{k-1})$; and
as $\Phi$ increases $k\rightarrow k-1$, hence, the number of
subgap modes contributing to the current decreases successively by
one. This process gives rise to a stepwise current profile and is
illustrated in Fig.~\ref{cartoon}. For $\tau\rightarrow 1$
($\tilde\varphi_0\rightarrow \pi$) the flux range where
$(k-1)I_{Z}(\tilde\varphi_0)$ needs to be considered
vanishes/shrinks, and only the two terms $kI_{Z}(\theta_k)$,
$(k-1)I_{Z}(\theta_{k-1})$ are important. When
$kI_{Z}(\theta_k)=(k-1)I_{Z}(\theta_{k-1})$ a kink point is
formed, and using Eq.~(\ref{theta}) with
\begin{equation}
I_{Z}(\varphi_0) = - \frac{e}{\hbar} \Delta \sin(\varphi_0/2),
\end{equation}
we can determine the corresponding flux value
\begin{equation}
\frac{\Phi}{\Phi_0} = \Delta(\Phi) \sqrt{  \frac{k^2 - (k-1)^2 }{
k^2 w^{2}_{k} - (k-1)^2 w^{2}_{k-1} } }.
\end{equation}
In the opposite limit, $\tau\rightarrow 0$
($\tilde\varphi_0\rightarrow \pi/2$), the flux range where
$(k-1)I_{Z}(\tilde\varphi_0)$ dominates is maximum and current
steps are clearly formed. A kink point is now formed when
$kI_{Z}(\theta_k)=(k-1)I_{Z}(\tilde\varphi_0)$ and the steps
become flatter as the flux dependence of
$(k-1)I_{Z}(\tilde\varphi_0)$, due to $\Delta(\Phi)$, weakens. In
this analysis, because of the special value $w_1=0$ we have
$I_{Z}(\theta_1)\rightarrow I_{Z}(\tilde{\varphi_0})$ for any
value of $\tau$. The $k=1$ mode does not shift with flux and the
number of modes drops to zero at the boundaries of the lobe
because $\Delta(\Phi)= 0$.

\begin{figure*}
\includegraphics[width=3.5cm, angle=270]{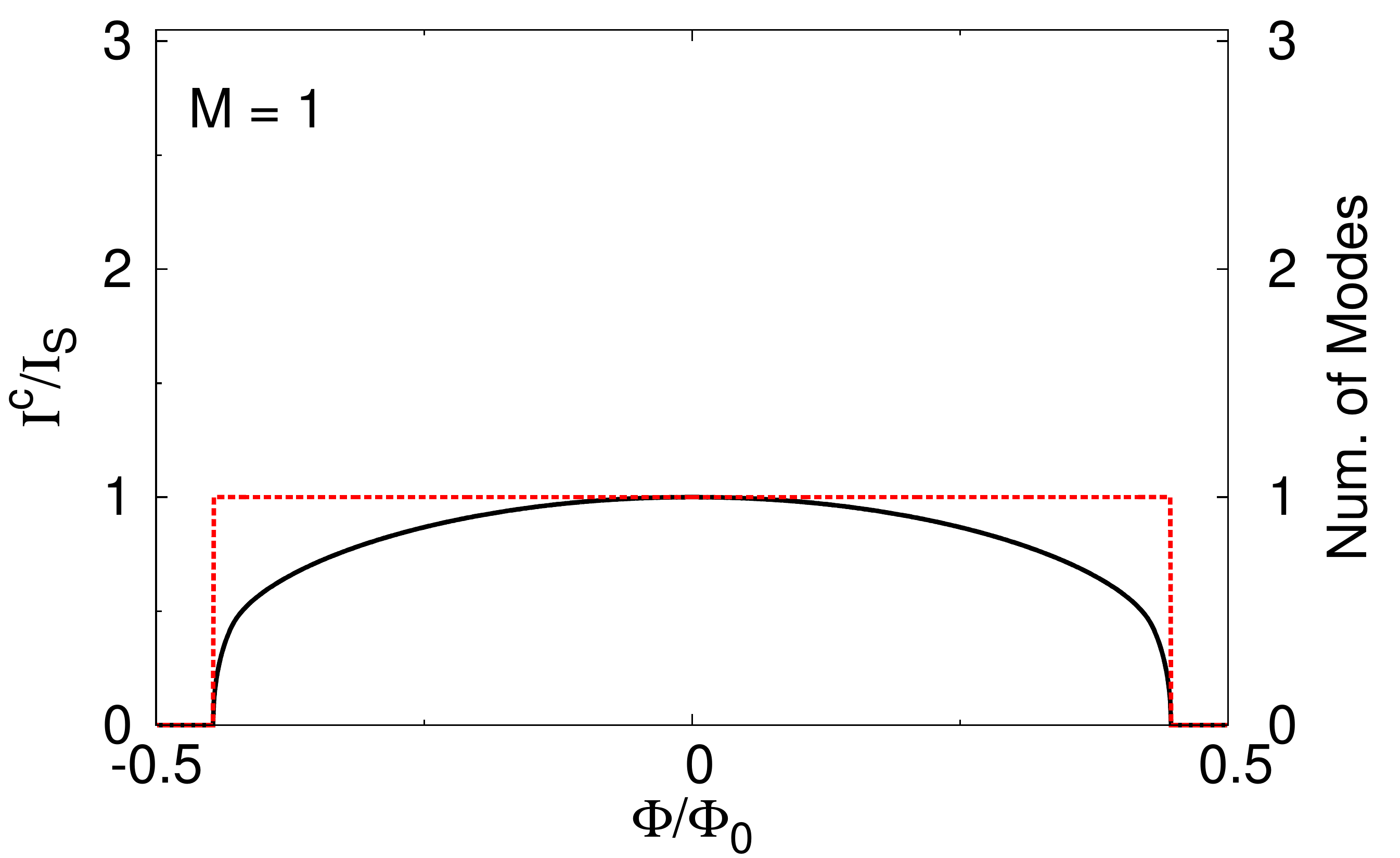}
\includegraphics[width=3.5cm, angle=270]{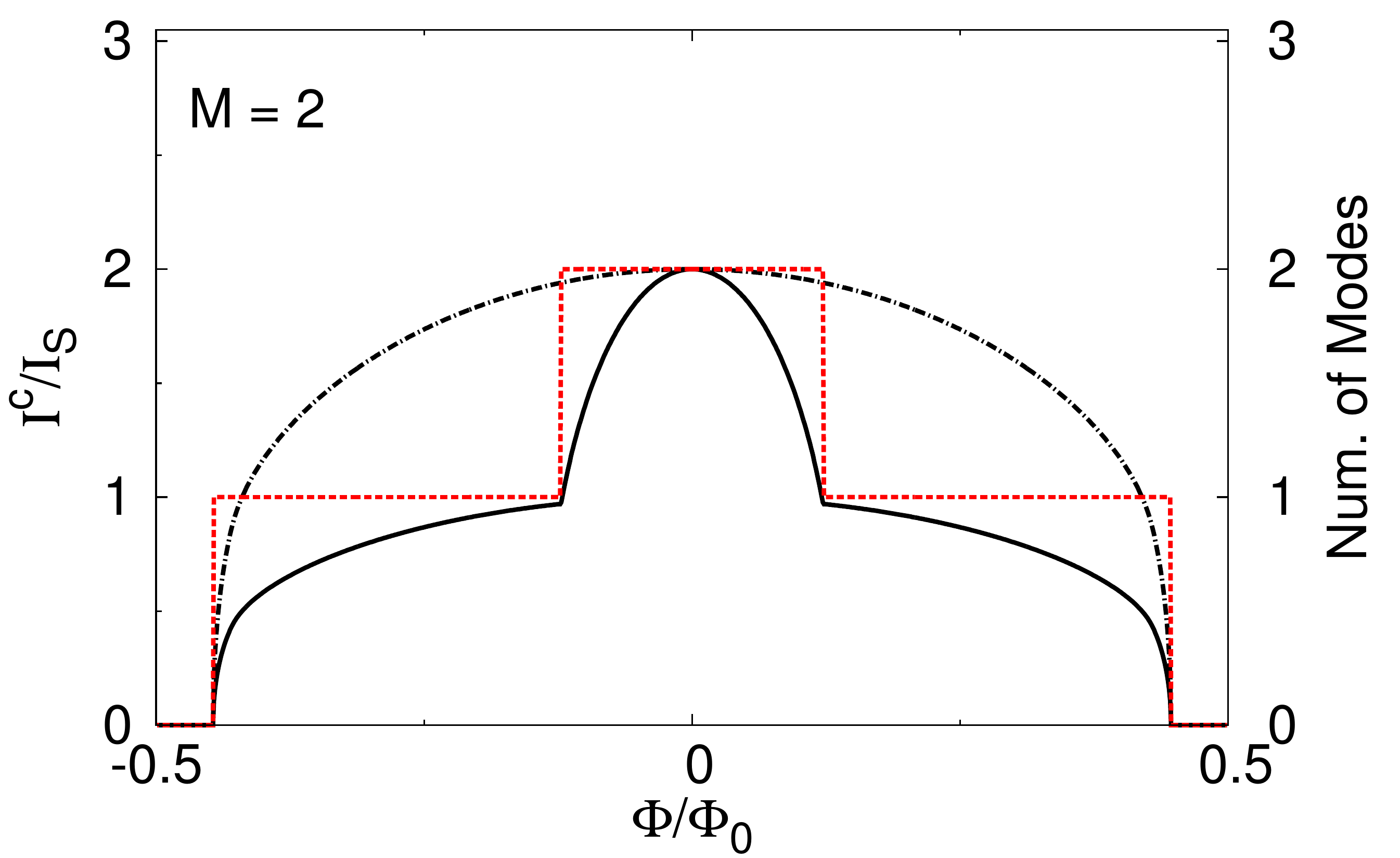}
\includegraphics[width=3.5cm, angle=270]{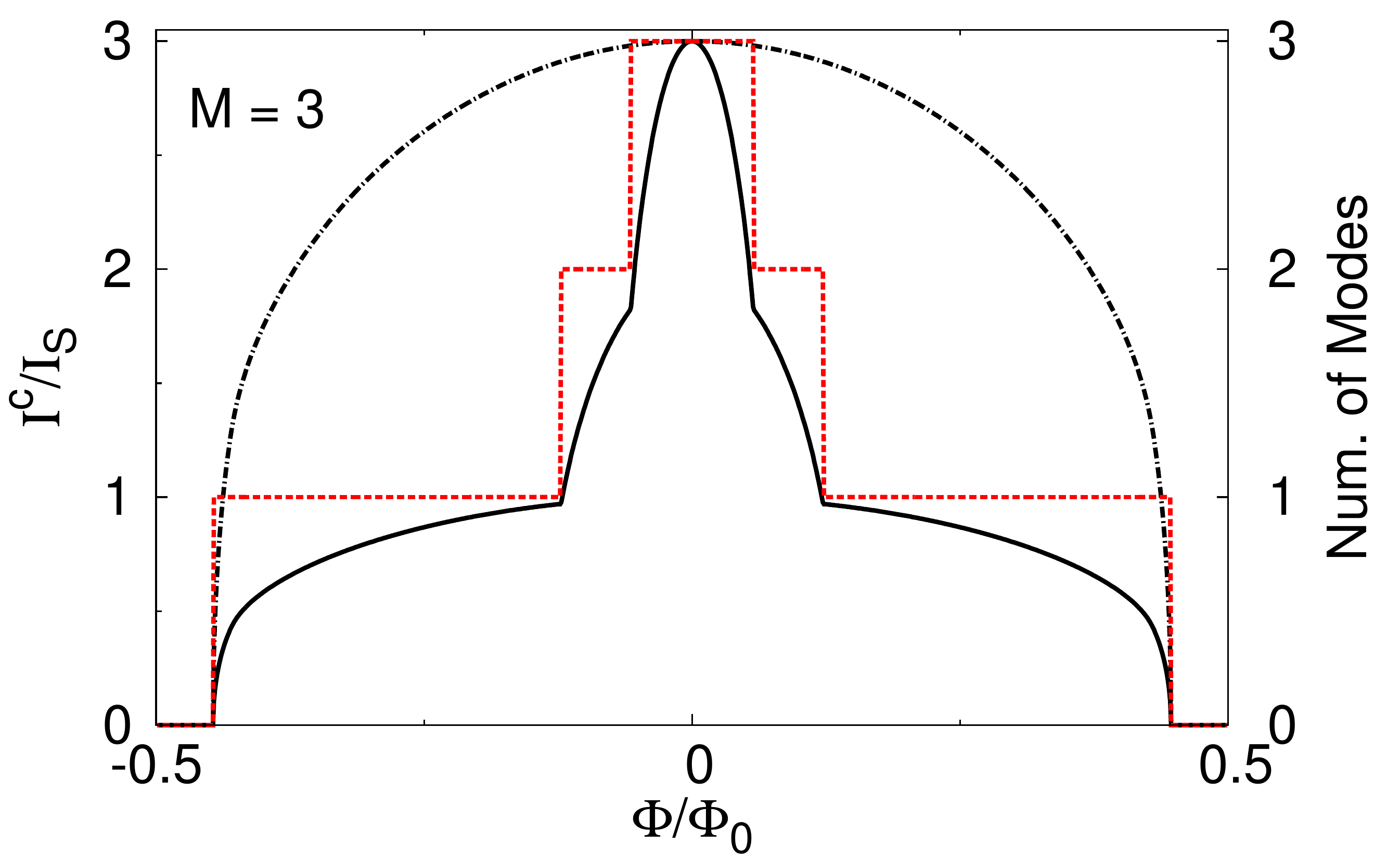}
\caption{Critical current and number of subgap modes (dotted
lines, right axis) derived from simplified model, Eq.~(\ref{ABS}),
with $\tau_k=\tau=1$, $R_0=43$ nm, $I_{S}=e\Delta_0/\hbar$.
Dashdotted curves show $I^{\text{c}}(0)\Delta/\Delta_0$; this
coincides with the critical current for $M=1$. $\Delta$ is
calculated from Eq.~(\ref{LP}) with $\xi=80$ nm,
$d_{\text{sc}}=0$, $\Delta_0=0.2$ meV.}\label{modetoy}
\end{figure*}

\begin{figure*}
\includegraphics[width=3.5cm, angle=270]{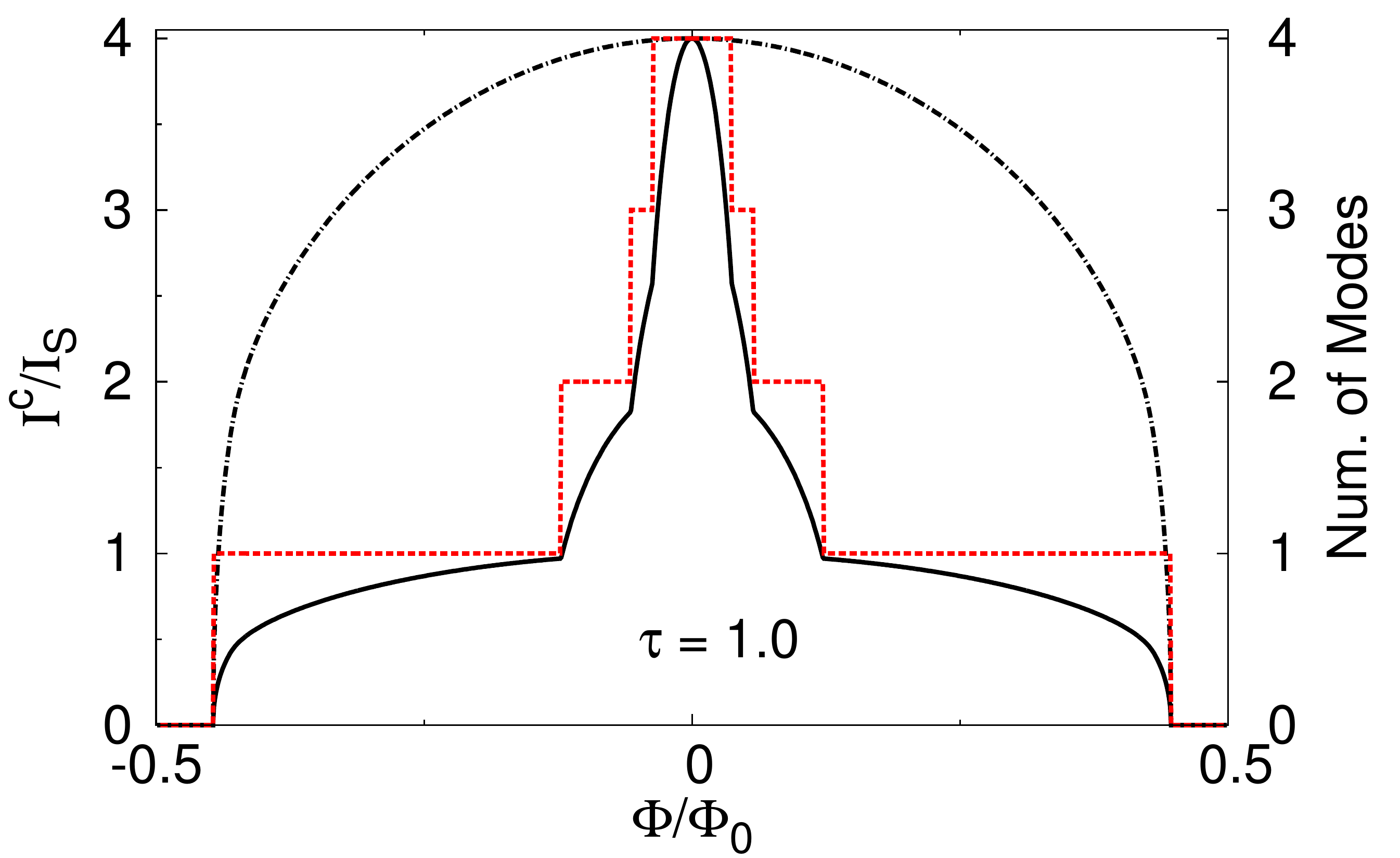}
\includegraphics[width=3.5cm, angle=270]{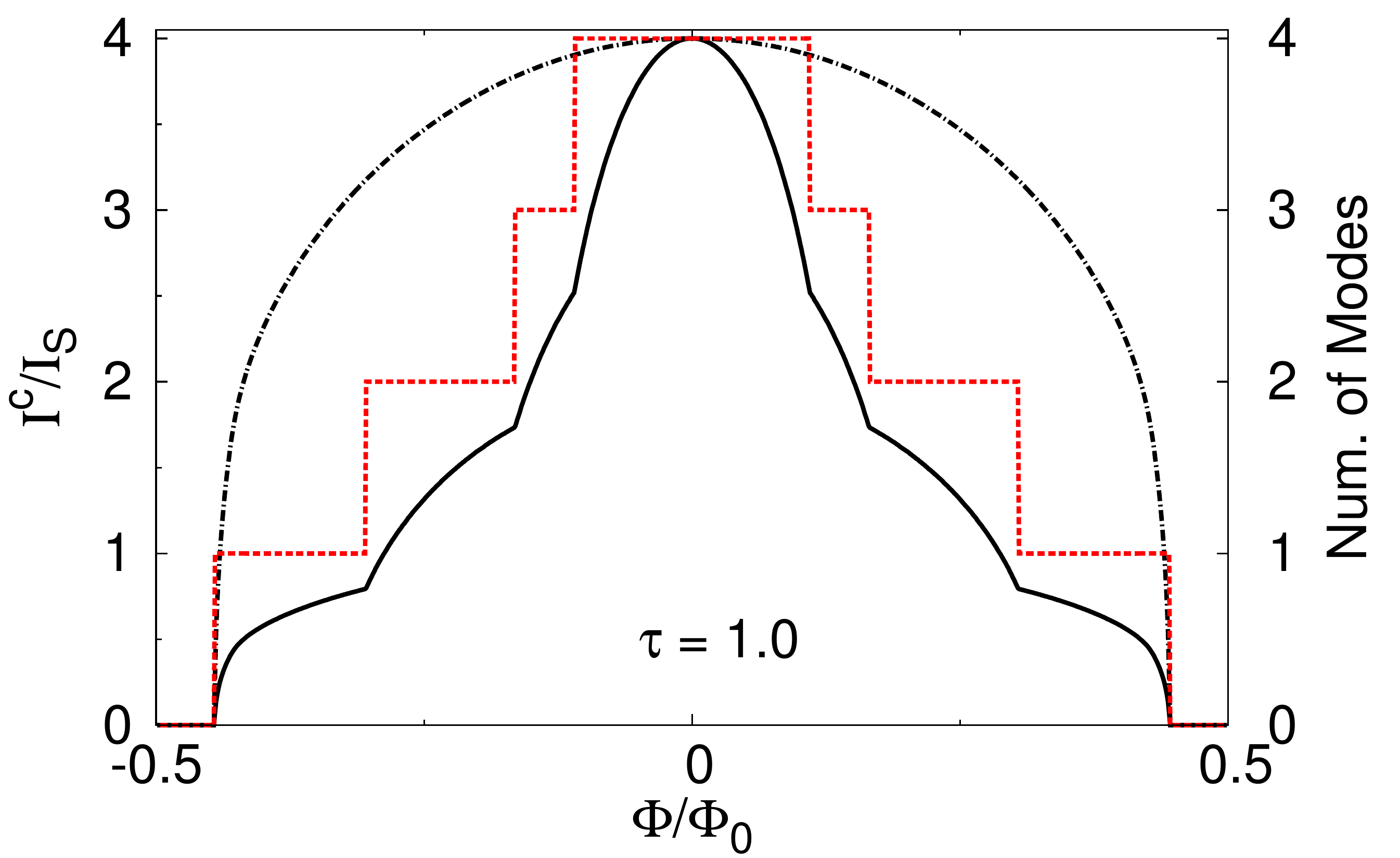}
\includegraphics[width=3.5cm, angle=270]{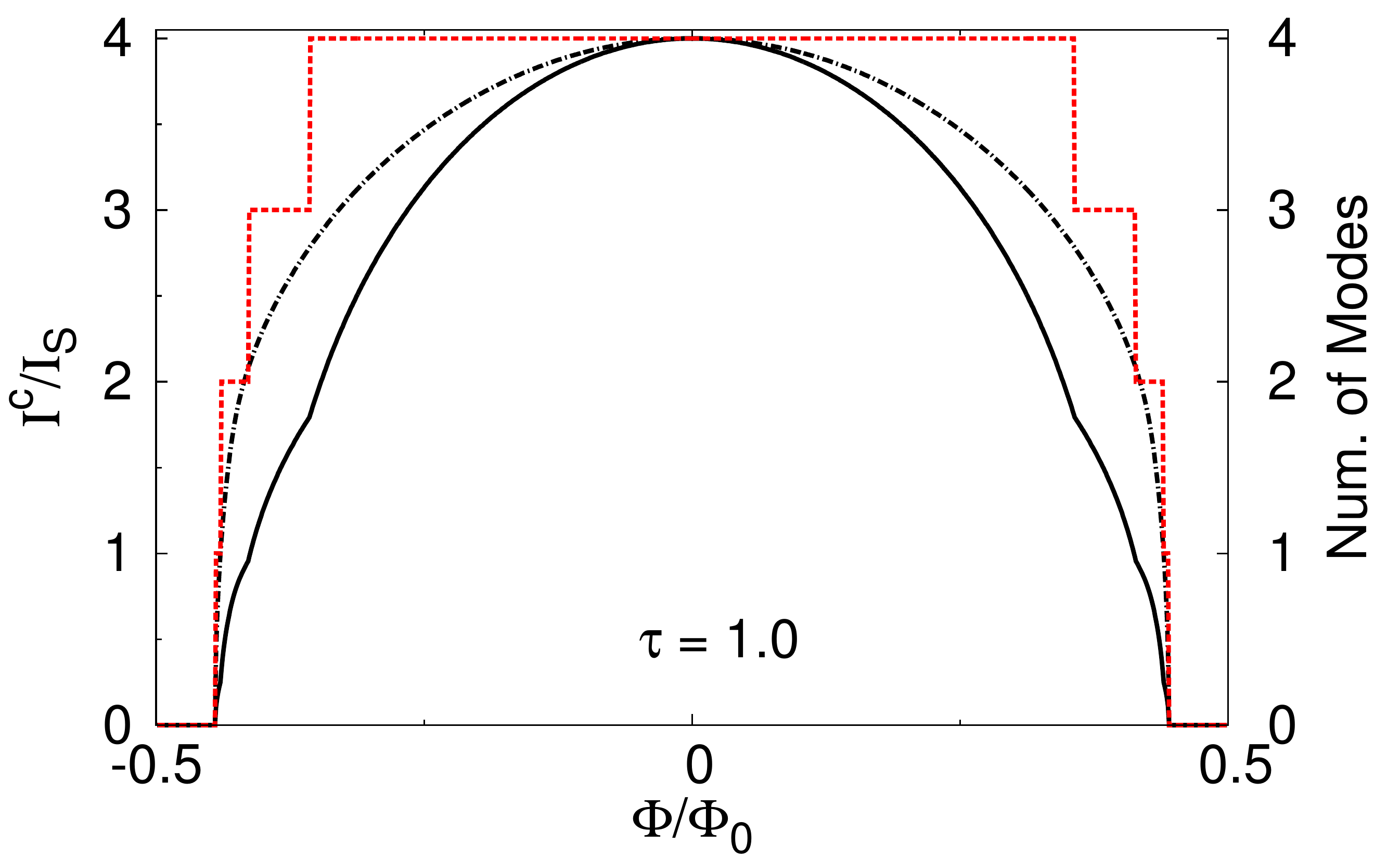}
\includegraphics[width=3.5cm, angle=270]{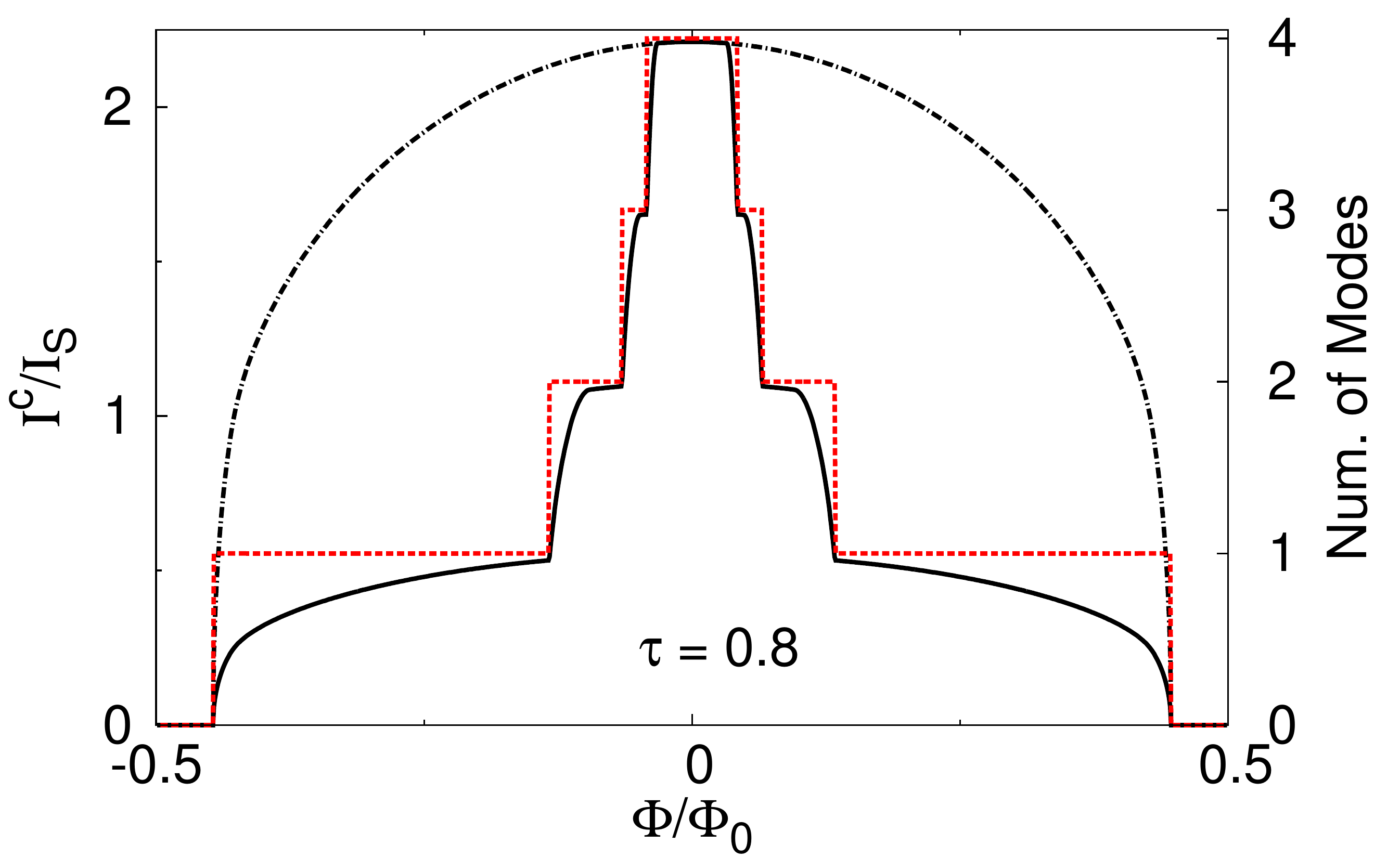}
\includegraphics[width=3.5cm, angle=270]{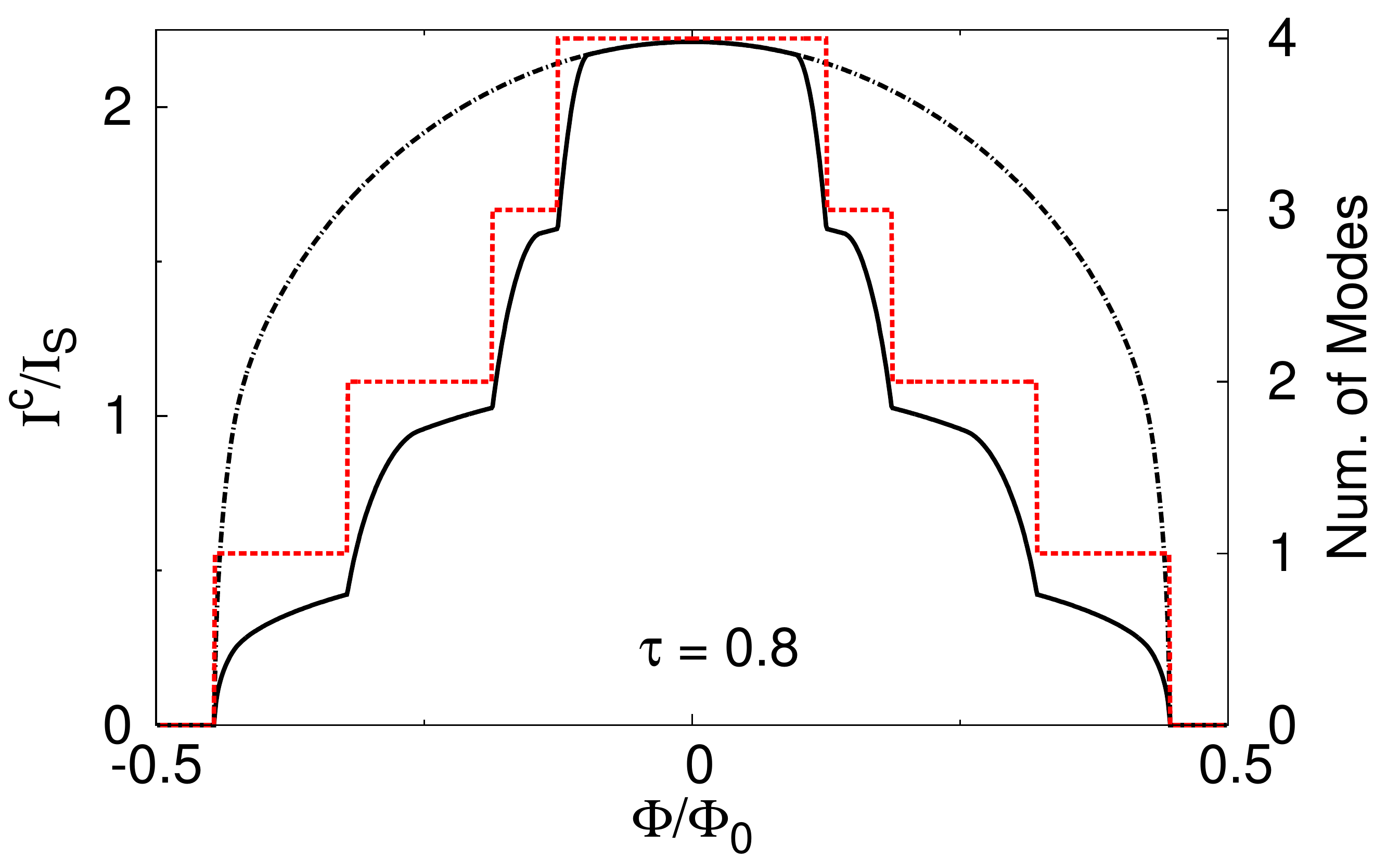}
\includegraphics[width=3.5cm, angle=270]{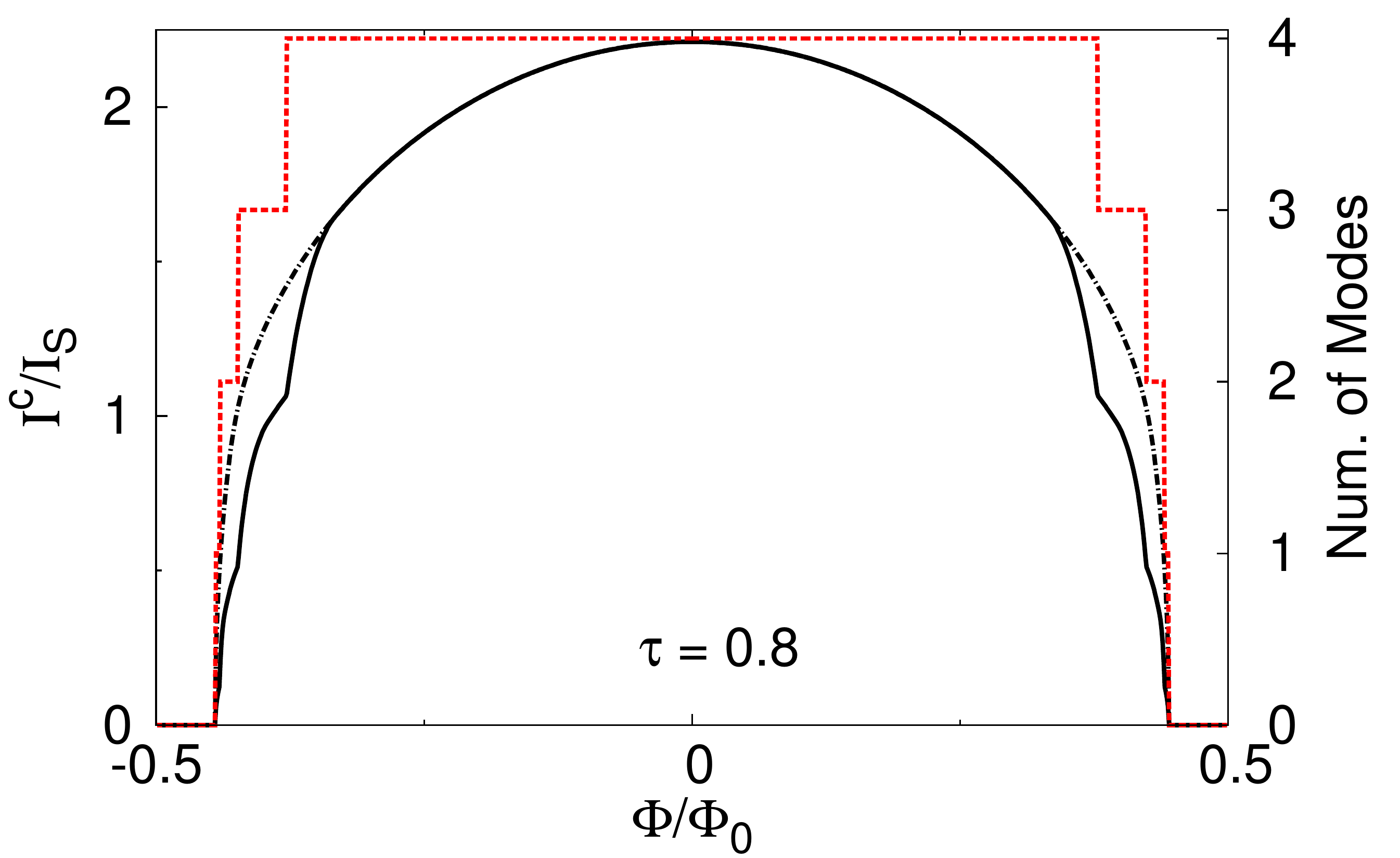}
\includegraphics[width=3.5cm, angle=270]{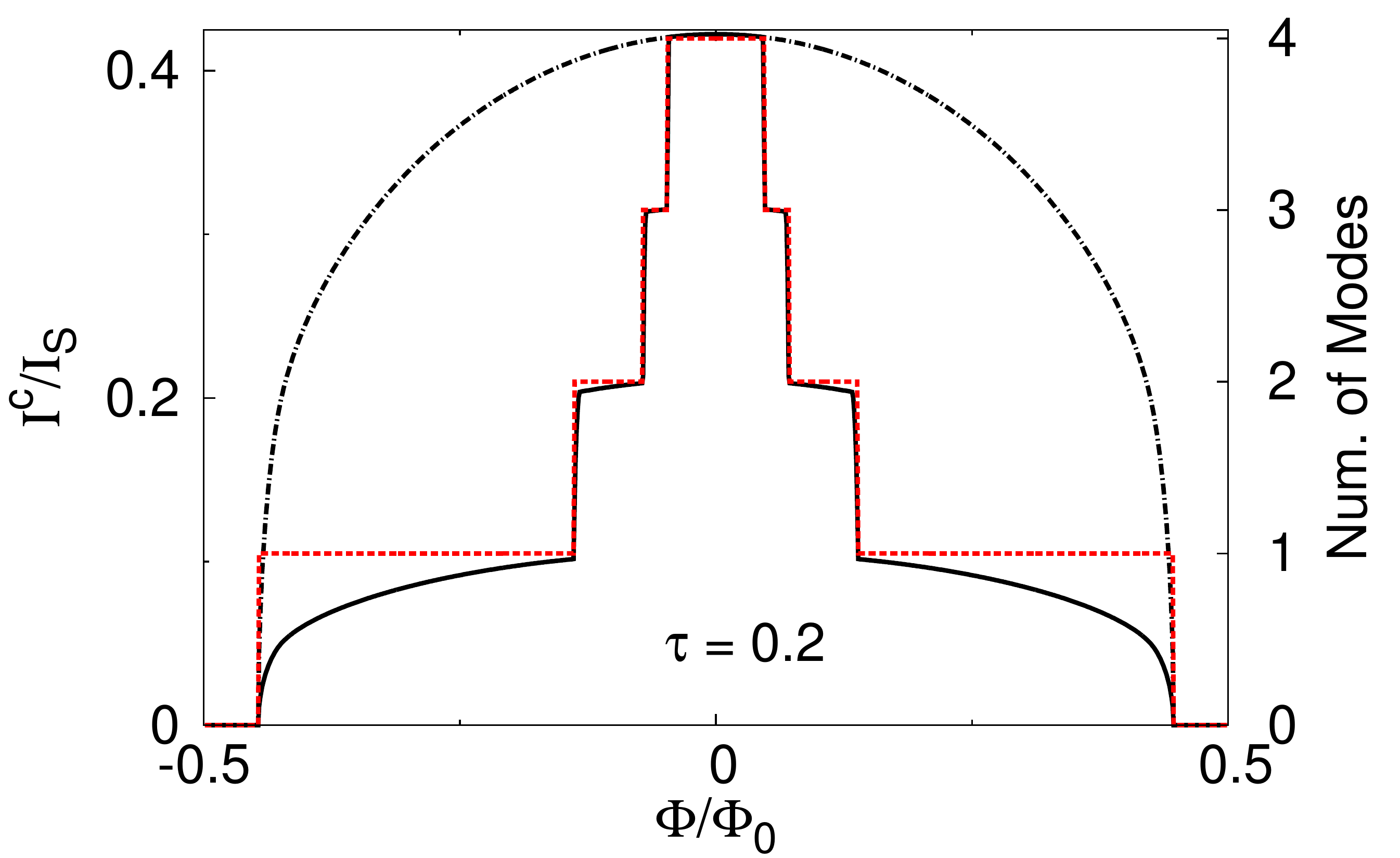}
\includegraphics[width=3.5cm, angle=270]{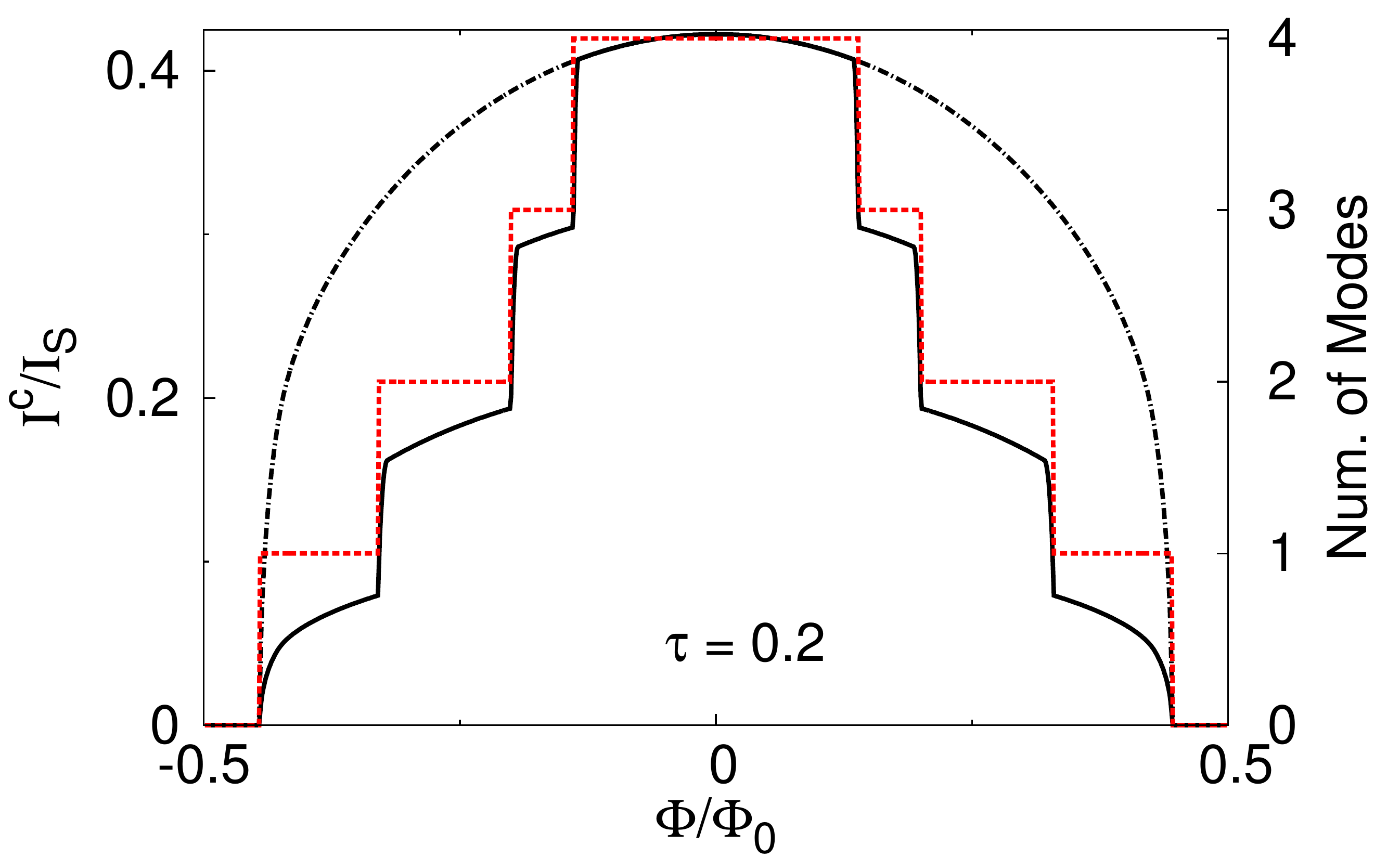}
\includegraphics[width=3.5cm, angle=270]{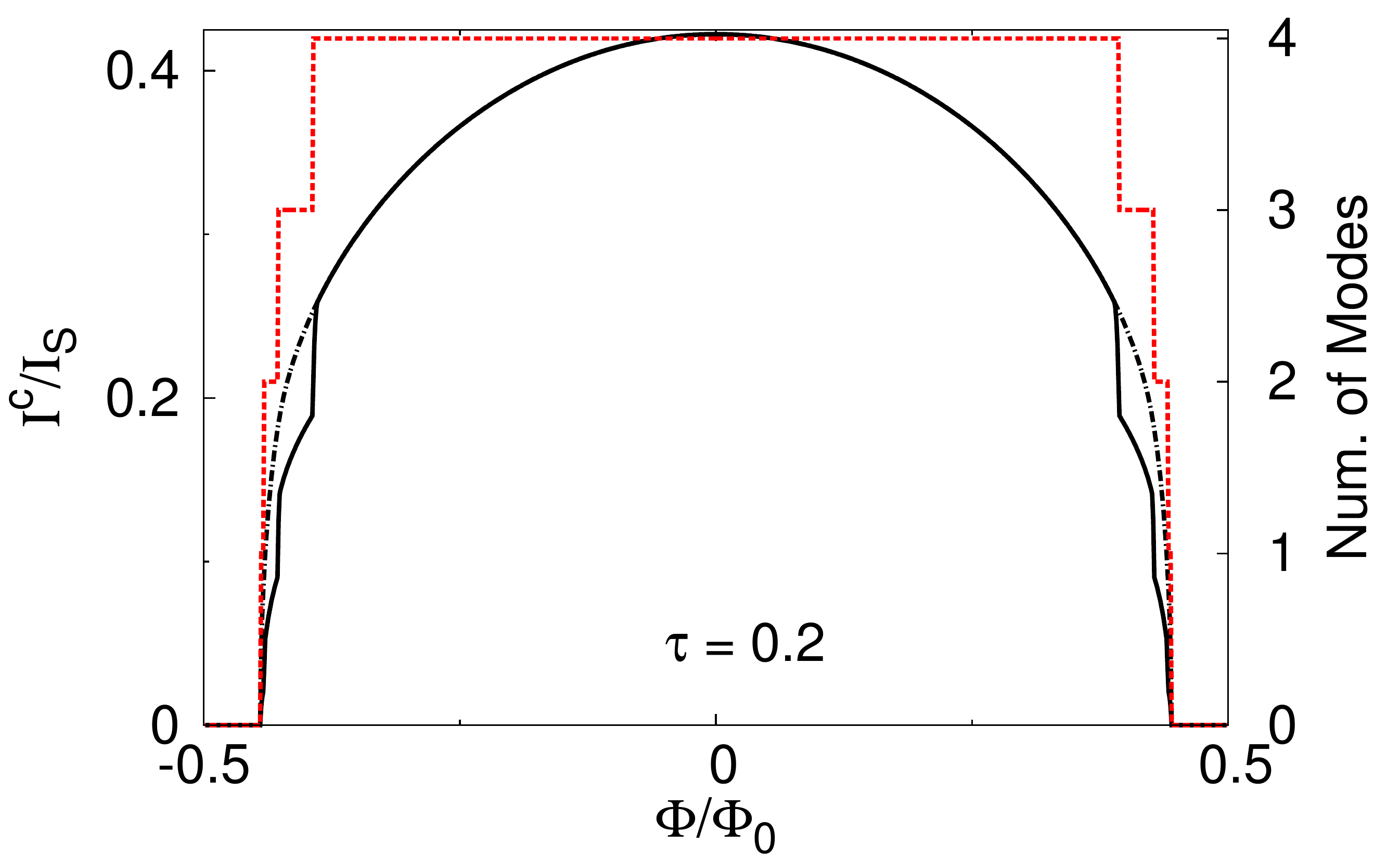}
\caption{Critical current and number of subgap modes (dotted
lines, right axis) derived from simplified model, Eq.~(\ref{ABS}),
with $M=4$, $\tau_k=\tau$, $I_{S}=e\Delta_0/\hbar$. Dashdotted
curves show $I^{\text{c}}(0)\Delta/\Delta_0$. First column:
$R_0=43$ nm, $\xi=80$ nm. Second column: $R_0=75$ nm, $\xi=140$
nm. Third column: $R_0=160$ nm, $\xi=299$ nm. $\Delta$ is
calculated from Eq.~(\ref{LP}) with $d_{\text{sc}}=0$,
$\Delta_0=0.2$ meV.}\label{tautoy}
\end{figure*}

In Fig.~\ref{modetoy} we plot the critical current for different
number of subgap modes $M$. By increasing $M$, extra steps/kink
points are formed. Most importantly, the overall current profile
is qualitatively the same as that derived from the exact BdG
Hamiltonian; see for example Fig.~4 in the main article. For
$M=1$, $w_1=0$ and the current can be described by the formula
$I^{\text{c}}(\Phi) = I^{\text{c}}(0) \Delta(\Phi)/\Delta_0$, in
stark contrast for $M>1$ this formula is inapplicable. In
Fig.~\ref{tautoy} we take $M=4$ and plot typical examples of the
critical current for three different radii $R_0$. For a better
comparison, we adjust the coherence length of the shell $\xi$ so
that for each $R_0$ the pairing potential vanishes when
$|\Phi/\Phi_0| \gtrsim 0.45$. For illustrative reasons, we also
present one case for an unrealistically large radius, $R_0=160$
nm. The purpose is to demonstrate how the size of $w_k$ affects
the overall profile of the current. The role of the terms $w_k$
weakens for larger values of $R_0$. In particular, when the ratio
$1/m^{*}R^{2}_{0}$ becomes vanishingly small the required flux to
induce a kink point lies nearly at the boundaries of the lobe. In
the regime, $w_k\approx0$, the flux dependence of the current can
be accurately described by the formula $I^{\text{c}}(\Phi) =
I^{\text{c}}(0) \Delta(\Phi)/\Delta_0$; when $\tau=1$ this equals
$M e \Delta(\Phi)/\hbar$. Another important observation is that
for smaller values of $w_{k}$ the flux dependence of $\theta_k$ is
weaker. This explains why for a given radius the current steps
formed at larger fluxes are in general broader
[Fig.~\ref{tautoy}].

So far in our analysis we have focused on $\tau_k=\tau$, but our
results can be easily generalized to the most general case when
the transparency, $\tau_k$, of each individual mode is different.
Some analytical expressions for the critical current can again be
derived, however, these are not particularly enlightening. The
important conclusion is that all the basic features presented in
Figs.~\ref{modetoy} and \ref{tautoy} are still observable in the
most general case. Our analysis is also applicable when the
$\varphi_0$-dispersion of the subgap modes is different from that
specified by Eq.~(\ref{ABS}). Numerically calculated subgap modes
derived from the exact BdG Hamiltonian can equally well
demonstrate the physics.

\begin{figure*}
\includegraphics[width=4.0cm, angle=270]{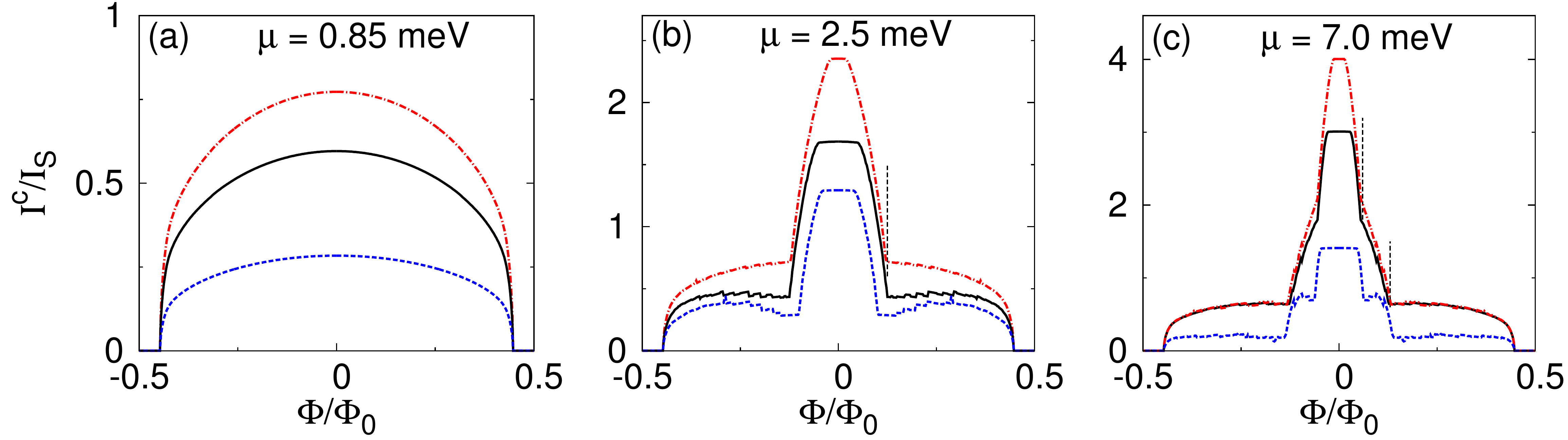}
\includegraphics[width=4.0cm, angle=270]{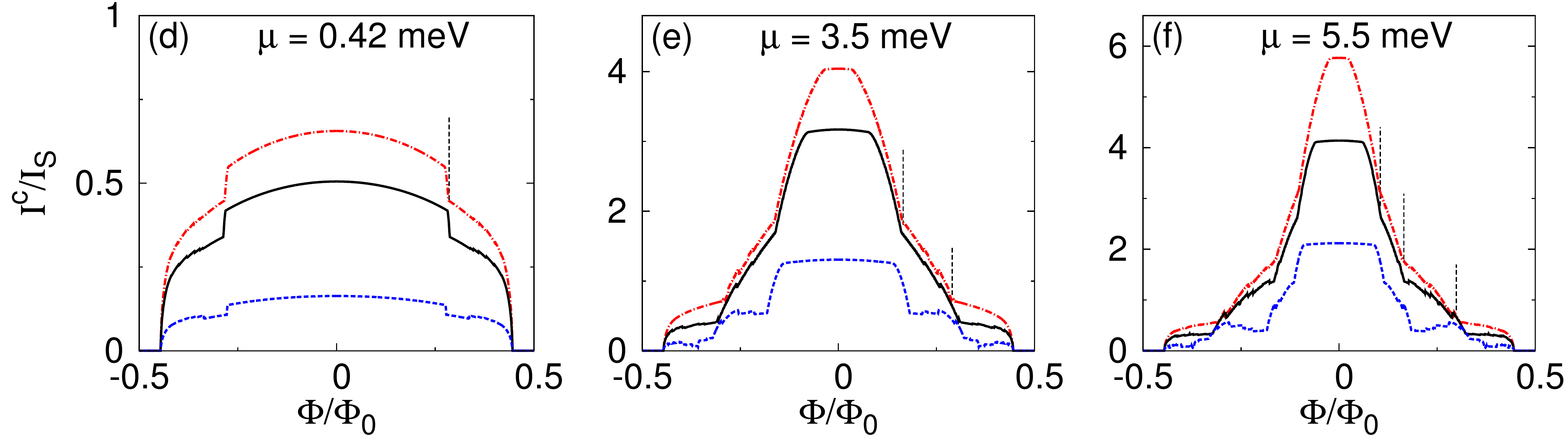}
\caption{Critical current as a function of magnetic flux
calculated with $\Delta$ given by Eq.~(\ref{LP}). From upper to
lower curve: $\tau_0=1$, 0.6, 0.2. Parameters: $L_S=2000$ nm,
$L_N=100$ nm, $\alpha=0$, $\Delta_0=0.2$ meV, $d_{\text{sc}}=0$,
$I_S = e \Delta_0/\hbar$. (a)-(c) $R_0=43$ nm, $\xi=80$ nm.
(d)-(f) $R_0=75$ nm, $\xi=140$ nm. Vertical lines indicate kink
points for $\tau_0=1$. As $\Phi$ increases the number of
nondegenerate subgap modes decreases by one at each kink point.
The latter, in general, shift for different $\tau_0$. At $\Phi=0$
the subgap modes contributing to the current are derived from: (a)
$m_j=1/2$ (b) $m_j=1/2$, 3/2 (c) $m_j=1/2$, 3/2, 5/2 (d)
$m_j=1/2$, 3/2 (e) $m_j=1/2$, 3/2, 5/2 (f) $m_j=1/2$, 3/2, 5/2,
7/2.}\label{tau}
\end{figure*}

\section{Flux tunable current in a reduced transparency junction}

In the main article we consider an ideal SNS junction, namely, a
transparent junction where there is no explicit physical mechanism
to suppress tunnelling between the S and N regions. As introduced
above, a spatially dependent chemical potential controls the
number of subgap levels as well as the current, but this control
is sensitive to the value of $m_j$. In this respect, it is
interesting to explore how the degree of transparency affects the
current when the number of subgap levels remains approximately
constant. This can be done using again Eq.~(\ref{mz}), but now the
parameter $D$ needs to be carefully optimized. This makes the
computational procedure inefficient and time consuming. For this
reason, we model the transparency of the junction
phenomenologically by employing a similar methodology to that
presented originally in Ref.~\onlinecite{Cayao:PRB17}.
Specifically, in the BdG Hamiltonian we introduce a dimensionless
parameter $\tau_0$, with $0< \tau_0 \le 1$, which scales the
kinetic terms along the z direction ($\tau_0 p^2_{z}/2m^{*}$). We
assume this scaling to take place within the N region and an
adjacent small part ($x_{S}\approx$ 20 nm $\ll L_S$) in the S
regions. The transparent limit (main article) corresponds to
$\tau_0=1$, whereas the opposite limit, $\tau_0 \approx 0$, is not
of interest here since the critical current is almost completely
suppressed. Thus, in this work we choose the lower limit to be
$\tau_0=0.2$ which allows us to capture all the essential
characteristics.

For the computations, the kinetic term along z is written as
($\hbar=1$)
\begin{equation}
T_{\text{kin}} = -\frac{1}{2m^{*}}\frac{d }{d z}\left(\tau_{0}(z)
\frac{d y(z)}{d z} \right),
\end{equation}
where $y(z)$ represents any of the four components of the BdG
Hamiltonian. Using centered differences and defining $\tau_0$ on
the mid lattice points the kinetic term is discretized as follows
\begin{equation}
T_{\text{kin}} \approx b^{i-1} y^{i-1} - [ b^{i-1} + b^{i} ] y^{i}
+ b^{i} y^{i+1},
\end{equation}
with
\begin{equation}
b^{i} = - \frac{1}{2 m^{*} \delta^{2}} \tau_{0}^{i+1/2}, \quad
b^{i-1} = - \frac{1}{2 m^{*} \delta^{2}}\tau_{0}^{i-1/2},
\end{equation}
and $\delta$ is the spacing between the lattice points. Here,
$b^{i}$ describes hopping between the lattice points $i$ and $i+1$
while $\tau_{0}^{i+1/2}$ is the value of the transparency between
these two lattice points. When $\tau_0$ is constant the usual
finite-difference approximation to the kinetic term is recovered.

We calculate the critical current using the approximation
described in the main article and show some representative
numerical results in Fig.~\ref{tau}. Note, that as happens with
the current suppression versus $\mu_{\text{pot}}$ in
Fig.~\ref{ImuB}, the current suppression versus $\tau_0$ at a
fixed chemical potential and flux is not always monotonic. As
$\tau_0$ decreases the basic flux tunable features are well-formed
at least for intermediate ($\tau_0 \approx 0.5$) and somewhat
smaller values of $\tau_0$. Our calculations within the exact BdG
Hamiltonian confirm that this behaviour is robust for different
sets of parameters: $R_0$, $\mu$ and $x_S$. They also indicate
that the form of the current steps does not necessarily improve
upon decreasing $\tau_0$. Although, some steps become more
pronounced this cannot be guaranteed to be the general rule. In
our SNS junction with energy levels above the gap contributing to
the current as well as with an explicit $\mu$ (and/or
$\mu_{\text{pot}}$) dependence, the underlying physics is expected
to deviate to some degree from the simplified model. Energy levels
which lie above the gap at $\Phi=0$ tend to make the steps more
`noisy' at $\Phi\ne0$, and the overall noise is sensitive to the
exact values of $\mu$ and $\tau_0$. These effects are not captured
by the simplified model. The quality of the steps is expected to
improve in SNS junctions with shorter N region. Another important
aspect, which might be relevant to experimental studies, is that
steps formed at larger fluxes can be almost completely suppressed
for a relatively large $\tau_0$. This effect can lead to the wrong
conclusion that the current suppression is due to the LP effect. A
rigorous method to disentangle the LP suppression from that caused
by $\tau_0$ deserves further investigation.

\begin{figure*}
\includegraphics[width=4.0cm, angle=270]{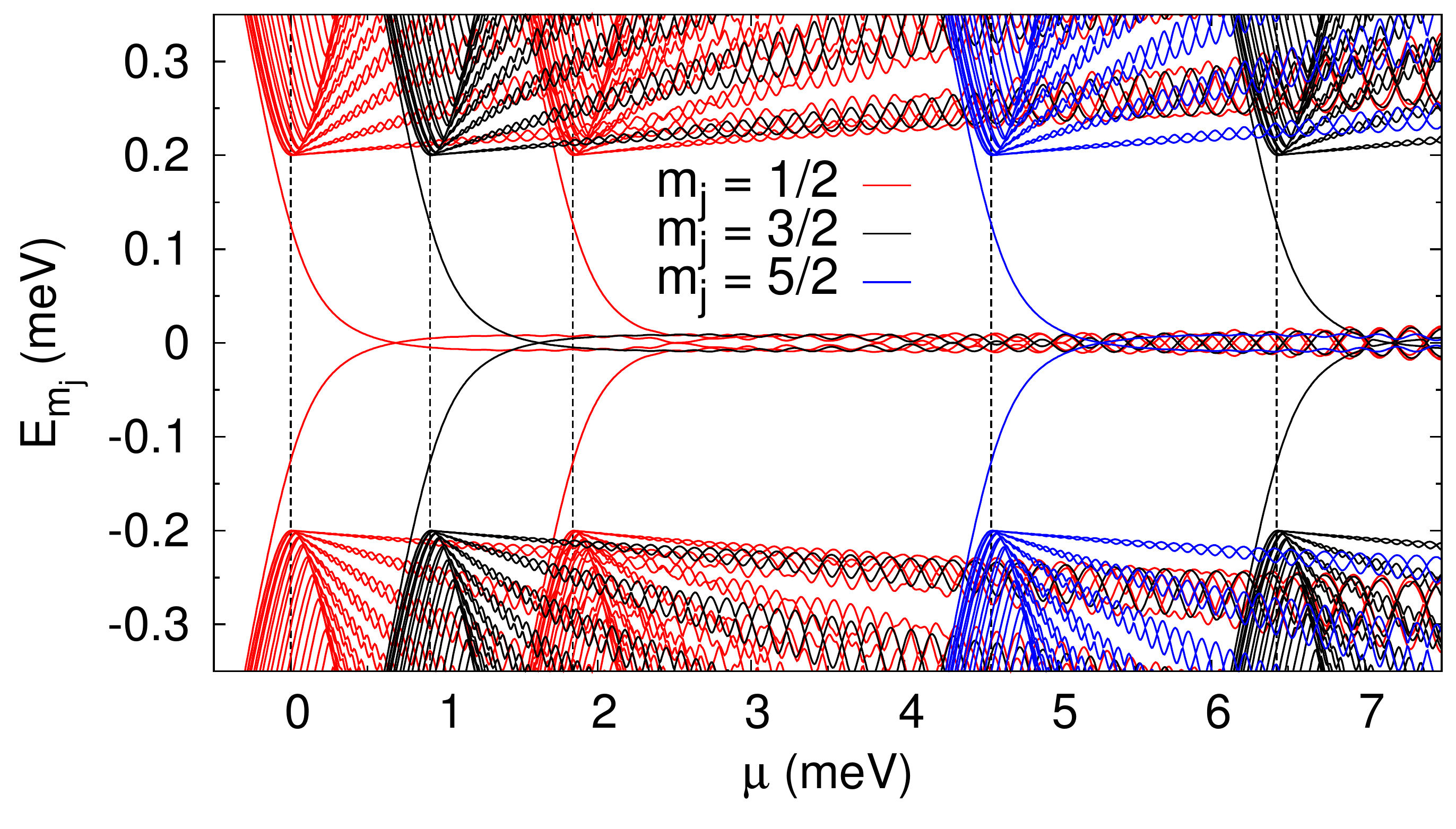}\\
\includegraphics[width=4.0cm, angle=270]{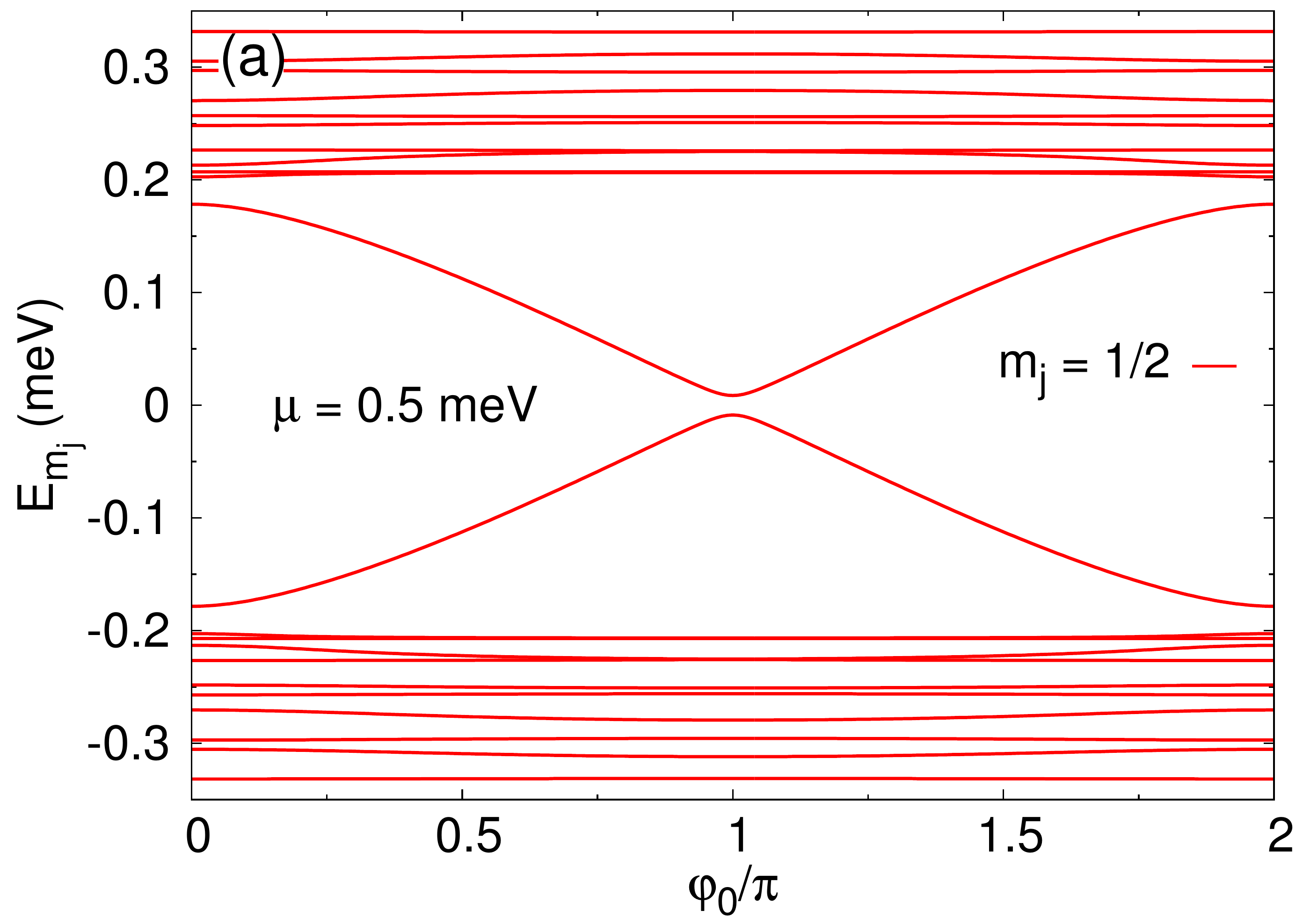}
\includegraphics[width=4.0cm, angle=270]{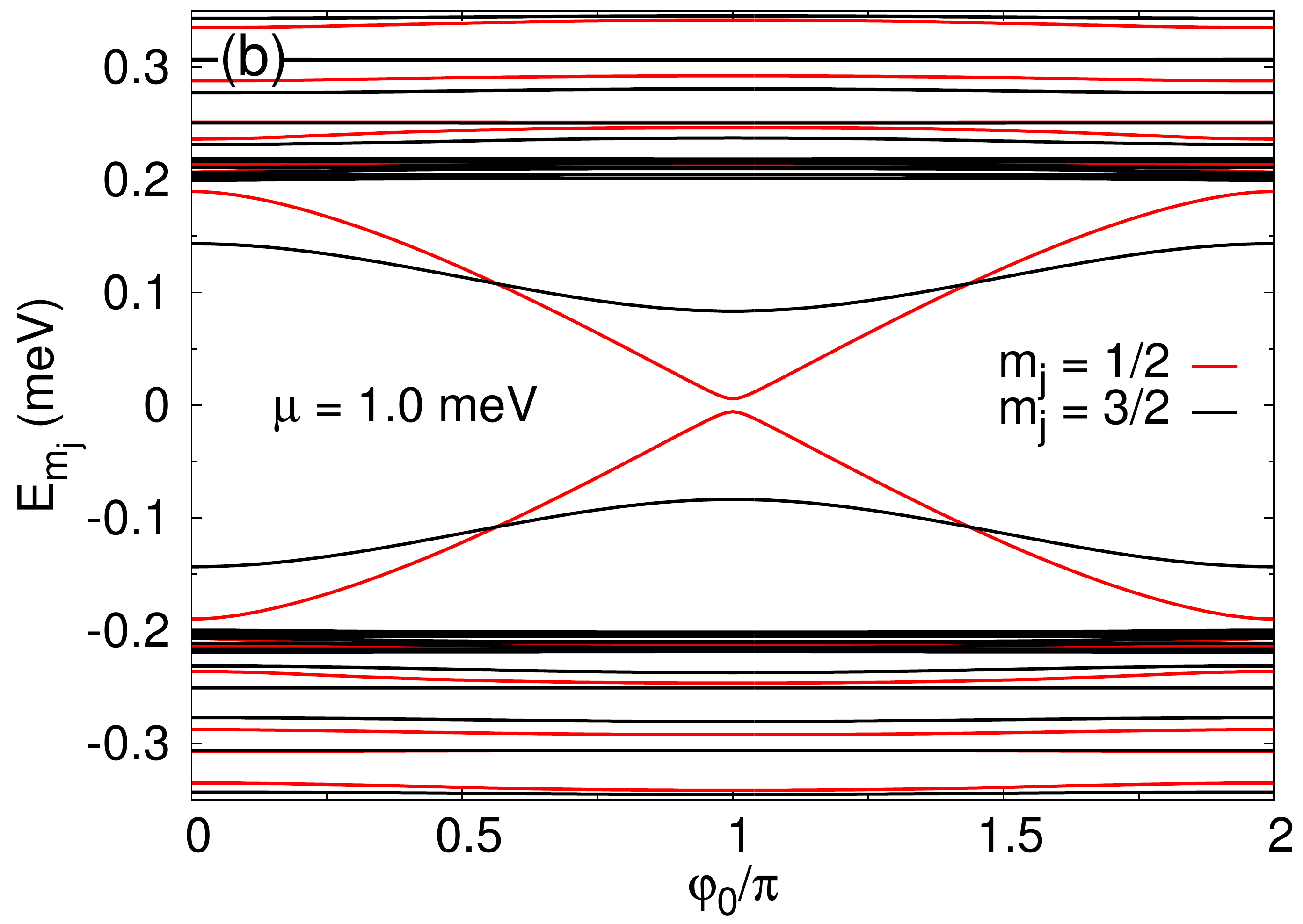}
\includegraphics[width=4.0cm, angle=270]{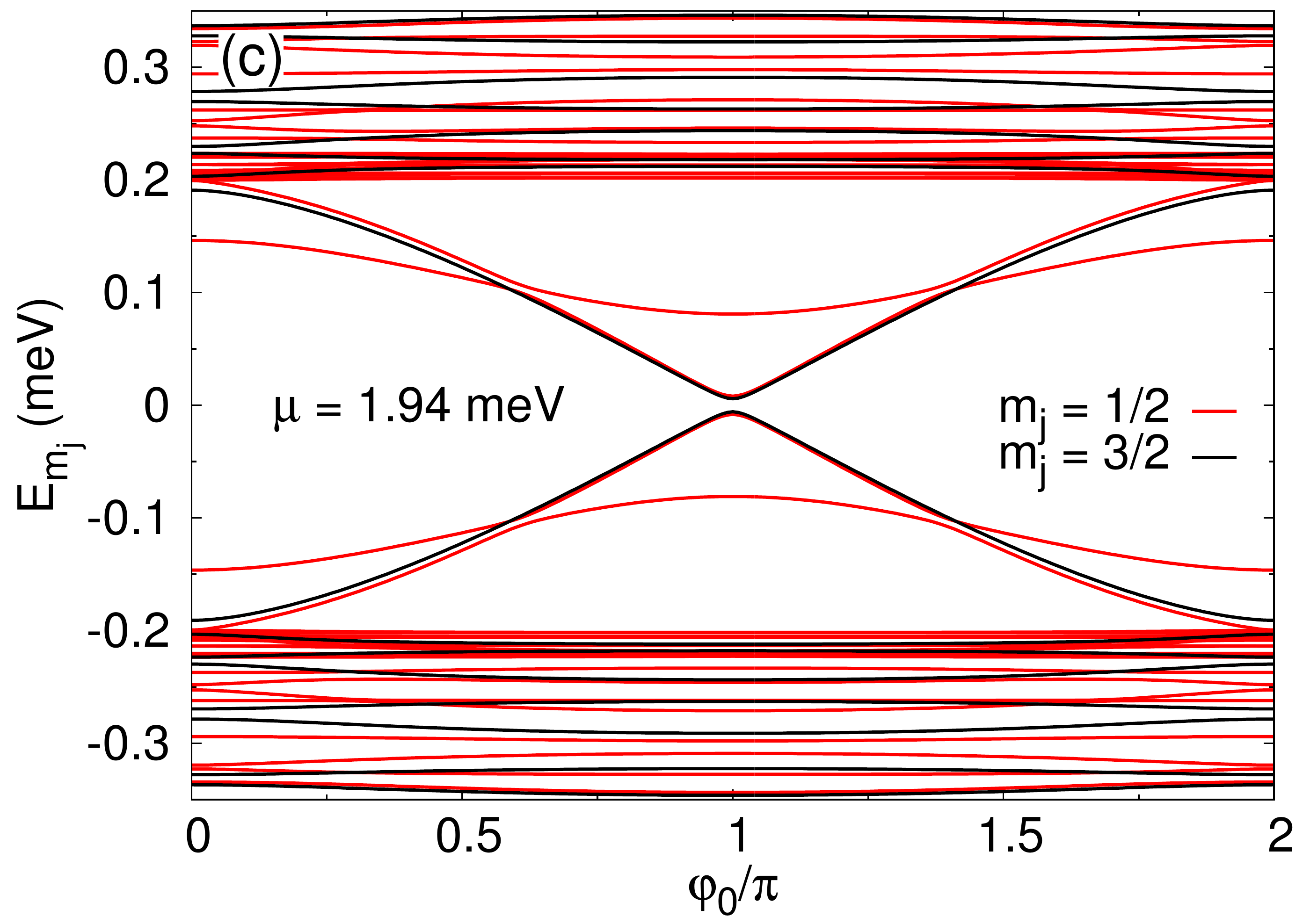}\\
\includegraphics[width=4.0cm, angle=270]{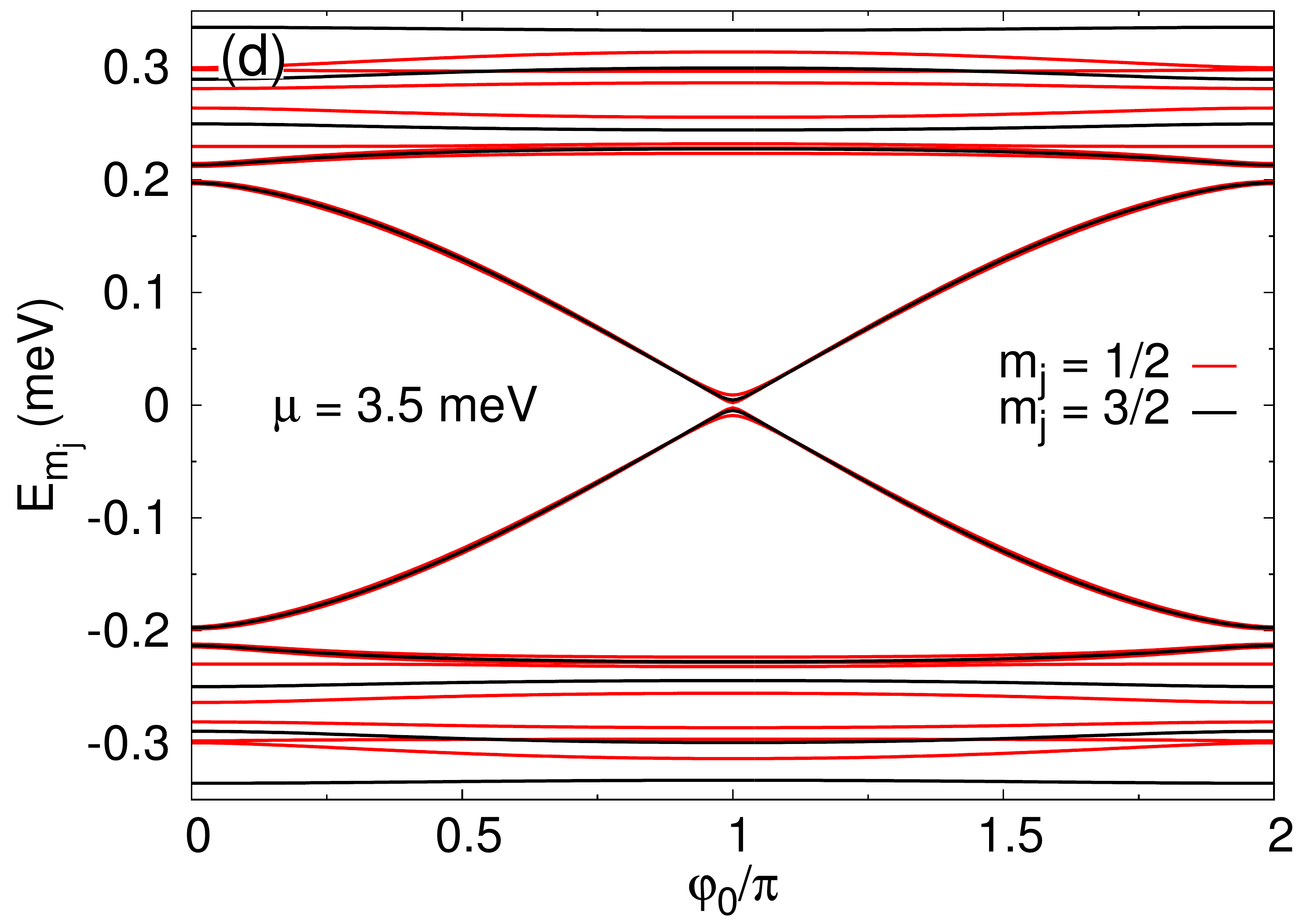}
\includegraphics[width=4.0cm, angle=270]{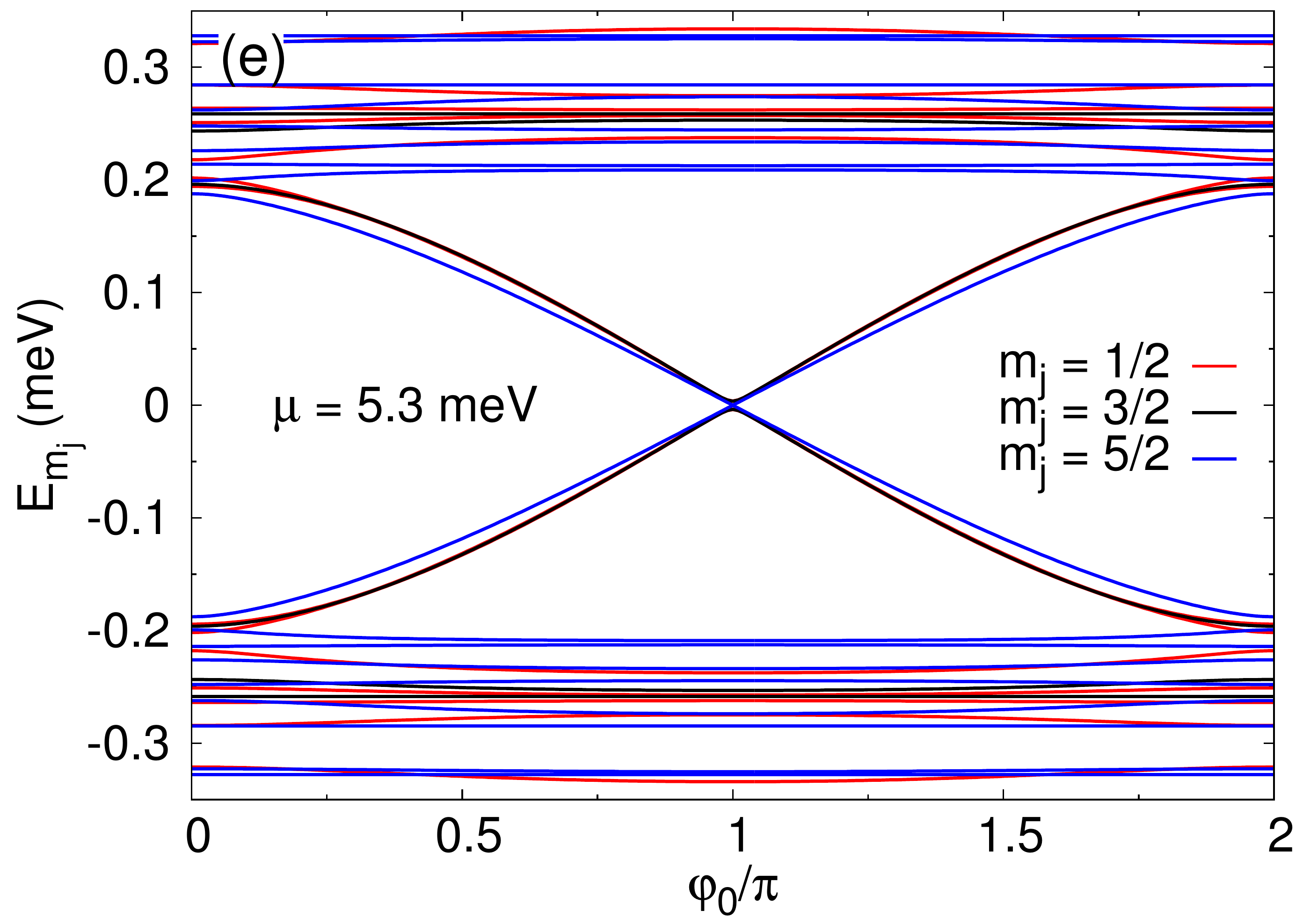}
\includegraphics[width=4.0cm, angle=270]{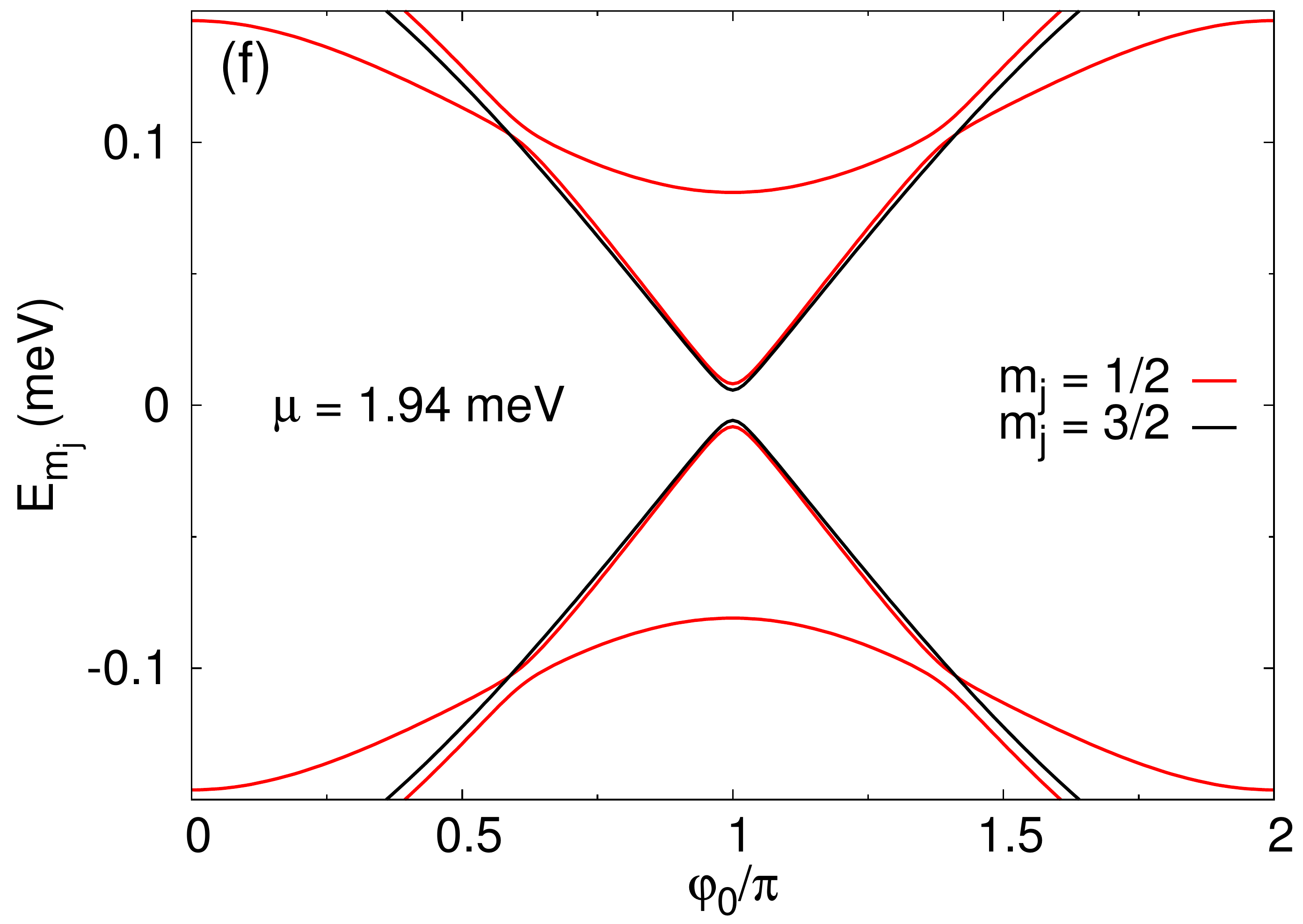}
\caption{Top frame: Zero-flux energy spectrum of SNS junction as a
function of chemical potential at phase difference
$\varphi_0=\pi$. Energies are derived from $H_{\text{BdG}}$
[Eq.~(2) main article] and satisfy $E_{-m_j}=-E_{mj}$. Vertical
lines define the effective potentials $V^{0}_1(m_j) \approx 0$ and
$V^{0}_3(m_j) \approx 0$ for $m_j = 1/2$, 3/2, 5/2. (a)-(e)
Energies as a function of phase difference at different chemical
potentials. Only energies lying within the chosen energy range are
plotted. (f) Zoom in of (c) showing anticrossing point for
$m_j=1/2$ due to the SO coupling. Parameters: $L_S=2000$ nm,
$L_N=100$ nm, $R_0=43$ nm, $\alpha=20$ meV nm,
$\Delta=\Delta_0=0.2$ meV.}\label{somu}
\end{figure*}

\begin{figure*}
\includegraphics[width=4.0cm, angle=270]{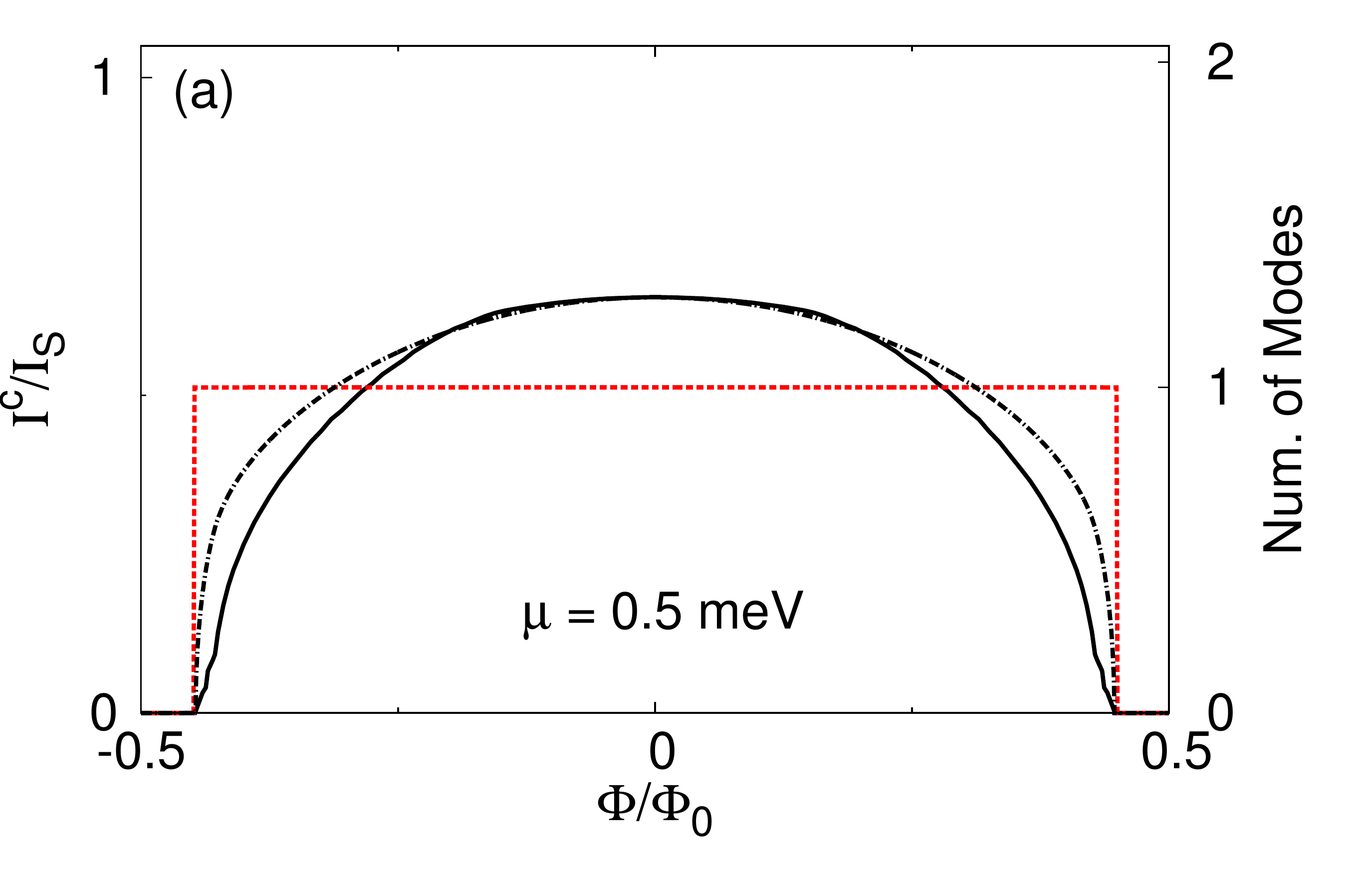}
\includegraphics[width=4.0cm, angle=270]{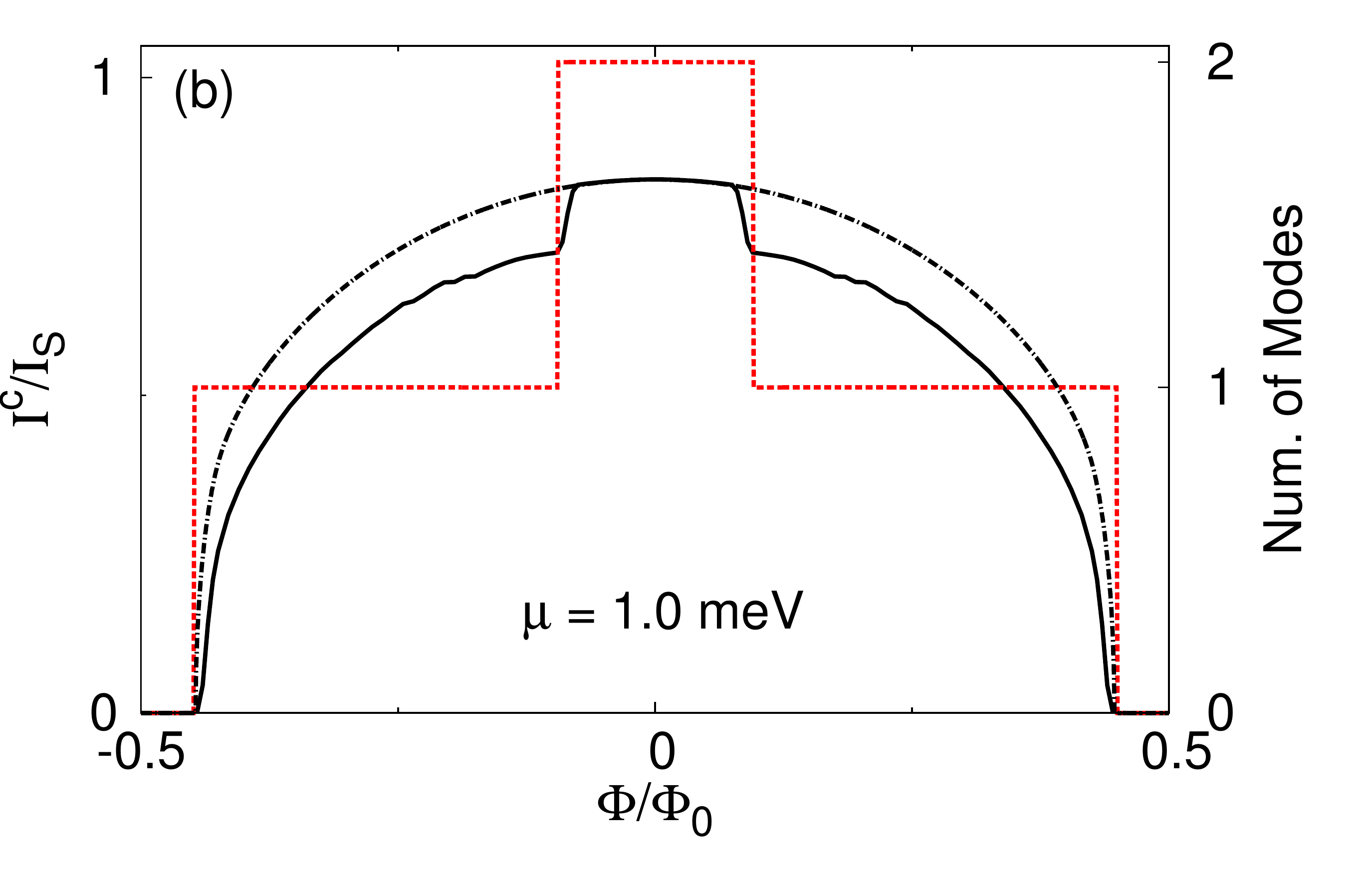}\\
\includegraphics[width=4.0cm, angle=270]{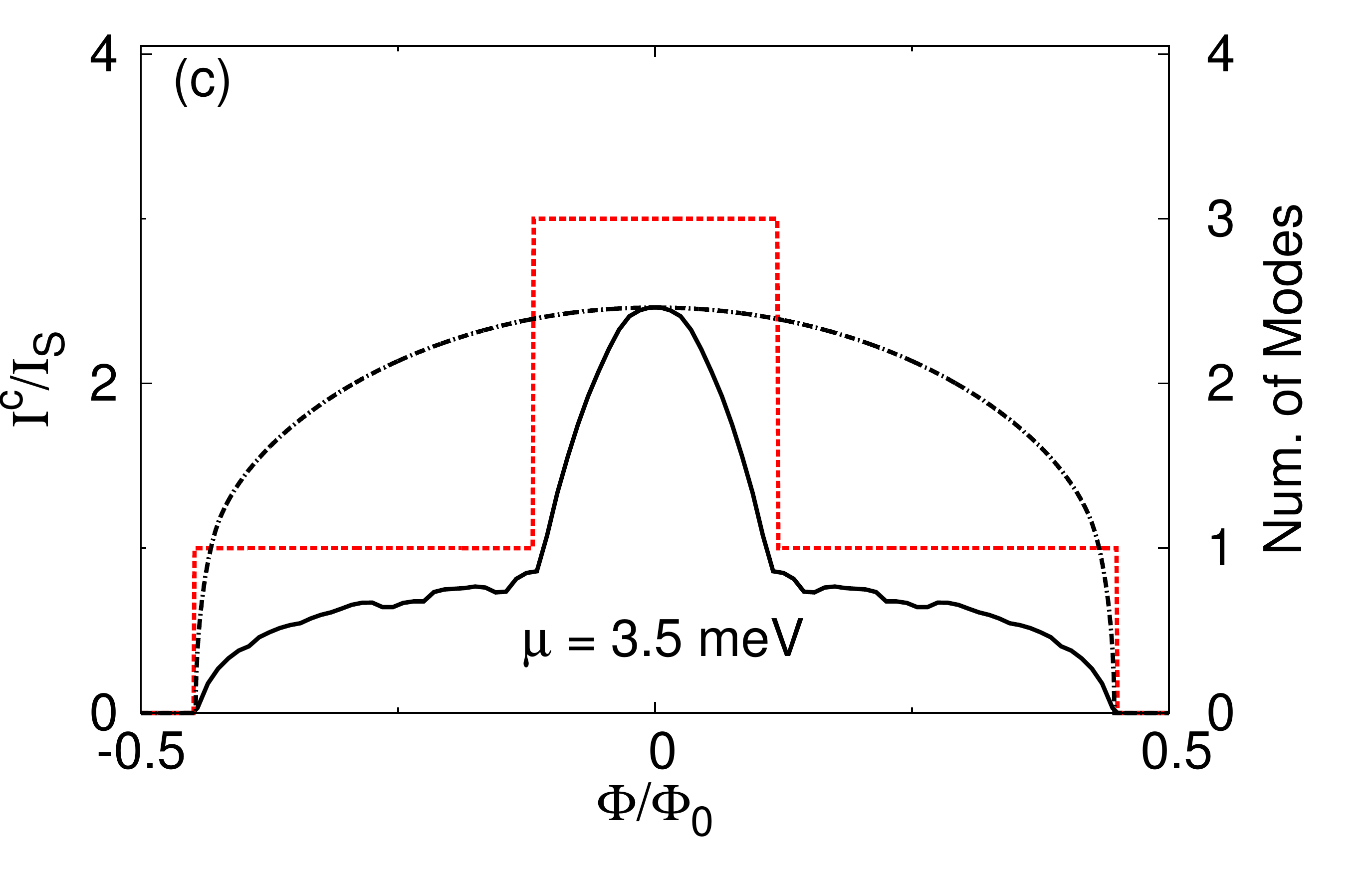}
\includegraphics[width=4.0cm, angle=270]{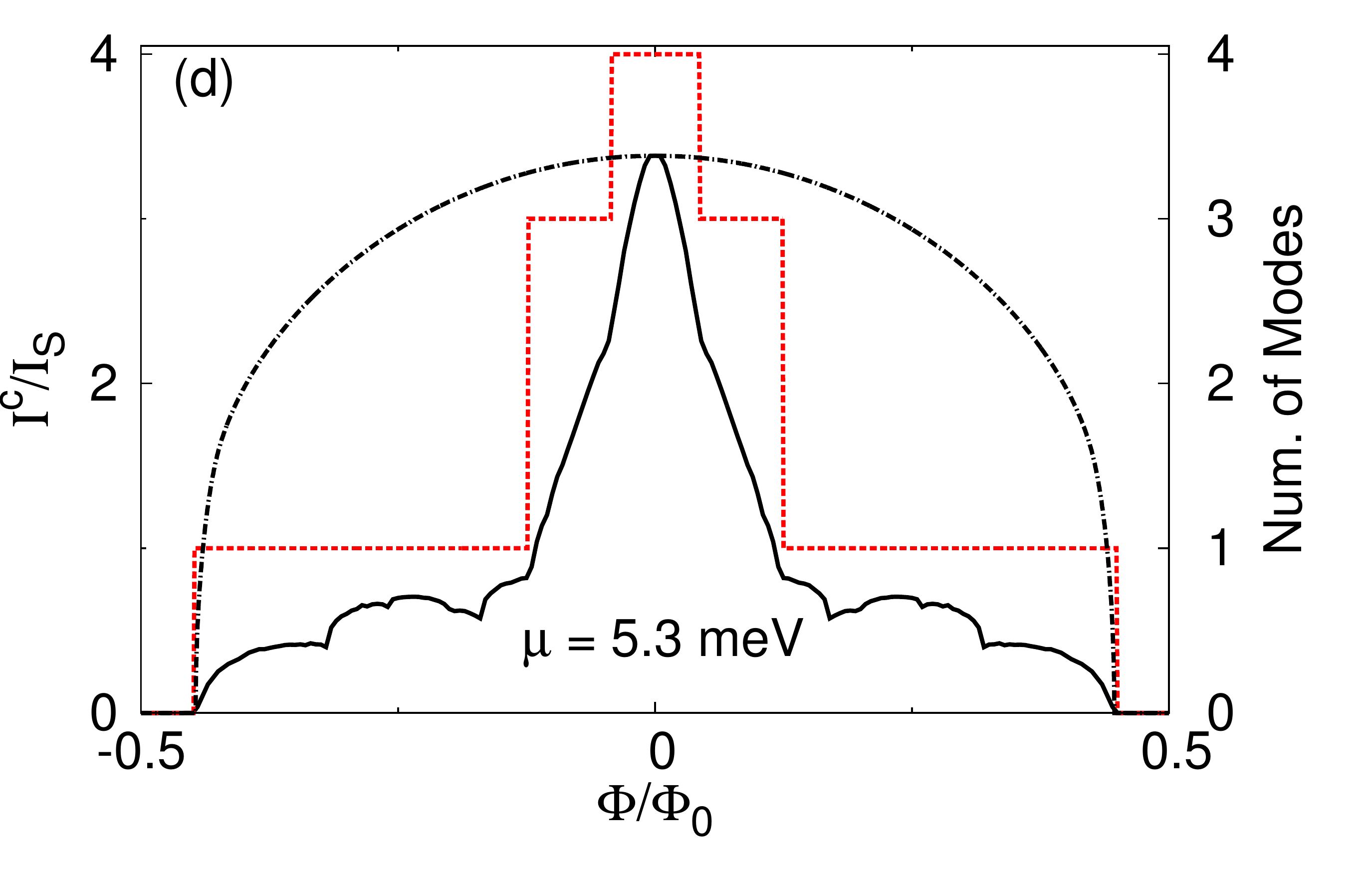}
\caption{Solid curves show exact critical current,
$I^\text{c}(\Phi)$, calculated with $\Delta$ given by
Eq.~(\ref{LP}). Dashdotted curves show
$I^\text{c}(0)\Delta/\Delta_0$. Dotted lines show the number of
subgap modes (right axis). Parameters: $L_S=2000$ nm, $L_N=100$
nm, $R_0=43$ nm, $\alpha=20$ meV nm, $\Delta_0=0.2$ meV, $\xi=80$
nm, $d_{\text{sc}}=0$, $I_S = e \Delta_0/\hbar$. At $\Phi=0$ the
subgap modes contributing to the current are derived from: (a)
$m_j=1/2$ (b) $m_j=1/2$, 3/2 (c) $m_j=1/2$, 3/2 (d) $m_j=1/2$,
3/2, 5/2.}\label{soflux}
\end{figure*}

\section{SNS junction with spin-orbit coupling}

In this section we examine the effect of the Rashba spin-orbit
(SO) coupling on the flux dependence of the critical current. Our
aim is to demonstrate that the current profile presented in Fig.~4
of the main article can still be observed in the presence of weak
SO coupling. Analyzing in detail SO effects is beyond the scope of
this work.

We consider the chemical potential, $\mu$, to be constant along
the SNS junction, thus, as explained above we focus on the
resonant case where the critical current is maximum. A nonuniform
$\mu$ simply results in a reduced current. A nonzero SO coupling,
$\alpha$, shifts further apart the values of $\mu$ at which the
energy levels of $H_A(m_j)$ and $H_B(m_j)$ respectively enter the
superconducting gap. This shift is of the order of
$2m_j\alpha/R_0$ as can be understood directly from Eqs.~(4) in
the main article. In addition, when $\alpha \ne 0$ the energies of
$H_A(m_j+1)$ and $H_B(m_j)$ are no longer degenerate, since
$V^{0}_{1}(m_j+1)\ne V^{0}_{3}(m_j)$, and when the parameters are
tuned so that $(\hbar=1)$
\begin{equation}
\alpha \approx \frac{2m_j-1}{4m^{*}R_{0}},
\end{equation}
the condition $V^{0}_1(m_j) \approx -\mu$ can be satisfied for
$m_j > 1/2$ (assuming $\Phi=0$). In this regime and at low
chemical potentials the critical current is no longer dominated by
$m_j=1/2$ only, therefore, an enhanced critical current can be
observed compared to that for $\alpha=0$. A subtle point is that
this enhancement is not due to the actual coupling between
$H_A(m_j)$ and $H_B(m_j)$ caused by
$H^z_{\text{SO}}=-\alpha\partial_z\tilde{\tau}_z$, but to the
rearrangement of the potentials terms $V^{0}_1(m_j)$ and
$V^{0}_3(m_j)$.

For the numerical calculations, we consider a realistic value for
the SO coupling in the relatively weak regime, $\alpha=20$ meV nm,
and assume that $\alpha$ is constant along the SNS junction. This
can be considered as a first approximation, since $\alpha$ may
have a spatial dependence and/or be anisotropic, for example, due
to local electric fields induced by gate electrodes. Additionally
the sign of $\alpha$ is in general unknown. All these effects
should depend on the details of the SNS junction, however, small
deviations from a constant SO coupling are not expected to change
the flux dependence of the critical current studied here. Our
numerical calculations confirm this argument when $\alpha$ is
assumed to be different in the N and S regions.

The zero-flux energies of the BdG Hamiltonian [Eq.~(2) main
article] are plotted in Fig.~\ref{somu}. For each $m_j$ we can
identify the approximate value of $\mu$ that shifts an energy
level, originally belonging to $H_A(m_j)$ or $H_B(m_j)$, in the
superconducting gap by setting $V^{0}_1(m_j)\approx 0$ or
$V^{0}_3(m_j)\approx 0$ respectively. The energies as a function
of the superconducting phase exhibit similar overall
characteristics to $\alpha=0$. However, an important difference is
the formation of anticrossing points between the energy levels of
$H_A(m_j)$ and $H_B(m_j)$ [Fig.~\ref{somu}(c) and (f)] as a result
of the SO Hamiltonian
$H^z_{\text{SO}}=-\alpha\partial_z\tilde{\tau}_z$. The
anticrossing point is formed at a phase (in general $\varphi_0\ne
\pi$) which is sensitive to the chemical potential and the same
sensitivity is observed for the corresponding value of the
anticrossing gap.

In Fig.~\ref{soflux} we present the critical current, derived from
the BdG Hamiltonian, as a function of the magnetic flux for
various chemical potentials. The basic characteristics are the
same as in Fig.~4 in the main article for $\alpha=0$. At a small
potential ($\mu=0.5$ meV), and to a very good approximation, only
$m_j=1/2$ is relevant contributing a single subgap mode, and the
usual formula $I^\text{c}(0)\Delta/\Delta_0$ is in good agreement
with the exact current. In contrast, this formula is no longer
valid for large values of $\mu$ when extra subgap modes contribute
to the current. The SO coupling modifies the flux dependence of
the effective potentials, $\delta^{\pm}_{m_j}$, [Eq.~(5) main
article] by introducing an additional shift $\pm \alpha/2R_0$; for
$\alpha=20$ meV nm this shift is small especially for larger $m_j$
modes. Therefore, within a simplified approach a finite flux
shifts the $\Phi=0$ subgap modes outside the superconducting gap
in a similar way to the $\alpha=0$ case. An exception occurs for
the subgap mode belonging to $H_A(m_j=1/2)$, for which
$\delta^{+}_{1/2}\ne0$ provided $\alpha\ne0$, but, numerical
calculations in the range of parameters considered here do not
indicate any significant differences in the current from
$\alpha=0$. The regime where only $H_A(m_j=1/2)$ is relevant is
the simplest one to probe the SO coupling; large values of
$\alpha$ should induce kink points well within the lobe and the
resulting flux dependence of $I^\text{c}$ should deviate from that
of $\Delta$. For the proper $\mu$ and $\alpha$, when both $H_A$
and $H_B$ are relevant, the SO-induced anticrossings are expected
to add some new features to the flux dependence of the current
(rather small dips can be seen in Fig.~\ref{soflux}(d) in the
single mode regime), however, this investigation is not pursued in
this work.

\bibliography{biblio}













\end{document}